\documentclass[]{rmaa}
\usepackage[dvips]{epsfig}
\usepackage{paralist}
\usepackage[latin1]{inputenc}
\def\be{$b$~}
\def\re{$r$~}
\def\vel{$v$~}

\def\ce{$c$~}

\title{Simulations of un-equal binary collisions embedded in a turbulent gas cloud far away
of a massive gravitational center$^*$}

\altaffiltext{1}{Departamento de Investigaci\'on en F\'{\i}sica\\
Universidad de Sonora. MEXICO.}

\author{Guillermo Arreaga-Garc\'{\i}a\altaffilmark{1}}

\fulladdresses{\item Guillermo Arreaga-Garc\'{\i}a: Departamento de Investigaci\'on en F\'{\i}sica\\
Universidad de Sonora. \\
Apdo. Postal 14740, C.P. 83000, Hermosillo, Sonora, Mexico.\\
guillermo.arreaga@unison.mx}

\shortauthor{Arreaga-Garc\'{\i}a}
\shorttitle{\it Simulations of un-equal binary collisions } 

\resumen{Simulamos el colapso de una nube turbulenta de gas usando el c\'odigo basado en particulas Gadget2.
Escogemos dos subnubes, formadas por aquellas part\'{\i}culas localizadas alrededor de los centros 
$\vec{r}_L$ and $\vec{r}_R$ y entre de los radios $r_L$ and $r_R$, respectivamente. Una velocidad 
traslacional $\vec{v}_L$ and $\vec{v}_R$ se agrega, tal que las subnubes se mueven una 
hacia la otra para chocar. El radio y la velocidad
pre-colision de las subnubes se escogen desiguales y tanto colisiones frontales como oblicuas se consideran.
Las simulaciones son todas calibradas para tener la misma masa total y la razon de energ\'{\i}a inicial
$\alpha=0.16$, la cual se define como la raz\'on de la energ\'{\i}a t\'ermica a la energ\'{\i}a gravitacional. 
Comparamos modelos de baja-$\beta$ con modelos de alta-$\beta$, en donde $\beta$ se define como la raz\'on de la 
energ\'{\i}a cin\'etica a la energ\'{\i}a gravitacional. Finalmente, consideramos que la nube turbulenta esta bajo 
la influencia gravitacional de un objeto localizado suficientemente lejos, con 
el prop\'osito de aproximar los efectos de marea por medio de una velocidad azimutal $V_{\rm cir}$ agregada a las 
part\'{\i}culas de la nube adicionalmente a las velocidades traslacional y turbulenta mencionadas arriba. Comparamos 
un modelo de baja-$V_{\rm cir}$ con uno de alta-$V_{\rm cir}$.}

\abstract{We simulate the collapse of a turbulent gas cloud with the particle-based code
Gadget2. We choose two sub-clouds, formed by those particles located around the
centers $\vec{r}_L$ and $\vec{r}_R$ and within the radii $r_L$ and $r_R$, respectively. 
A translational velocity $\vec{v}_L$ or $\vec{v}_R$ is added, so that the sub-clouds move 
towards each other to collide. The radius
and pre-collision velocity of the sub-clouds are chosen to be un-equal, and both head-on and
oblique collisions are considered. The simulations are all calibrated to have the same 
total mass and the initial energy ratio $\alpha=0.16$, which is defined 
as the ratio of thermal energy to gravitational energy. We compare low-$\beta$ models 
to a high-$\beta$ models, where $\beta$ is defined as 
the ratio of kinetic energy to gravitational energy. Finally, we consider the turbulent
cloud to be under the gravitational influence of an object located
far enough, in order to approximate the tidal effects by means of an azimuthal velocity $V_{\rm cir}$
added to the cloud particles in addition to the translational and turbulent velocities mentioned
above. We compare a low-$V_{\rm cir}$ model with a high-$V_{\rm cir}$ one.}

\addkeyword{--stars: formation} 
\addkeyword{--physical processes: gravitational collapse; hydrodynamics} 
\addkeyword{--methods: numerical}

\begin{document}

\maketitle

%%%%%%%%%%%%%%%%%%%%%%%%%%%%%%%%%%%%%%%%%%%%%%%%%%%%%%%%%%%%%%%%%%%%%%%%%%%%%%%%%%%%%%%%%%%%%%%%%%%%%%%%%%%%%%%% 
\section{Introduction}
\label{sec:int}

There is ample observational evidence of the occurrence of cloud-cloud
collisions, see \citet{testi}, \citet{churchwell}, \citet{furukawa}, \citet{torii}, \citet{takahira}
and \citet{yamada}. Regions such as RCW49, Westerlund2, and NGC 3603
are examples of cloud-cloud collisions. Collision between clouds are widely
expected, because clouds moving at random directions have been observed in the plane of the
Milky Way, see \citet{roslowsky} and \citet{bolatto}. As a star formation mechanism, 
some observations indicate that cloud-cloud collisions are the external agent to trigger
the initial gas condensation at the interface of the colliding clouds. This star formation mechanism seems to
be a very important step to explain the formation of high-mass stars and clusters of stars. For
low-mass stars, the most relevant mechanism of formation seems to be the gravitational collapse of cloud
cores, that is induced by an internal agent, such as the expansion of HII regions, see \citet{scoville}.

The G0.253 + 0.016 molecular cloud, which is also called the Brick, has attracted a lot of 
attention and is belived to be formed by a cloud-cloud collision.  This cloud is a 
massive ($\approx \, 10^5\, M_{\odot}$) and
compact $3$ pc, cloud in the Central Molecular Zone of the Milky Way (CMZ). The Brick
can be considered to be a possible progenitor cloud of a young massive cluster (YMC) of stars; that is,
the Brick represents the initial conditions out of which high-mass proto-stars can be formed
through gravitational collapse, see \citet{petkova}.

\citet{longmore} presented deep, multiple-filter, near-IR observations of the
Brick, to ascertain its dynamical state.  \citet{longmore} noted
that large-scale emission from shocked-gas was detected toward the Brick, which indicates that
this cloud could have been formed by the convergence of large-scale flows of gas or
by a cloud-cloud collision.

Using ALMA line emission observations of sulfur monoxide, \citet{higuchi} compared the
filamentary structures observed in the cloud G0.253+0.016 with a cloud collision model. Consequently, the
shell structure was obtained theoretically which is similar to that shell-like structure observed in
the G0.253+0.016 cloud. The model proposed by \citet{higuchi} considered that
the giant G0.253+0.016 molecular cloud may have formed due to a
cloud collision between two un-equal clouds. The small cloud has a radius of $1.5$ pc
and a mass of $0.5 \, \times 10^5 \, M_{\odot}$; the larger cloud, has a radius of $3$ pc and a mass of
$2 \, \times \, 10^5 \, M_{\odot} $. Their approaching pre-collision velocity is within a from $30$ to $60$ km/s.

Using the Combined Array for Research in Millimeter-wave Astronomy (CARMA),
\citet{kauffmann} presented high-resolution interferometric molecular line and dust
emission maps for the G0.253+0.016 cloud. \citet{kauffmann} estimated the virial parameter of the
the G0.253+0.016 cloud, which yields a value of $\alpha_{\rm vir} < 0.8$. In addition, \citet{rathborne} used
ALMA observations of the Brick to investigate its physical conditions.

Many surveys have reported the physical conditions of gas structures of the ISM, that are on
the verge of collapse, see for instance, \citet{caselli} and \citet{jijina}. The dimensionless
ratios $\alpha$ and $\beta$, which are defined as the ratio of thermal energy to gravitational energy and the ratio
of kinetic energy to gravitational energy, respectively- are very useful to characterize the physical
state of these gas structures. Observations seem to favor the statistical occurrence of
low-$\beta$ clumps. However, recent observations have found a gas cloud with a high value
of $\beta$, see for example \citet{jackson}. In addition, for clouds in the CMZ, the gas is observed
to be highly turbulent, with high non-thermal line-widths in the range from 20 to 50 km/s. 
Consequently, considering an isothermal sound speed within the range from 0.3 to 0.6 km/s for gas
temperatures from 30 to 100 K, the typical Mach numbers are in the range from 10 to 60, which is
a highly-supersonic turbulence, see \citet{bally} and \citet{mills}.

From the theoretical side, a lot of simulations that aim to study a cloud-cloud collision process
have appeared in the last three decades, for instance, \citet{hausman}, \citet{lattanzio}, \citet{kimura},
\citet{klein} and \citet{marinho}. However, these early simulations were done with low resolution. Simulations
with much better resolution were done more recently by \citet{burkert} and \citet{pindika1}.
Colliding gas structures starting from hydrodynamical equilibrium were considered by
\citet{kitsionas} and \citet{pindika2}. \citet{pindika3} considered collisions between
un-equal gas structures. ~\citet{gom} and ~\citet{evaz} and other authors, studied the
generation of turbulence at the shock front of head-on collisions.

\citet{lis} proposed a model that was based on a cloud-cloud collision that aimed to explain far-infrared
continuum emission observations of the G0.253+0.016 molecular cloud. The simulation
presented by \citet{habe} assumed that the mass ratio of the non-identical
colliding clouds is 1:4 and the radius ratio is 1:2.

\citet{dale19} and \citet{Kruijssen19} have proposed hydrodynamical simulations of a gas cloud orbiting in
the gravitational potential of the CMZ, in the radial range from 1 to 300 pc. In these simulations, each SPH
particle has been given an additional external force to take the external potential of the CMZ into account. 

\citet{pindika3} studied a head-on collision between
two clouds of different sizes: one cloud was modeled as a Bonnor-Ebert sphere and the second cloud was 
modeled as a uniform density sphere. The cloud's pre-collision translation velocities are also un-equal: 
one moves at 10 km/s while the second cloud moves at -15 km/s. The formation of a bow-shock is the
main outcome of these simulations. The bow-shock continues collapsing, so that the models
showed a lot of fragmentation while other models with slow collision velocities showed
no sign of fragmentation. In addition, the author noted that this behavior (whether or not fragmentation is
presented) also depends on the simulation resolution.

In this paper we aim to study the collision process of several un-equal sub-clouds, which are
initially embedded within a parent turbulent cloud. The set up of this paper is similar to the physical
conditions mentioned by \citet{pindika3}, \citet{higuchi} and \citet{kauffmann}, so
that the system of interest resembles the Brick. In addition, we want to see what role the initial turbulence
of the cloud can have on the overall collision process. While \citet{pindika3} and other authors have
considered two separate clouds that collide, we emphasize that in this paper that the initial cloud
entirely contains the two sub-clouds that collide. We consider both kinds of clouds to simulate: 
low and high $\beta$ turbulent models, so that $\beta=0.5$ or $\beta=50$. In both cases, the
cloud will collapse once the initial turbulence has been dissipated. It has been shown
by \citet{miAAS2018} that the $\beta$ ratio can reach very high values, and yet the
simulations of these clouds show that they still collapse globally.

The models considered in this paper are clearly incomplete given that they do not take into account
the environment of the cloud, so that the models are taken as isolated systems, which is a common practice in 
numerical simulations of cloud collapse and evolution. In the case of the
Brick, or in general of a cloud located in the CMZ, a tidal force will be exerted upon
the clouds from a massive central object, see \citet{Molinari}.

For this reason, we introduce an approximate model to mimic the gravitational influence of a central massive
object on the cloud, in addition to the collision process described earlier. In this approximate model, an
additional velocity is added to each SPH particle of the cloud, so that this velocity is directly
related to the escape velocity induced by the massive central object on the cloud. The result obtained with this
simple model has allowed us to conclude that the collapse of the cloud is accelerated by the presence
of the external object, as was already pointed out by \citet{dale19} and \citet{Kruijssen19}, using a more
complete model.

It must be emphasized that this simple approximate model produces a central condensation during the
very early evolution of the cloud, which is decisive in the subsequent evolution. The results obtained 
from these simulations are in agreement with observations (\citet{hillenbrand}) and simulations (\citet{kirk}),
which indicate that the most massive member of a star cluster is always located
in the center of the cluster.

It must be noted that the gas particles of all the models include three types of velocities, which are
the turbulent velocity spectrum, the translational velocity and the azimuthal velocity, all of them
are given as initial conditions of the SPH particles. The particles are then allowed to evolve as gas
described by the Navier-Stokes hydrodynamic equations under the influence of their own
gravitational interaction.

The outline of this paper is as follows. In Section
\ref{sec:phy-sys} we describe the initial cloud, within which
all the collision models will take place. The initial conditions given to the simulation
particles are explained in Sections \ref{subsec:initvel} and \ref{subs:energies}.
We define the azimuthal velocity in Section \ref{subs:Vcir}. Then,
in Section \ref{subs:models} we give the details of the collision geometry and
define the models to be studied. We describe the
GADGET2 code, the resolution of the simulations and the equation of state in
Sections~ \ref{subs:code}, \ref{subs:res} and \ref{subs:eos}, respectively. In
Section \ref{sec:results}, we describe the
most important features of the time evolution of our simulations by
means of two-dimensional ($2D$) and three-dimensional ($3D$) plots.
A dynamical characterization of the simulation outcomes
is undertaken in Section \ref{subsec:charac}. Finally, in
Sections \ref{sec:dis} and \ref{sec:conclu} we discuss the relevance of our
results in view of those reported by previous papers and we make some concluding remarks.

%%%%%%%%%%%%%%%%%%%%%%%%%%%%%%%%%%%%%%%%%%%%%%%%%%%%%%%%%%%%%%%%%%%%%%%%%%%%%%%%%%%%%%%%%%%%
\section{The physical system and computational considerations}
\label{sec:phy-sys}

The gas cloud that is considered in this paper is a uniform sphere with a radius of $R_0= 3.0$ pc
and a mass of $M_0=1.0 \, \times 10^{5} \, M_{\odot}$. The average density and the
free-fall time of this cloud are $\rho_0=5.9 \, \times 10^{-20}$ g cm$^{-3}$ and
$t_{ff}=8.64 \times 10^{12} \,$ s or 0.27 Myr (2.7 $\times 10^{5}\,$ yr),
respectively. The values of $R_0$ and $M_0$ have been taken from \citet{kauffmann}
and \citet{higuchi}, to draw comparisons with their models of the Brick. The number
density of the cloud considered here is $n_0=15352$ particles per cm$^{3}$, in
which a mean molecular weight of 2.4 for the hydrogen molecule is assumed. Therefore, its 
mean mass is $3.9 \, \times \, 10^{-24}$ g.

It should be emphasized that these physical properties of density, mass and radius are typical of
the gas structures so-called "clumps" in the cloud classification
framework of \citet{jijina} and \citet{bergin} of the ISM, with a number density within
the range $10^{3}-10^{4}$ cm$^{3}$. With respect to the mass, the gas structure of this
paper would better correspond to a "cloud", because the typical mass of clouds is within the range
$10^{3}-10^{4} \, M_{\odot}$ while that of the clumps is in the range $50-500\, M_{\odot}$.

We therefore use the term cloud to refer to the gas structure considered here, although it is clearly
much denser that a typical cloud. For a cloud structure near the CMZ, the physical properties are
observed to be more extreme, so that the density number and the temperature are in general
higher than the clouds of ISM in the galactic disc, see \citet{longmore13a}.

%%%%%%%%%%%%%%%%%%%%%%%%%%%%%%%%%%%%%%%%%%%%%%%%%%%%%%%%%%%%%%%%%%%%%%%%%%%%%%%%%%%%%%%%%%%%
%%%%%%%%%%%%%%%%%%%%%%%%%%%%%%%%%%%%%%%%%%%%%%%%%%%%%%%%%%%%%%%%%%%%%%%%%%%%%%%%%%%%%%%%%%%%
\subsection{The initial conditions of the simulation particles}
\label{subsec:initvel}

\subsubsection{The initial positions}
\label{subsec:pos}

The gas particles are initially located in a simulation volume, which is divided
into small cubic elements, with a volume given by $\Delta x\, \Delta y\, \Delta z $. A gas particle 
is placed at the center of each cubic element. Next, each particle is displaced a distance of
the order $\Delta /4.0$ in a random spatial direction within each cubic
element. The total number of particles is 13,366,240. Therefore, the mass of a simulation
particle is given by $m_p= 7.48 \, \times \, 10^{-3} \, M_{\odot}$.

\subsubsection{The initial turbulent velocity spectrum}
\label{subsec:turvel}

To generate the turbulent velocity spectrum, we set up a mesh with a side length equal to 
2 times the cloud radius, $L_0=2\,R_0$, so that the size of
each grid element of this mesh is $\delta = L_0/N_g$ and the 
mesh partition is determined by $N_g=64$. In Fourier space, the partition 
is given by $\delta K= 1/L_0$, so that each wave-number vector $\vec{K}$ has the components
$K_x=i_{x}  \delta K$, $K_y=i_{y}  \delta K$ and $K_z=i_{z}  \delta K$,
where the indices $i_x,i_y,i_z$ in Eq.\ref{velPhi} take integer values in 
the range $[-N_g/2,N_g/2]$ to cover all of the mesh.

A velocity vector $\vec{v}(\vec{r})= (v_x,v_y,v_z)$ must be assigned for a SPH particle located 
at position $\vec{r}=(x,y,z)$, which is given by

\begin{equation}
\vec{v}(\vec{r})  \approx \Sigma_{i_x,i_y,i_z} \left| \vec{K}\right|^{\frac{-n-2}{2}} \; \vec{K} \;
\sin \left( \vec{K}\cdot \vec{r} + \Phi_K \right)
\label{velPhi}
\end{equation}
\noindent where $n$ is the spectral index. It must be noted that this kind of
turbulent velocity spectrum is known as a curl-free (CF) type. A method to 
obtain a divergence-free (DF) type of turbulence spectrum has been shown 
in \citet{dobbs}. \citet{arreaga2017} examined the effects on the collapse of 
cores due to variation of the number and size of the Fourier modes, for each 
turbulence type, whether divergence-free or curl-free. \citet{arreaga2017} demonstrated that 
the results of the core collape are not substantially different. \citet{miAAS2018} presented simulations in which 
the velocity vector given to each SPH particle was formed by a combination of the two 
types of turbulent spectra $\vec{v}= \frac{1}{2} \vec{v}_{DF} + \frac{1}{2} \vec{v}_{CF}$.

The initial power of the velocity field, for both types of turbulence is given
by:

\begin{equation}
P(\vec{K}) = < \left| v(\vec{K} ) \right|^2 > = \left| \vec{K} \right|^{-n}
\label{power}
\end{equation}
\noindent The spectral index has been fixed in 
our simulations to the value $n = -1$ and thus we will 
have $P \approx K$  and $v^{2} \approx K^{-1}$. Other authors have used other values
of the spectral index, for instance $n = 2$, so that their power and velocity 
go as $P \approx K^{-2}$ and $v^{2} \approx K^{-2}$, respectively, see \citet{dobbs}.  

Finally, the level of turbulence can be adjusted by introducing a multiplicative 
constant in front of the right-hand side of Eq. \ref{velPhi}, whose value is fixed, as we explain 
it in the next Section \ref{subs:energies}. Later, we will show that the velocity spectrum that is proposed in this 
Section \ref{subsec:turvel} has some of the well-known characteristics of turbulence, see Section \ref{subsec:nubeaislada}.
  
%%%%%%%%%%%%%%%%%%%%%%%%%%%%%%%%%%%%%%%%%%%%%%%%%%%%%%%%%%%%%%%%%%%%%%%%%%%%%%%%%%%%%%%%%%%%%%%%%%%%%
\subsection{Initial energies}
\label{subs:energies}

In a particle-based simulation, the thermal, kinetic and gravitational energies
are given by

\begin{equation}
\begin{array}{l}
E_{\rm ther}=\frac{3}{2}\sum_{i} \, m_{i}\frac{P_{i}}{\rho _{i}}\\
E_{\rm kin}=\frac{1}{2}\sum_{i} \, m_{i} v_i^{2},\\
E_{\rm grav}=\frac{1}{2}\sum_{i} \, m_{i}\Phi_{i}
\label{energiespart}
\end{array}
\end{equation}
\noindent where $P_i$ is the pressure and $\Phi _{i}$ is the gravitational
potential at the location of particle $i$, with velocity $v_i$ and mass $m_i$. It should be emphasized
that all of the SPH particles of a simulation must be used in the summation of Eq.~\ref{energiespart}.

Let $\alpha$ be defined as the ratio of the thermal energy to the gravitational
energy and let $\beta$ be the ratio of the kinetic energy
to the gravitational energy, so that

\begin{equation}
\alpha \equiv \frac{E_{ther}}{\left|E_{grav}\right|}
\label{defalpha}
\end{equation}
\noindent and

\begin{equation}
\beta \equiv \frac{E_{kin}}{\left|E_{grav}\right|}.
\label{defbeta}
\end{equation}

The value of the speed of sound $c_0$ has been fixed at 225,000 cm/s, so that the initial turbulent cloud
has the $\alpha_0$ ratio given by 0.16, for all the collision models. The multiplicative constant
mentioned in Section \ref{subsec:turvel} has been adjusted so that the initial turbulent cloud has a
$\beta_0$ ratio given by 0.5.\footnote{We also consider models with a very high value of the ratio
of the kinetic energy to the gravitational energy; in addition, there is observational
interest in these kinds of model, see Section \ref{justi} below.}  \citet{higuchi} presented
line emission observations of the Brick using the Atacama Large Millimeter/Submillimeter Array
and considered values of $\beta_0=0.1$ and $\alpha_0=0.02$ taken into account a cloud mass of
$2 \, \times 10^{5} \, M_{\odot}$, radius of $2.8$ pc, a temperature of 20 K and a
one-dimensional velocity dispersion of 4 km/s.

The virial parameter is very useful when characterizing the physical
state of a gas structure, which is defined observationally by

\begin{equation}
\beta_{\rm vir} \equiv \frac{5 \, \sigma_{1D}^2 \, R}{G \, M }
\label{defbetavir}
\end{equation}
\noindent where $G$ is Newton's gravitational constant, $M$ and $R$ are the mass and radius
of the gas structure, and $\sigma_{1D}$ is the intrinsic one-dimensional velocity dispersion of
the hydrogen molecule. Assuming isotropic motions, a 3D velocity dispersion
can be simply related by $\sigma_{3D}= \sqrt{3} \, \sigma_{1D}$. It should be noted that a gas
structure in virial equilibrium would have $\beta_{\rm vir}=1$.

The empirical relation between the virial parameter is $\beta_{\rm vir}= 2 \, a \, \beta$,
where $\beta$ is the dimensionless ratio defined in Eq.~\ref{defbeta} and $a$ is a numerical
factor that is empirically included to take modifications of non-homogeneous
and non-spherical density distributions into account. According to this empirical relation, the virial parameter
of the simulation of this paper is approximately 1.

Later, the virial theorem will be useful to show the level of virialization of the
simulation outcome. In general terms, for a gas structure at virial equilibrium, the energy ratios 
defined above in Eqs.\ref{defalpha} and \ref{defbeta} satisfy the relation

\begin{equation}
\alpha + \beta =\frac{1}{2}\;.
\label{abvirial}
\end{equation}
\noindent It is expected that if a gaseous system
has $\alpha + \beta > 1/2$, then it will expand; in
the other case, if $\alpha + \beta < 1/2$, then
the system will collapse. It must be mentioned that \citet{miyama}, \citet{hachisu1} and \citet{hachisu2} obtained
a criterion of the type $\alpha \times \beta <  0.2 $ to predict the output of a given simulation.
%%%%%%%%%%%%%%%%%%%%%%%%%%%%%%%%%%%%%%%%%%%%%%%%%%%%%%%%%%%%%%%%%%%%%%%%%%%%%%%%%%%%%%%%%%%%%%%%%%%%%%%%%%%%%%%%%%%
\subsubsection{Observational evidence for models with extreme initial kinetic energy.}
\label{justi}

High kinetic energy molecular clouds have recently been observed. For example, \citet{jackson}, reported
unusually large line-widths of the G337.342-0.119 gas structure, which is also known as the Pebble. These kinds
of clouds are expected not to collapse in terms of the virial theorem, because a gas structure such as the Pebble
reaches a virial parameter of 3.7.\footnote{Recall that the virial theorem states that a gas structure
with a virial parameter less than 1 will collapse; otherwise, if the virial parameter is greater
than 1, then a gas structure will not collapse.}

In spite of this, numerical simulations have shown that
there are gas structures with a high kinetic energy, so that their virial parameter is
around or greater than 2, and are in a state of global collapse, see for instance \citet{ballesteros}. In
addition, \citet{miAAS2018} determined the extreme kinetic energy allowed for a turbulent core to collapse under the
influence of its own self-gravity. The results that these authors found are given in terms of the
ratio $\beta$, which is defined here in Eq.~\ref{defbeta} of Section~\ref{subs:energies}, so that
a turbulent core can have an initial $\beta$ as high as $2-8$ and with an initial Mach number
within $3-9$ and still finish its evolution in a collapsed state.

For clouds in the CMZ, the gas is observed
to be highly turbulent, with high non-thermal line-widths in the range from 20 to 50 km/s. Considering 
an isothermal sound speed within the range from 0.3 to 0.6 km/s for gas
temperatures from 30 to 100 K, the typical Mach numbers are in the range from 10 to 60, which is
a highly supersonic turbulence, see \citet{bally} and \citet{mills}.
%%%%%%%%%%%%%%%%%%%%%%%%%%%%%%%%%%%%%%%%%%%%%%%%%%%%%%%%%%%%%%%%%%%%%%%%%%%%%%%%%%%%%%%%%%%%%%%%%%
\subsection{The azimuthal velocity}
\label{subs:Vcir}

Let us consider a massive agent, such as a dwarf spheroidal galaxy, which is located in the origin of a
coordinate system. Let us place the molecular cloud of interest here to be
in the z-axis, at a distance $Z_C$. This massive agent induces an escape velocity at
each radius $R$ (with respect to the center of the massive agent), so that

\begin{equation}
V_{\rm cir}=\sqrt{2\, G \, M(R) / R}
\label{escape}
\end{equation}
\noindent where $M(R)$ is the mass contained up to
the radius $R$ and $G$ is Newton's gravitational constant. In spherical coordinates
$R$,$\theta$ and $\phi$, the velocity vector $\vec{V}$ has components $V_R$, $V_{\theta}$
and $V_{\phi}$, respectively. These spherical
components are related to the Cartesian components of velocity $V_x$, $V_y$ and $V_z$ by three simultaneous
equations, whose coefficients are given in terms of the sine and cosine of the polar and azimuthal
angles $\theta$ and $\phi$, as follows:

\begin{equation}
\begin{array}{c}
V_R = V_x \, \sin(\theta) \, \cos(\phi) + V_y \, \sin(\theta) \, \sin(\phi)+ V_z  \, \cos(\theta) \\
V_{\theta} = V_x\, \cos(\theta) \, \cos(\phi) + V_y \, \cos(\theta) \, \sin(\phi) - V_z \, \sin(\theta) \\
V_{\phi}=  -V_x \, \sin(\phi) + V_y \, \cos(\phi)  \\
\label{velcomponents}
\end{array}
\end{equation}

Let us assume that $Z_C$ and the radius of the cloud ( which is defined in Section \ref{sec:phy-sys} $R_0=3$ pc
with respect to the center of the cloud), satisfy the relation $R_0/Z_C \, \ll 1$, then as
$\sin(\theta)= R_0/R \approx \, R_0/Z_C \ll \, 1$ then $\theta \approx 0$, for which $\cos(\theta)  \approx 1$
and $\sin(\theta)  \approx 0$. In addition, in the particular
case that the cloud follows a circular orbit around the massive center at radius $R$, then the velocity vector 
would only have a non-zero polar velocity component, $V_{\theta}$, while the radial and azimuthal 
components $V_R$ and $V_{\phi}$
are zero. Let us denote this velocity as $V_{\theta}=-V_{\rm cir}$, where the minus sign indicates that
the assumed rotation of the cloud is counter-close-wise in its orbit around the massive agent. Under these 
simplifications, the relations \ref{velcomponents} between Cartesian and spherical components of velocity are reduced to

\begin{equation}
\begin{array}{c}
V_x = - V_{\rm cir} \, \cos(\phi)\\
V_y = - V_{\rm cir} \, \sin(\phi)\\
V_z=  0
\label{velcomponentssol}
\end{array}
\end{equation}
\noindent so that the magnitude of the velocity of a particle located at any radius $R$ is therefore
always given by $\sqrt{V_x^2+V_y^2}=V_{\rm cir}$. These Cartesian components of the velocity will be added
to the particle velocity defined in Eq.\ref{velPhi} to generate four new models in which this
approximation will be implemented. It should be noted that the azimuthal angles have the same projection
in both coordinate systems: the first is based on the cloud center and the second is located at 
the gravitational agent center.

We will call this velocity the "azimuthal velocity" because it is given only in
terms of the azimuthal angle $\phi$. Then, the approximation that replaces the tidal force
by an azimuthal velocity, as described in
Eq. \ref{velcomponentssol}, does not depend explicitly on the distance of the massive center to the
cloud as long as ratio between the cloud radius to this distance is quite small.

Following with the model of a cloud of the CMZ, the mass of this massive agent has been fixed at
$M_{B}=3.6 \, \times \, 10^{6} \, M_{\odot}$, which corresponds to the black hole located at the
center of the Milky Way. In this case, the escape velocity at $500$ pc is 5.57 km/s. To
compare this velocity with those displayed at Fig.\ref{VelDist}, in terms of the speed of sound $c_0$ defined in
Section \ref{subs:energies}, we have a Mach number of $V_{\rm cir}=2.47$. It must be noted that the magnitude of this
circular velocity $V_{\rm cir}$ is quite small as compared to that proposed by \citet{Molinari}, in which a model
for the orbit of the gas stream near the massive center Sgr B2 is around 80 km/s. For this reason, we have also
included a second set of models in which the massive agent is considered to be molecular gas concentrated in
the nuclear bulge of the Milky Way, see \citet{laun}, so that the total mass is
$M_{H}=8.4 \, \times \, 10^{8} \, M_{\odot}$, see also \citet{mills}, for which the escape velocity at $500$ pc
is $85$ km/s, such that the normalized velocity in terms of the speed of sound 37.82.

%%%%%%%%%%%%%%%%%%%%%%%%%%%%%%%%%%%%%%%%%%%%%%%%%%%%%%%%%%%%%%%%%%%%%%%%%%%%%%%%%%%%%%%%%%%%%%%%%%%%%%%%%%%%%%%%%%%%%%
\subsection{The collision models}
\label{subs:models}

It is important to emphasize that the cloud entirely contains
the two subsets of particles that are going to collide. Let us call these subsets the
pre-collision sub-clouds. They are located initially along the X-axis, so that the centers
are: for the left-hand clump (-2.55,0,0) pc and for the right-hand clump (2.55,0,0) pc.

The radius of the pre-collision sub-clouds are chosen to be equal for two models, and different for
other two models. The former models are head-on collisions. An impact 
parameter $b$ has also been considered for the latter models, so that they are oblique collisions, in which
the $b$ takes the value 1.5 pc along the Y-axis. \citet{bekki} have demonstrated observationally that
the most likely impact parameter $b$ in cloud collisions in the Large Magellanic Cloud and the Small Magellanic Cloud
is $0.5 \, D <b < D$ where $D$ is the diameter of the cloud. For the
radius $R_0$ of the cloud considered in this paper, $b$ has been chosen such that $b=0.25 \, D$.

\begin{figure}
\begin{center}
\begin{tabular}{cc}
\includegraphics[width=2.25 in]{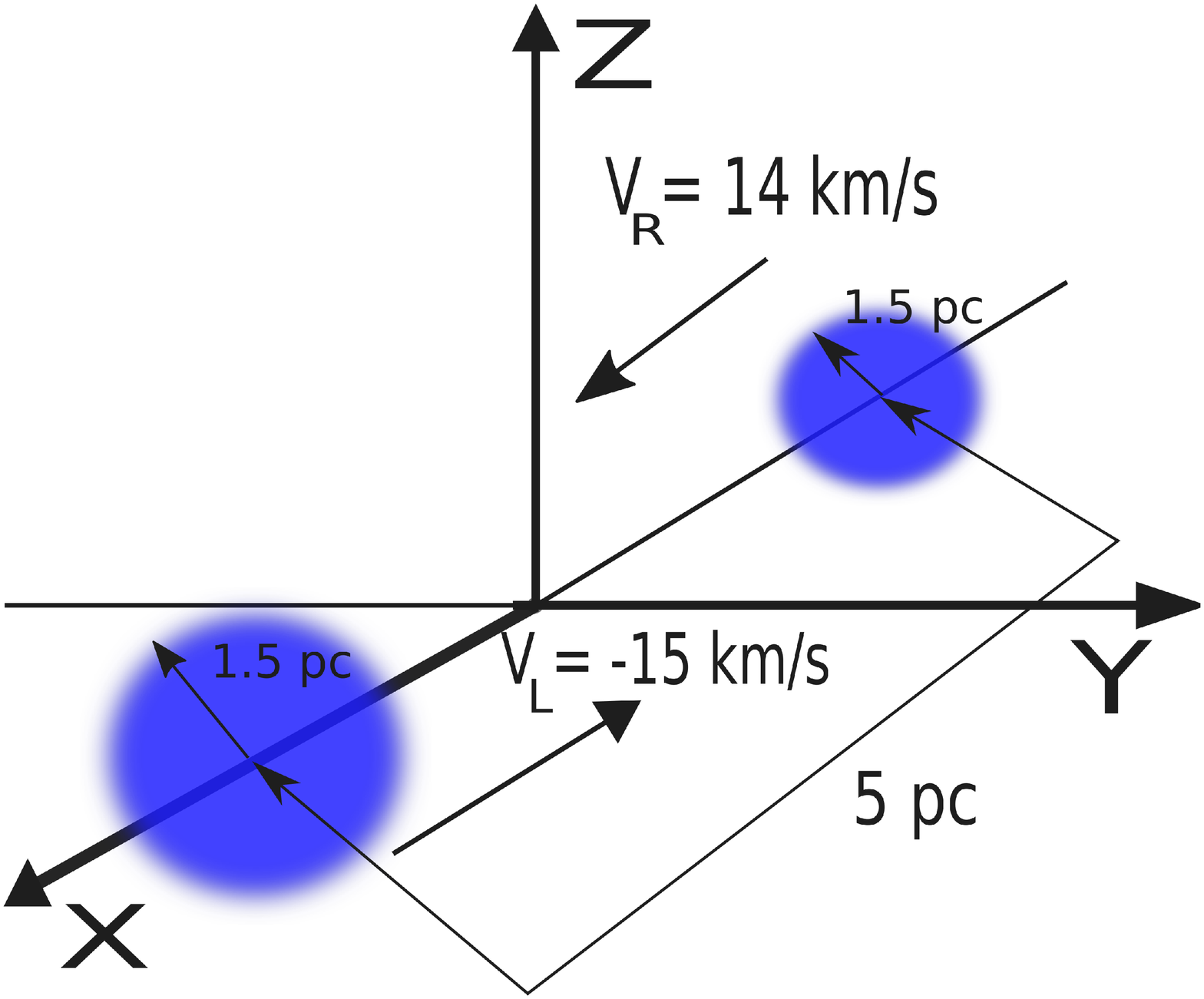} & \includegraphics[width=2.25 in]{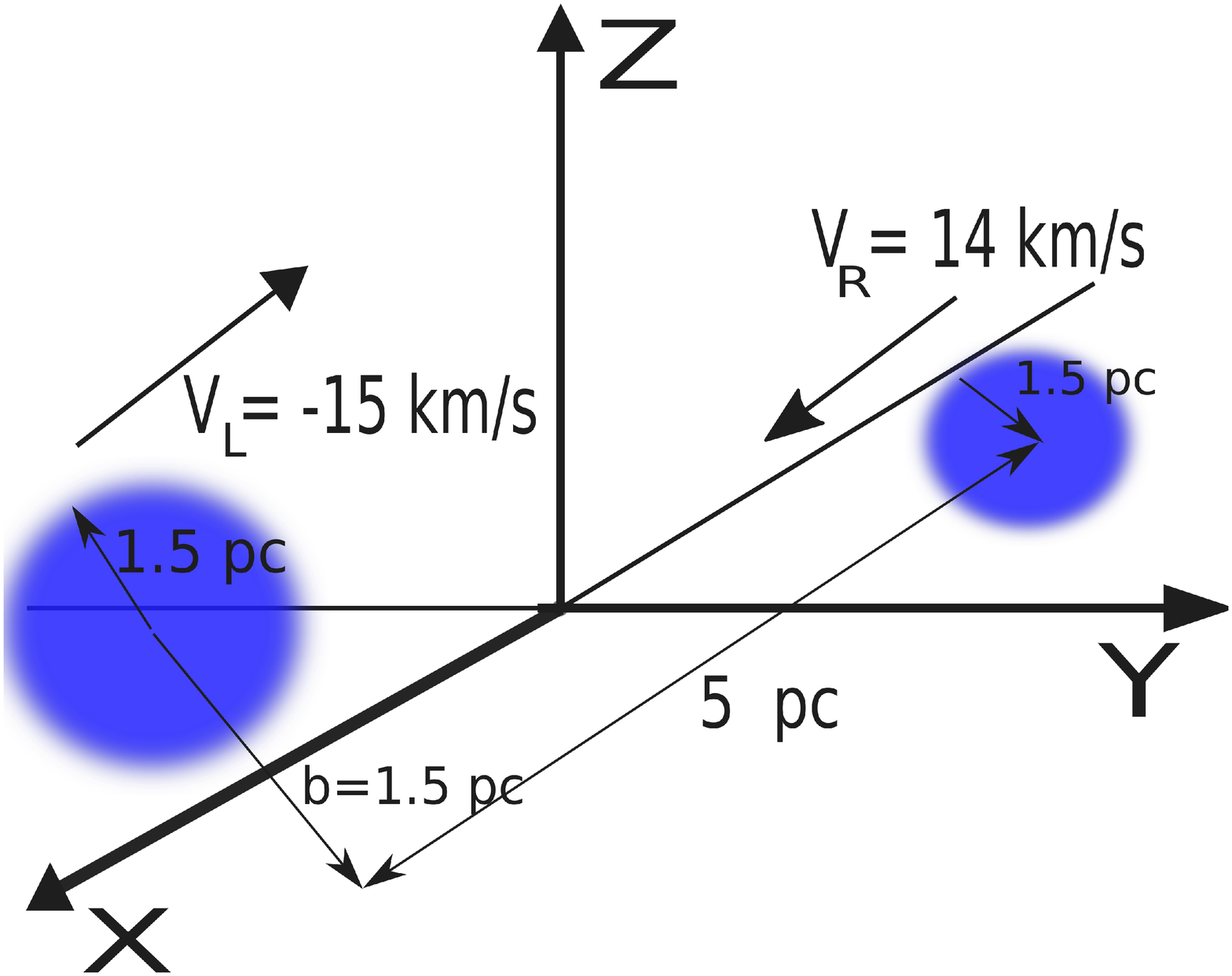}
\end{tabular}
\caption{\label{Cartoon} Schematic diagram of the pre-collision geometry for 
(left-hand panel) a head-on collision and (right-hand panel) an oblique collision.}
\end{center}
\end{figure}

We show all these models in Table~\ref{tab:models}. The label is shown in column one.
The impact parameter value is shown in column two.  In columns three and four, the relationship of 
the radius and translational velocities are shown, for the left-hand and right-hand sub-clouds, respectively. 
It is important to emphasize that the
relative pre-collision velocity of the sub-clouds is 29 km/s and is formed for non-identical velocities
for the left-hand and right-hand sub-clouds. In column five, the value of the ratio of the kinetic energy
to the gravitational energy for the initial configuration of particles is shown, see
Eq.\ref{defbeta}. Finally, column six gives the value of the azimuthal velocity added
to the cloud particles, see Section \ref{subs:Vcir}. It must be clarified
that these models are a sample from a larger set of models that was considered in a first 
manuscript, so that the numbers of the labels do not show any ordering.

\begin{table}[ph]
\caption{The collision models; $\be$ is the impact parameter; $\re_L:\re_R$ is the initial relation of the 
colliding sub-cloud radii; $\vel_L:\vel_R$ is the relation of the translational velocities or 
pre-collision velocities; $\beta_0$ is the initial ratio of kinetic energy to gravitational energy; and 
$V_{\rm cir}/\ce_0$ is the ratio between the azimuthal velocity and the speed of sound
.}
{\begin{tabular}{|c|c|c|c|c|c|} \hline
model & $b$ [pc] &$r_L:r_R$ [pc]& $v_{L}:v_{R}$ [km/s] & $\beta_0$ & $V_{\rm cir}/c_0$\\
\hline
\hline
U5       & 0   &  0.75:1.5 & 14:-15 & 0.5 & 0\\
\hline
U13      & 0   &  1.5:1.5 & 14:-15 & 0.5  & 0\\

\hline
U9       & 1.5 &  0.75:1.5 & 14:-15  & 0.5 & 0\\
\hline
U11      & 1.5 &  1.5:1.5 & 14:-15 & 0.5   & 0\\
\hline
\hline
U5b       & 0   &  0.75:1.5 & 14:-15 & 50  & 0\\
\hline
U13b      & 0   &  1.5:1.5 & 14:-15 & 50   & 0\\
\hline
U9b       & 1.5 &  0.75:1.5 & 14:-15  & 50 & 0\\
\hline
U11b      & 1.5 &  1.5:1.5 & 14:-15 & 50   & 0\\
\hline
\hline
U5r       & 0   &  0.75:1.5 & 14:-15 & 0.5 & 2.47 \\
\hline
U13r      & 0   &  1.5:1.5 & 14:-15 & 0.5  & 2.47 \\
\hline
U9r       & 1.5 &  0.75:1.5 & 14:-15  & 0.5 & 2.47\\
\hline
U11r      & 1.5 &  1.5:1.5 & 14:-15 & 0.5   & 2.47 \\
\hline
\hline
U5rb       & 0   &  0.75:1.5 & 14:-15 & 0.5 &  37.82 \\
\hline
U13rb      & 0   &  1.5:1.5 & 14:-15 & 0.5  &  37.82 \\
\hline
U9rb       & 1.5 &  0.75:1.5 & 14:-15  & 0.5 & 37.82 \\
\hline
U11rb      & 1.5 &  1.5:1.5 & 14:-15 & 0.5   &  37.82 \\
\hline
\hline
\end{tabular} }
\label{tab:models}
\end{table}

\subsubsection{A note on the physical parameters chosen for the sub-clouds.}
\label{justitras}

As we mentioned in Section \ref{sec:int}, the idea that the Brick could be formed by a
non-identical cloud-cloud collision has been explored for some time. \citet{habe} assumed
that the mass ratio of the colliding clouds is 1:4 and the radius ratio is 1:2. Later, \citet{lis}
followed this collision model, so that their pre-collision velocities of the clouds are taken
for this paper exactly as these author introduced them.

More recently, using observations, \citet{kauffmann} estimated that the virial parameter of the
Brick is around $\beta_{\rm vir} \, \le \, 0.8$ and considered the same geometry of un-equal
clouds at the same relation proposed by \citet{habe} and \citet{lis}. In this paper, we have taken
the value of $\beta_0=0.5$, so that we expect to have a value of the virial parameter of $\approx \, 1$, see
Section \ref{subs:energies}.

Shortly after, \citet{higuchi} reconsidered this idea and continued the exploration
of a cloud-cloud collision model in which the relative speed of colliding clouds was within the
range from 30 to 60 km/s, and the radii were of $1.5$ and $3$ pc, for the small and big clouds, respectively.

In this paper, the translation velocity shown in Table \ref{tab:models}, $v_{L}:v_{R}$, is given
in terms of the sound speed $c_0$ by 6.2:6.6 Mach, so that the relative velocity of approach is
little greater than 12 Mach.

To allow comparison of the results of the present paper with these authors \citet{lis},
\citet{kauffmann} and \citet{higuchi}, we use here the values for the mass, radius and translation velocity
of the cloud-cloud collision model that were used by these authors.

It must be noted that the gas particles of all the models described in Table \ref{tab:models} include
the Cartesian components of velocity described in Eq.\ref{velPhi}, which are the turbulent velocity spectrum and
the translation velocity $v_{L}:v_{R}$ of the sub-clouds. However, only the last four models
include a third set of velocity components already described Eq.\ref{velcomponentssol}, which are
needed to implement the approximation that replaces the tidal force by an azimuthal velocity. All of these
three types of velocities enter as initial conditions of the SPH particles, as we will describe in
Section \ref{subs:velocities}.

%%%%%%%%%%%%%%%%%%%%%%%%%%%%%%%%%%%%%%%%%%%%%%%%%%%%%%%%%%%%%%%%%%%%%%%%%%%%%%%%%%%%%%%%%%%%%%%%%
\subsection{Characterization of the initial turbulence}
\label{subs:velocities}

To show the nature of the turbulence that is implemetd in Section \ref{subsec:turvel}, we consider the 
distribution functions of the initial velocity.

\begin{figure}
\begin{center}
\begin{tabular}{cc}
\includegraphics[width=2.5 in]{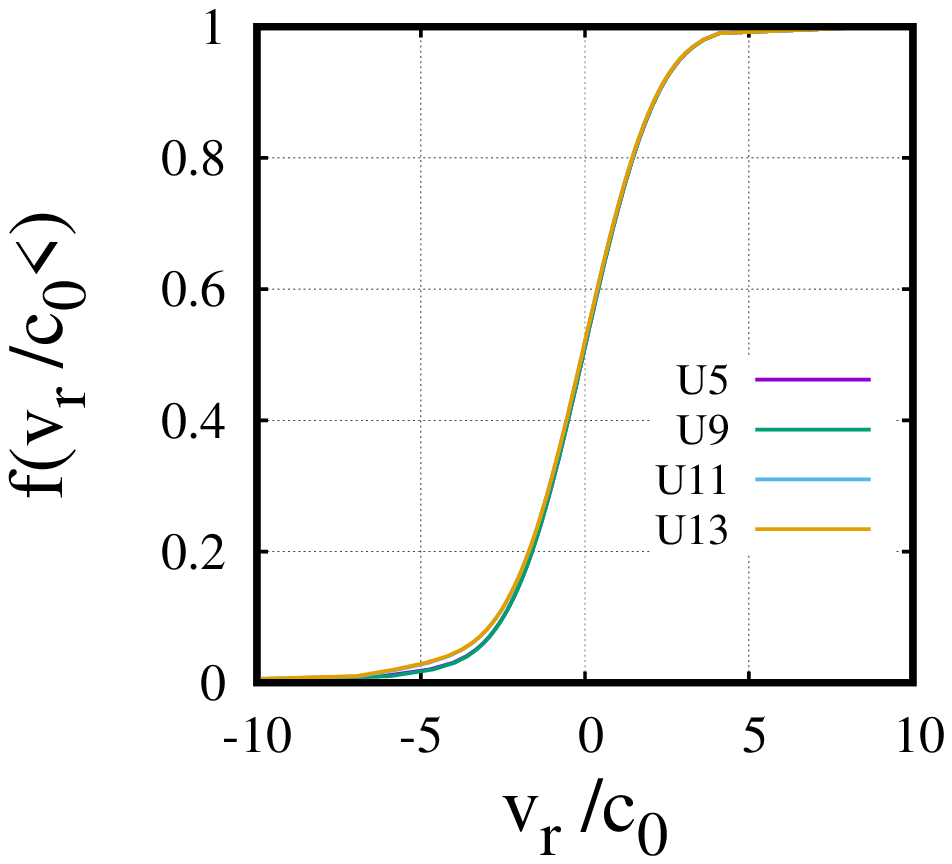} & \includegraphics[width=2.5 in]{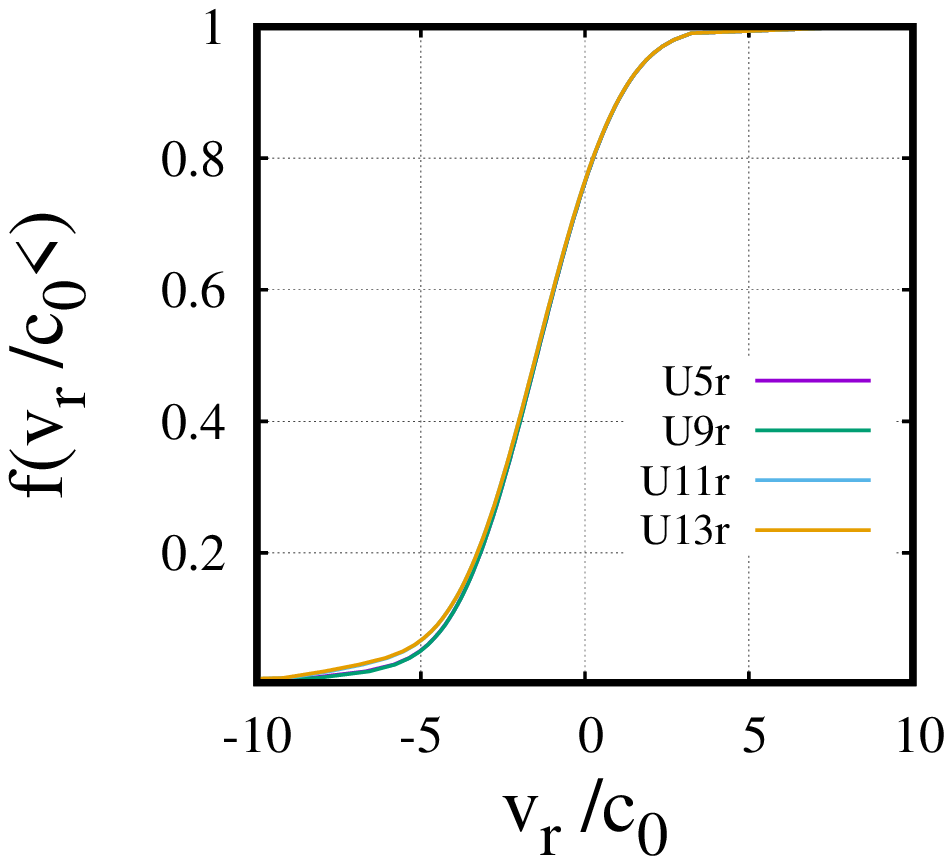}\\
\includegraphics[width=2.5 in]{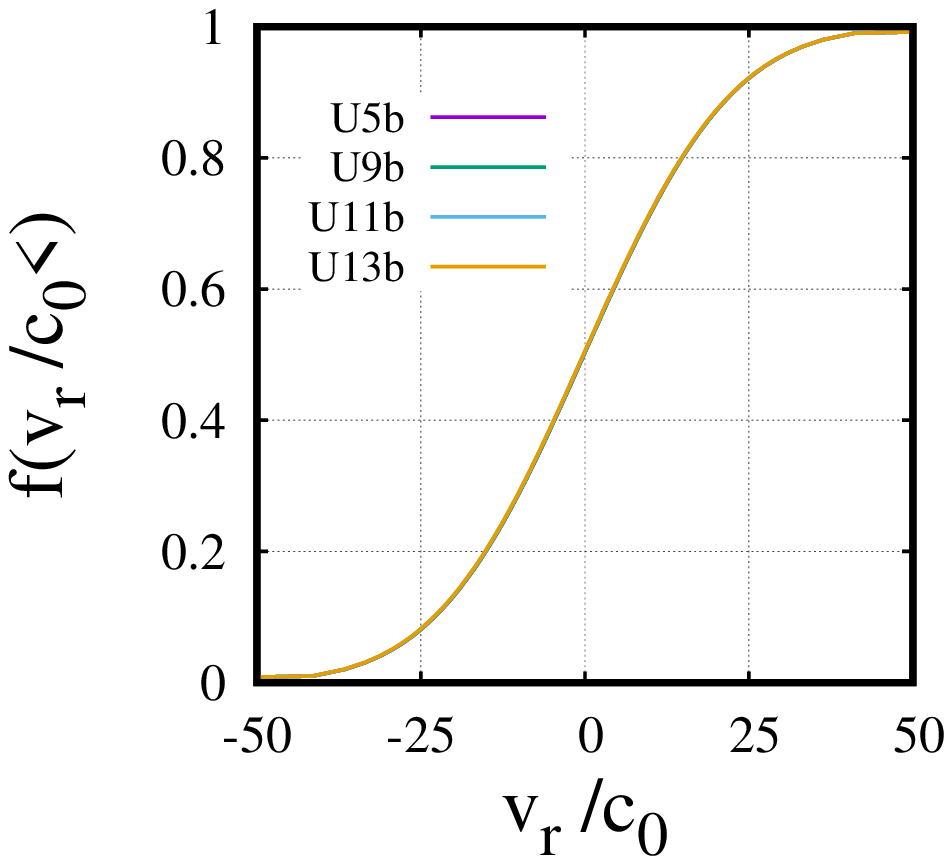} & \includegraphics[width=2.5 in]{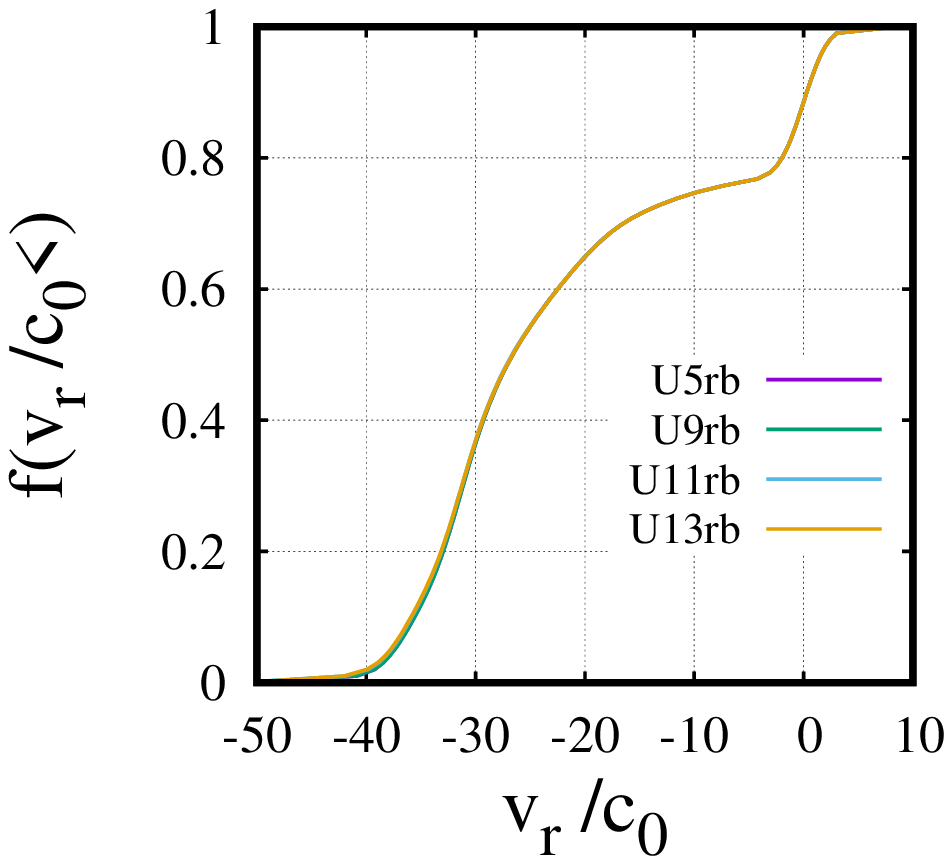} \\
\end{tabular}
\caption{\label{VelDist} Distribution function for the radial 
component of the velocity at the
initial snapshot at $t/t_{ff}=0$, for the model
(top left-hand) $U$ with a low level of turbulence;
(top right-hand) $Ur$ with a low azimuthal velocity;
(bottom left-hand) $Ub$ with a high level of turbulence and
(bottom right-hand) $Urb$ with a high azimuthal velocity.
$f$ is the fraction of particles whose magnitude of the 
velocity $v_r/c_0$ (normalized with the sound speed)
is smaller than that value shown in the horizontal axis.}
\end{center}
\end{figure}

In Fig.\ref{VelDist}, we show the distribution functions of the radial component of the
velocity at the initial snapshot, so that in the vertical axis the fraction $f$ of
the simulation particles whose magnitude of the velocity is smaller than that value shown
in the horizontal axis. The radial component has been calculated with respect to the origin of
the coordinates located in the center of the cloud, which is located at the center of the
simulation box.

According to the left-hand column panels of Fig.\ref{VelDist} that is, for
models $U$ and $Ub$, half of the simulation particles have negative radial velocity component,
while the other half have a positive radial component. This symmetry is
expected from the random process described in Section \ref{subsec:initvel} to generate
the direction of the velocity vectors.

It must be emphasized that for the right-hand column panels of Fig.\ref{VelDist} that is, for models
$Ur$ and $Urb$, which include an azimuthal velocity, the symmetry of the curves with respect to the
positive and negative radial components has been lost. These panels indicate that the azimuthal
velocity favor that 80 percent of the particles have negative radial component
of the velocity.

It must also be emphasized that both types of models $U$, $Ur$ and $Urb$ have the same initial turbulent
velocity spectrum with the same level of energy, as defined in Section \ref{subsec:initvel}
and Section \ref{subs:energies}, and have the same translational velocity. The only difference between them is
whether or not they include the azimuthal velocity, as described in Section \ref{subs:Vcir}. It is observed
in Fig.\ref{VelDist} that the distribution function of the models $U$ shows a magnitude of the velocity
12 percent smaller than that of models $Ur$.

Later, we will compare these curves at the initial snapshot with curves obtained for an snapshot of the
final evolution stage.

A brief description of the dynamics of the isolated cloud is given in Section \ref{subsec:nubeaislada}.
%%%%%%%%%%%%%%%%%%%%%%%%%%%%%%%%%%%%%%%%%%%%%%%%%%%%%%%%%%%%%%%%%%%%%%%%%%%%%%%%%%%%%%%%%%%%%%%%%%%
\subsection{The evolution code}
\label{subs:code}

The simulations of this paper are evolved using the particle-based Gadget2 code, which 
implements the SPH method to solve the Euler equations of
hydrodynamics; see~\citet{gadget2}. Gadget2 has a Monaghan-Balsara form
for the artificial viscosity, see \citet{balsara1995}, so that the strength of the
viscosity is regulated by setting the parameter $\alpha_{\nu} = 0.75$ and
$\beta_{\nu}=\frac{1}{2}.
\times \alpha_v$, see Equations 11 and 14
in~\citet{gadget2}. The Courant factor has been fixed at $0.1$.

The SPH sums are evaluated using the spherically symmetric M4 kernel
and so gravity is spline-softened with this same kernel.
The smoothing length $h$ establishes the compact support, so
that only a finite number of neighbors to each
particle contribute to the SPH sums. The smoothing length changes with time for each particle,
so that the mass contained in the kernel volume is a constant for the estimated density. Particles also
have gravity softening lengths $\epsilon$, which change step by step
with the smoothing length $h$, so that the ratio $\epsilon/h$ is of
order unity. In Gadget2, $\epsilon$ is set equal to the minimum
smoothing length $h_{\rm min}$, which is calculated over all particles at the end of each time
step.

%%%%%%%%%%%%%%%%%%%%%%%%%%%%%%%%%%%%%%%%%%%%%%%%%%%%%%%%%%%%%%%%%%%%%%%%%%%%%%%%%%%%%%%%%%%%%%%%%%%%%%%%
\subsection{Resolution}
\label{subs:res}

\citet{truelove} demonstrated that the resolution requirement of a hydrodynamic simulation
can be expressed in terms of the Jeans wavelength $\lambda_J$, which is given by

\begin{equation}
\lambda_J=\sqrt{ \frac{\pi \, c^2}{G\, \rho}} \; ,
\label{ljeans}
\end{equation}
\noindent where $G$ is Newton's gravitation constant, $c$
is the instantaneous sound speed and $\rho$ is the local density, so that a mesh-based simulation
must always have its grid length scale $l$ such that  $l<\lambda_J/4$.

\citet{bateburkert97} demonstrated that the resolution requirement for a particle-based code, the
Jeans wavelength $\lambda_J$ is better written in terms of the spherical Jeans
mass $M_J$, which is defined by

\begin{equation}
M_J \equiv \frac{4}{3}\pi \; \rho \left(\frac{ \lambda_J}{2}
\right)^3 = \frac{ \pi^\frac{5}{2} }{6} \frac{c^3}{ \sqrt{G^3 \,
\rho} } \;. \label{mjeans}
\end{equation}
\noindent so that an SPH code will produce correct
results as long as the minimum resolvable mass $m_r$ is always less than the Jeans
mass $M_J$. The mass $m_r$ is given by $m_r \approx M_J / (2 N_{neigh})$, where
$N_{neigh}$ is the number of particles included in the
SPH kernel (i.e., the number of neighbors). Therefore, our simulations will comply with this
resolution requirement if the particle mass $m_p$ is such that $m_p/m_r<1$.

As we mentioned in Section \ref{subsec:initvel}, we have $N=13 366 240$ SPH particles
in each simulation and therefore $m_p= 7.4 \, \times \, 10^{-3} \, M_{\odot}$. Now, if we consider
that the highest peak density in our collision models is
$\rho_{\rm max}=5.0\, \times \, 10^{-12}$ g/cm$^3$, then the minimum Jeans mass
would be given by $\left( M_J\right)_{\rm min} \approx 0.594048 M_\odot$,
so that, we obtain $m_r=7.4 \times 10^{-3} \, M_\odot$. Thus, for that peak density the
ratio $m_p/m_r \leq 1$, and the Jeans
resolution requirement is satisfied. In this sense, we are sure to avoid
the growth of numerical instabilities or the occurrence of artificial fragmentation in
all our simulations up to densities smaller or equal than $\rho_{\rm max}$.
In the next sections, we will present our results in terms of a normalized density, so that
$\log \left( \rho_{\rm max}/\rho_0 \right)$ is given by 7.9, where $\rho_0$ is the average density
of the initial cloud, as we mentioned in Section \ref{sec:phy-sys}.

%%%%%%%%%%%%%%%%%%%%%%%%%%%%%%%%%%%%%%%%%%%%%%%%%%%%%%%%%%%%%%%%%%%%%%%%%%%%%%%%%%%%%%%
\subsection{Equation of state}
\label{subs:eos}

Most simulations in the field of collapse used an
ideal equation of state or a barotropic equation of
state (BEOS), as was proposed by~\citet{boss2000}:

\begin{equation}
p= c_0^2 \, \rho \left[ 1 + \left(
\frac{\rho}{\rho_{crit}}\right)^{\gamma -1 } \, \right] ,
\label{beos}
\end{equation}
\noindent where $\gamma\,\equiv 5/3$ and $\rho_{\rm crit}$ is a critical density, a parameter which we
explain now. This BEOS takes into account the increase in temperature of the gas as
it begins to heat once that gravity has produced a substantial
contraction of the cloud. In this paper, we also use this BEOS scheme
for simplicity with a critical density $\rho_{\rm crit}=5.0 \, \times \,10^{-14}$ g/cm$^{3}$, which
is 100 times smaller than the peak density considered in Section \ref{subs:res} for the
resolution requirement estimate; that is, $\rho_{\rm max}$. However, it should be emphasized that it is only an
approximation. Consequently, to describe correctly the transition from the ideal to the adiabatic
regime, one needs to solve the radiative transfer problem coupled to gravity in a
self-consistent way.

%%%%%%%%%%%%%%%%%%%%%%%%%%%%%%%%%%%%%%%%%%%%%%%%%%%%%%%%%%%%%%%%%%%%%%%%%%%%%%%%%%%%%%%
%%%%%%%%%%%%%%%%%%%%%%%%%%%%%%%%%%%%%%%%%%%%%%%%%%%%%%%%%%%%%%%%%%%%%%%%%%%%%%%%%%%%%%%
%%%%%%%%%%%%%%%%%%%%%%%%%%%%%%%%%%%%%%%%%%%%%%%%%%%%%%%%%%%%%%%%%%%%%%%%%%%%%%%%%%%%%%%
%%%%%%%%%%%%%%%%%%%%%%%%%%%%%%%%%%%%%%%%%%%%%%%%%%%%%%%%%%%%%%%%%%%%%%%%%%%%%%%%%%%%%%%
\section{Results}
\label{sec:results}

%%%%%%%%%%%%%%%%%%%%%%%%%%%%%%%%%%%%%%%%%%%%%%%%%%%%%%%%%%%%%%%%%%%%%%%%%%%%%%%%%%%%%%
\subsection{Evolution of the density peak}
\label{subsec:peak}

\begin{figure}
\begin{center}
\begin{tabular}{cc}
\includegraphics[width=2.5 in]{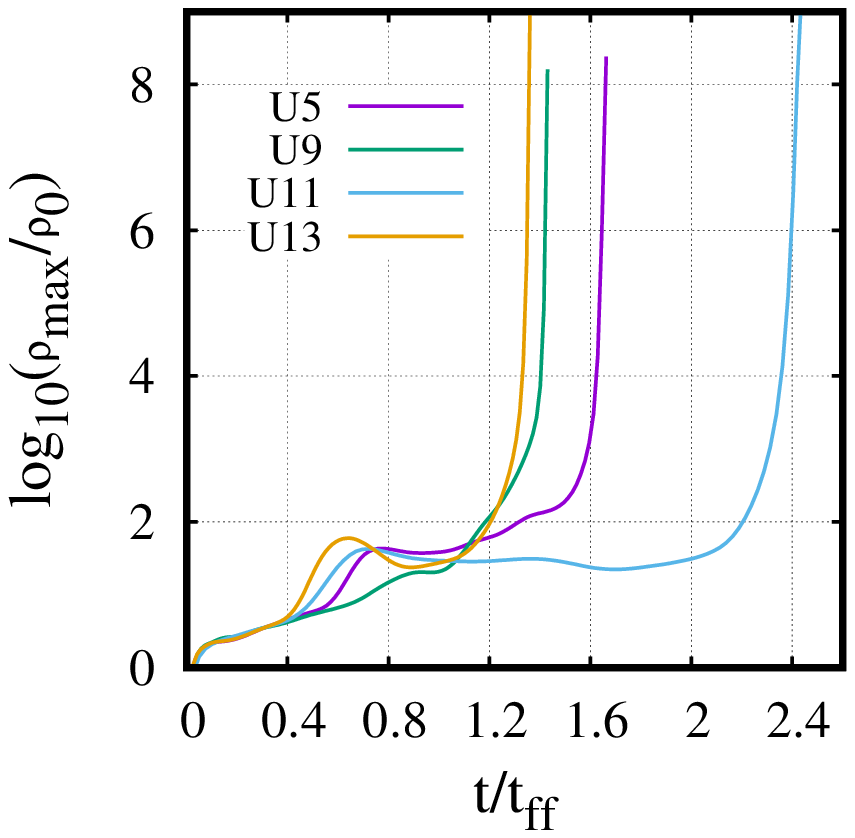} & \includegraphics[width=2.5 in]{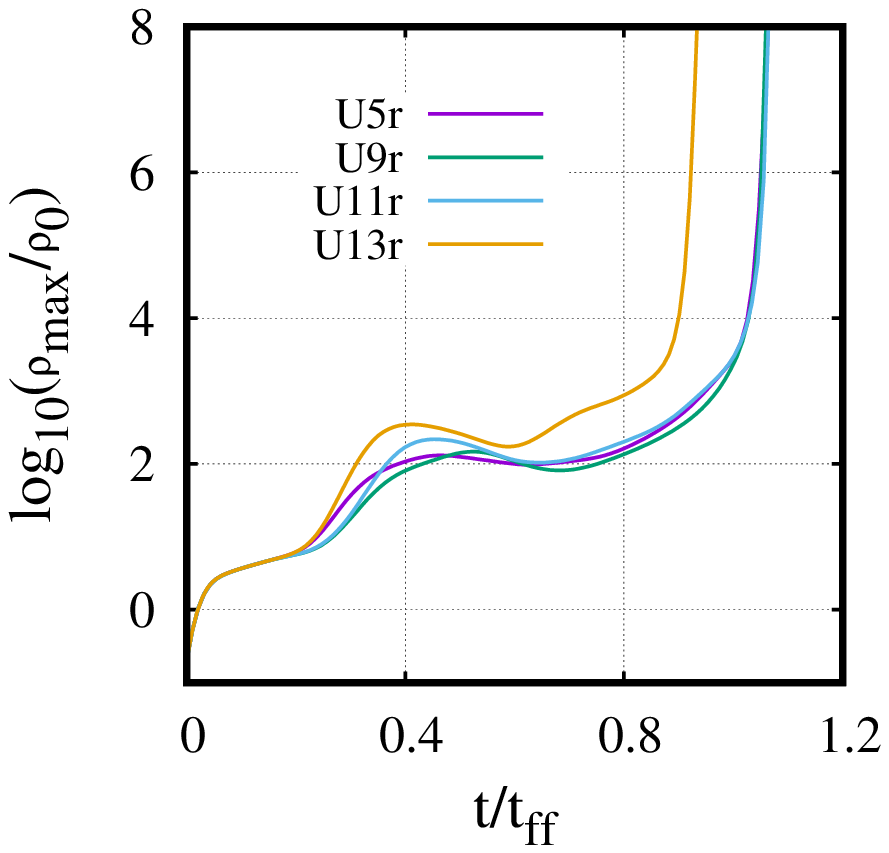}\\
\includegraphics[width=2.5 in]{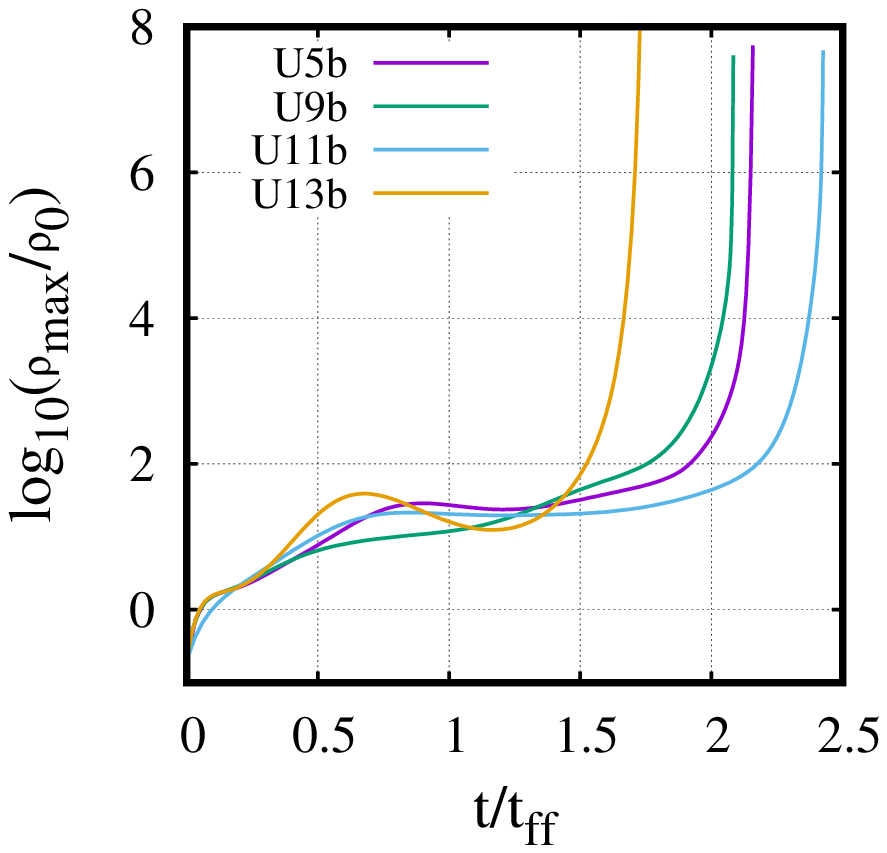} & \includegraphics[width=2.5 in]{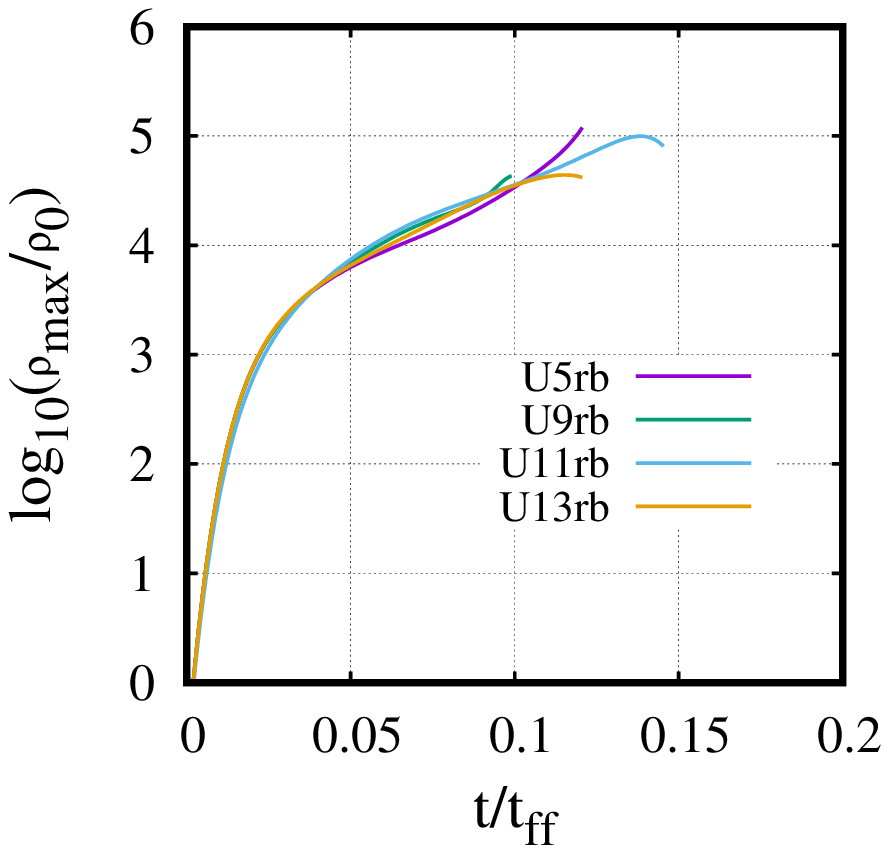} \\
\end{tabular}
\caption{\label{fig:DenMax} Evolution of the density peak for all the models
(top left-hand) $U$ with low level of turbulence;
(top right-hand) $Ur$ with a low azimuthal velocity
(bottom left-hand) $Ub$ with high level of turbulence and
(bottom right-hand) $Urb$ with a high azimuthal velocity.}
\end{center}
\end{figure}

In Fig.\ref{fig:DenMax} we show the time evolution of the global density peak, irrespective
of where the particle with the highest density is located in the simulation volume. As can be seen in this
figure, all of the models collapse at different times (as expected).

%%%%%%%%%%%%%%%%%%%%%%%%%%%%%%%%%%%%%%%%%%%%%%%%%%%%%%%%%%%%%%%%%%%%%%%%%%%%%%%%%%%%%

Let us consider the top left-hand panel of Fig.\ref{fig:DenMax}, the models with a low level
of turbulence. The fastest collapse is that of
model $U13$; followed by the collapse of models $U9$ and
$U5$, respectively. The slowest collapse is that
of model $U11$.

This ordering seems to be a consequence of the gas dragging, which is caused by the
asymmetry in radius and velocity. Consequently, as the gas flows, it is more difficult to
condense by the action of the gravity. Model $U13$ does not show any sign of gas dragging,
because the gas remains of the head-on collision is still around of the pre-collision center. This is
the reason why this model collapses first.

The density peak curve of the model $U9$ very closely follows to that of model $U13$; This happens because the
right-hand sub-cloud acts as a primary member in the binary system formed, which is only slightly perturbed by
the left-hand sub-cloud, that acts as a secondary member of the binary. The collapse takes place first
in the primary, which is more massive.

The density peak curve of the model $U5$ is the third to reach the collapse. This happens because the
right-hand sub-cloud entirely swallows the left-hand sub-cloud during the collision, so it produces a mass perturbation
in the central region, which must first settle down for the collapse to continue. The slowest collapse
is that of model $U11$. This happens because the dragging of the colliding sub-clouds is maximum, so the
mass does not stack easily.

Let us now consider the bottom left-hand panel of Fig.\ref{fig:DenMax},
the models $Ub$ with a high level of turbulence. In this case, the behavior of all
of the curves is the same as that explained earlier for the top panel, but for the panel
of model $Ub$ there is a slight shift to the right to longer
evolution times, above all for the models $U5b$,$U9b$ and $U13b$. The collapse time
of model $U11b$ is almost similar to that observed in model $U11$.

We observe a very significant change in the time scale for the right column panels
of Fig.\ref{fig:DenMax}; in the top right-hand panel, the models $Ur$ that include a
low azimuthal velocity, the collapse is accelerated, so that
the collapse time is shorter than the previous models $U$ and $Ub$
by 40 percent, approximately. In the bottom right-hand panel, the models $Urb$
that include a high azimuthal velocity, the collapse is extremely fast and the time scale
has been reduced to a range from 0.1 to 0.15 $t/t_{ff}$. In addition, the curves for
the models $Urb$ do not show any sign of the early random collisions between the
SPH particles, due to the turbulence spectrum induced on each particle velocity.

%%%%%%%%%%%%%%%%%%%%%%%%%%%%%%%%%%%%%%%%%%%%%%%%%%%%%%%%%%%%%%%%%%%%%%%%%%%%%%%%%%%%%%%
\subsection{Column density plots}
\label{subsec:col}

The main outcome of the collision models is shown by means of column density
plots of a thin slice of gas, parallel to the x-y plane. To make a proper comparison
between the different models, we have selected for each model a snapshot whose peak density is such
that $\log \left( \rho_{\rm max}/\rho_0 \right)  \approx 5$.

%%%%%%%%%%%%%%%%%%%%%%%%%%%%%%%%%%%%%%%%%%%%%%%%%%%%%%%%%%%%%%%
\begin{figure}
\begin{center}
\begin{tabular}{cc}
\includegraphics[width=2.0 in]{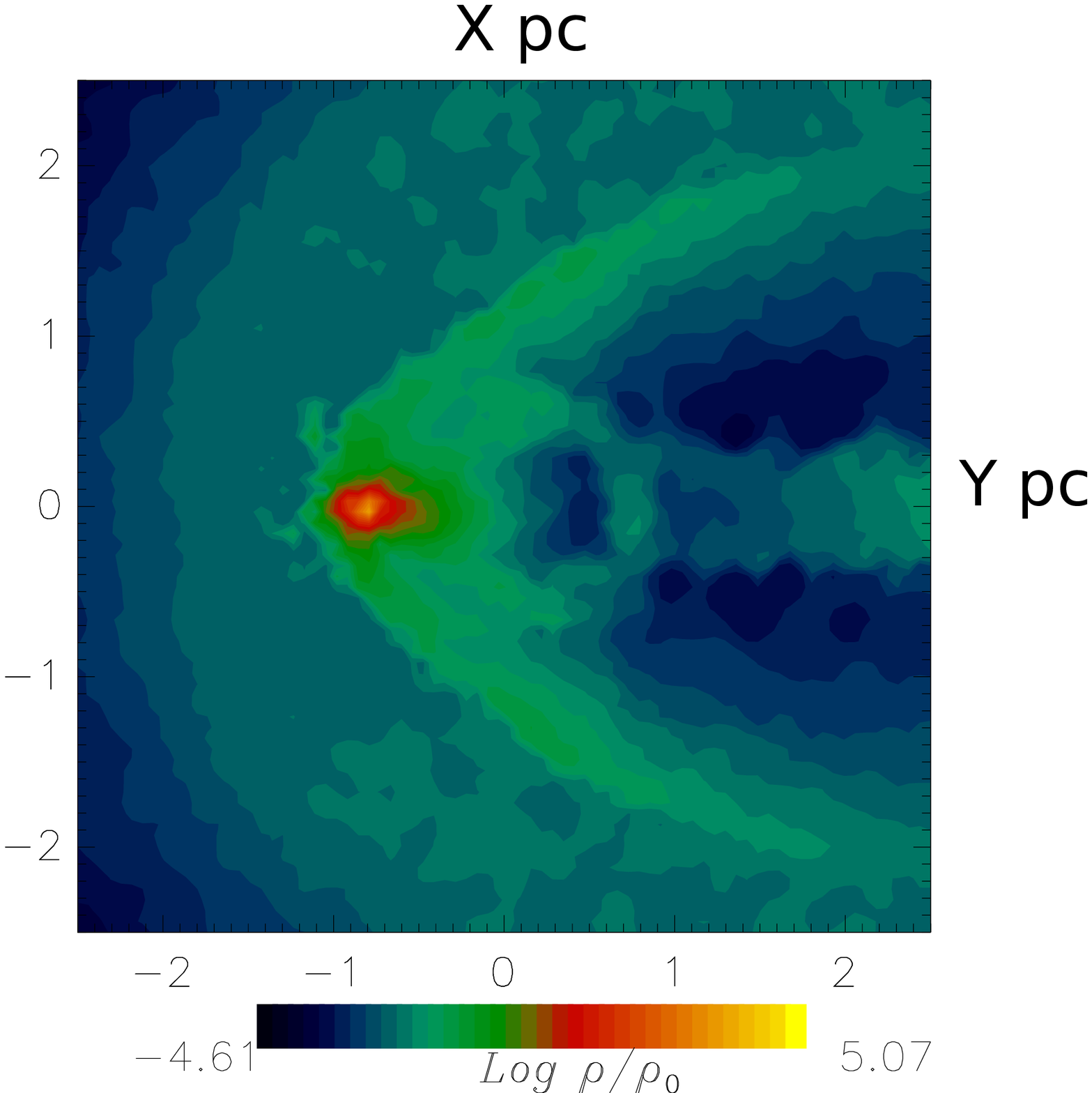} &
\includegraphics[width=2.0 in]{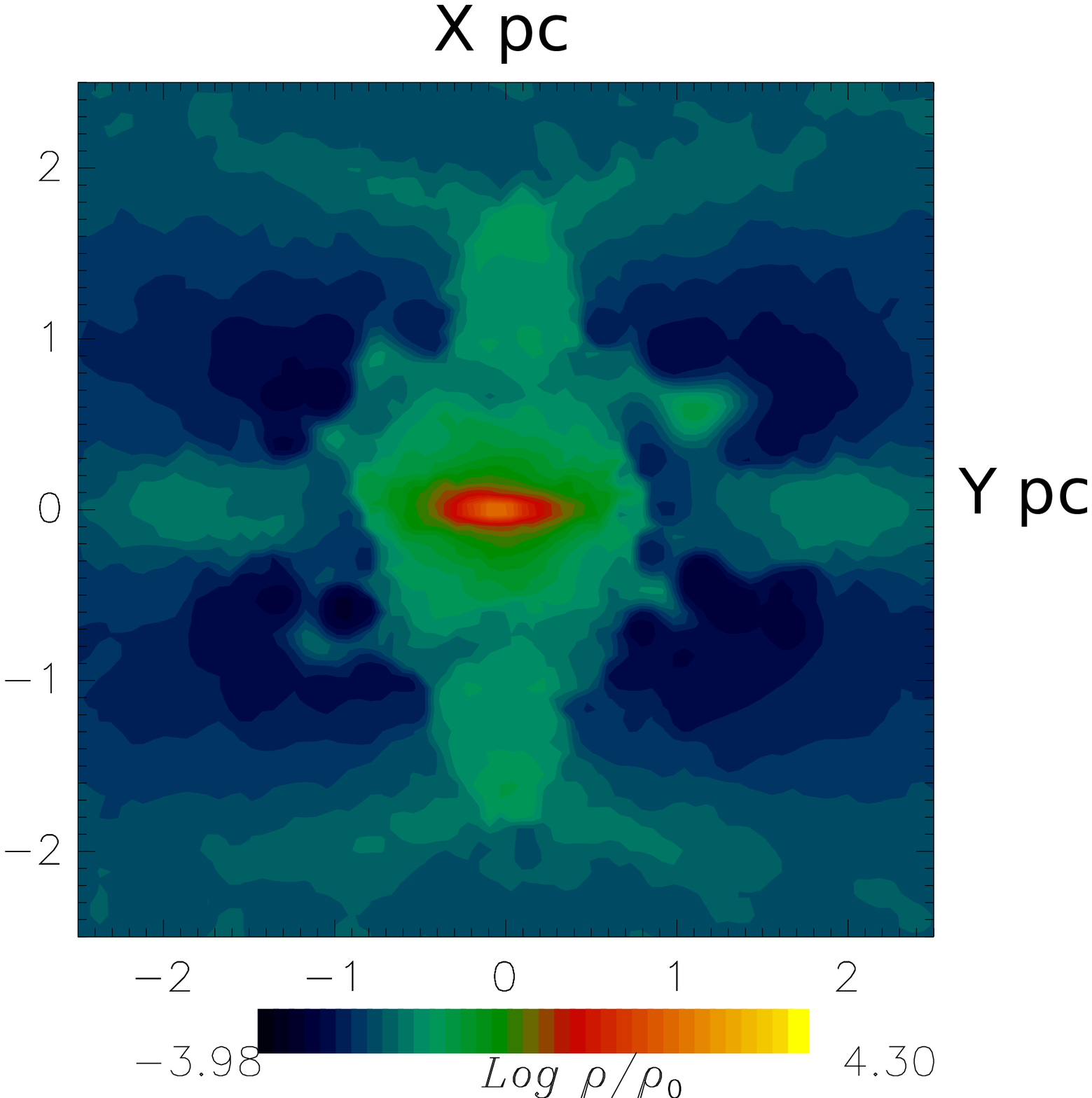}\\
\includegraphics[width=2.0 in]{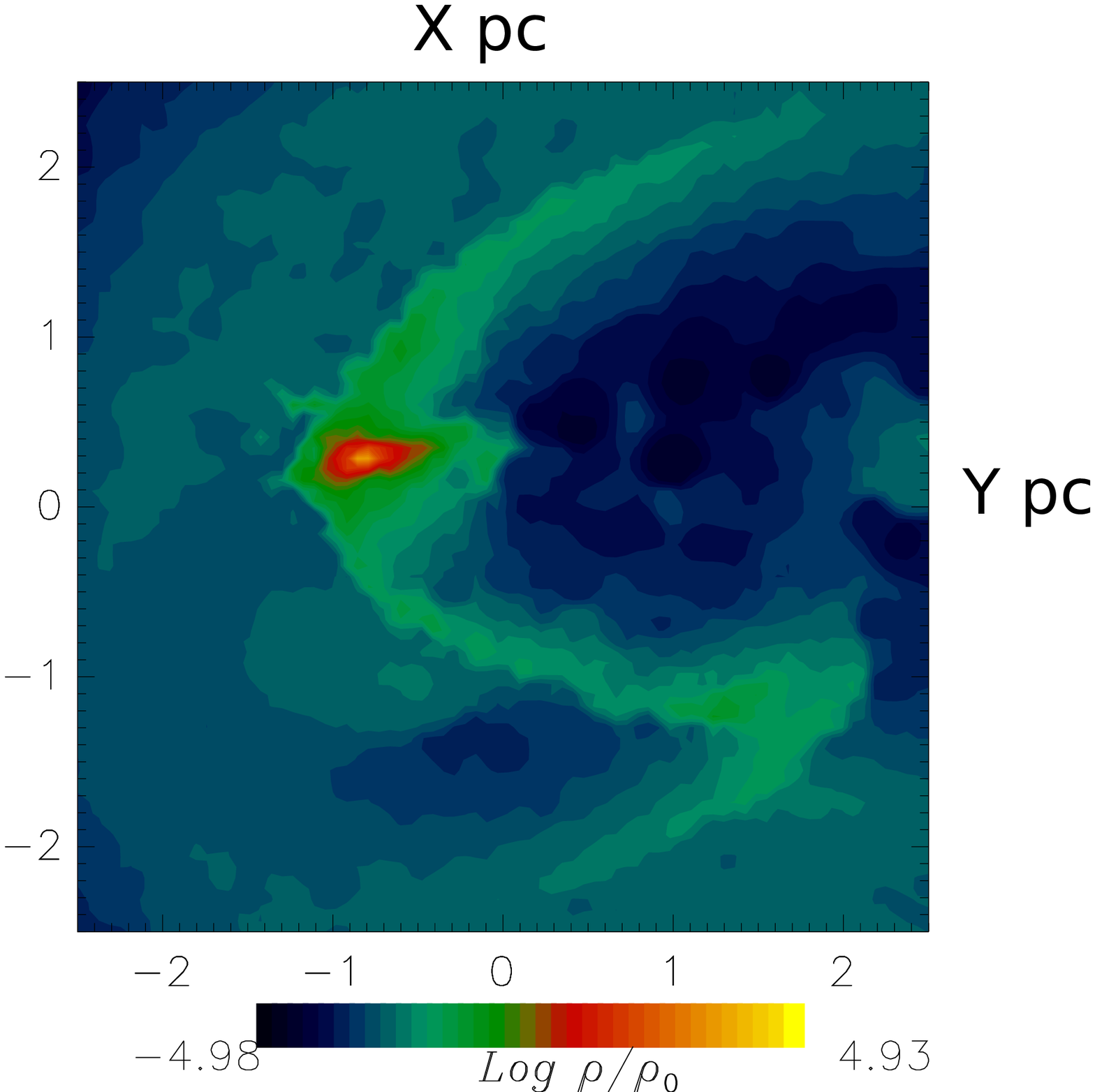} &
\includegraphics[width=2.0 in]{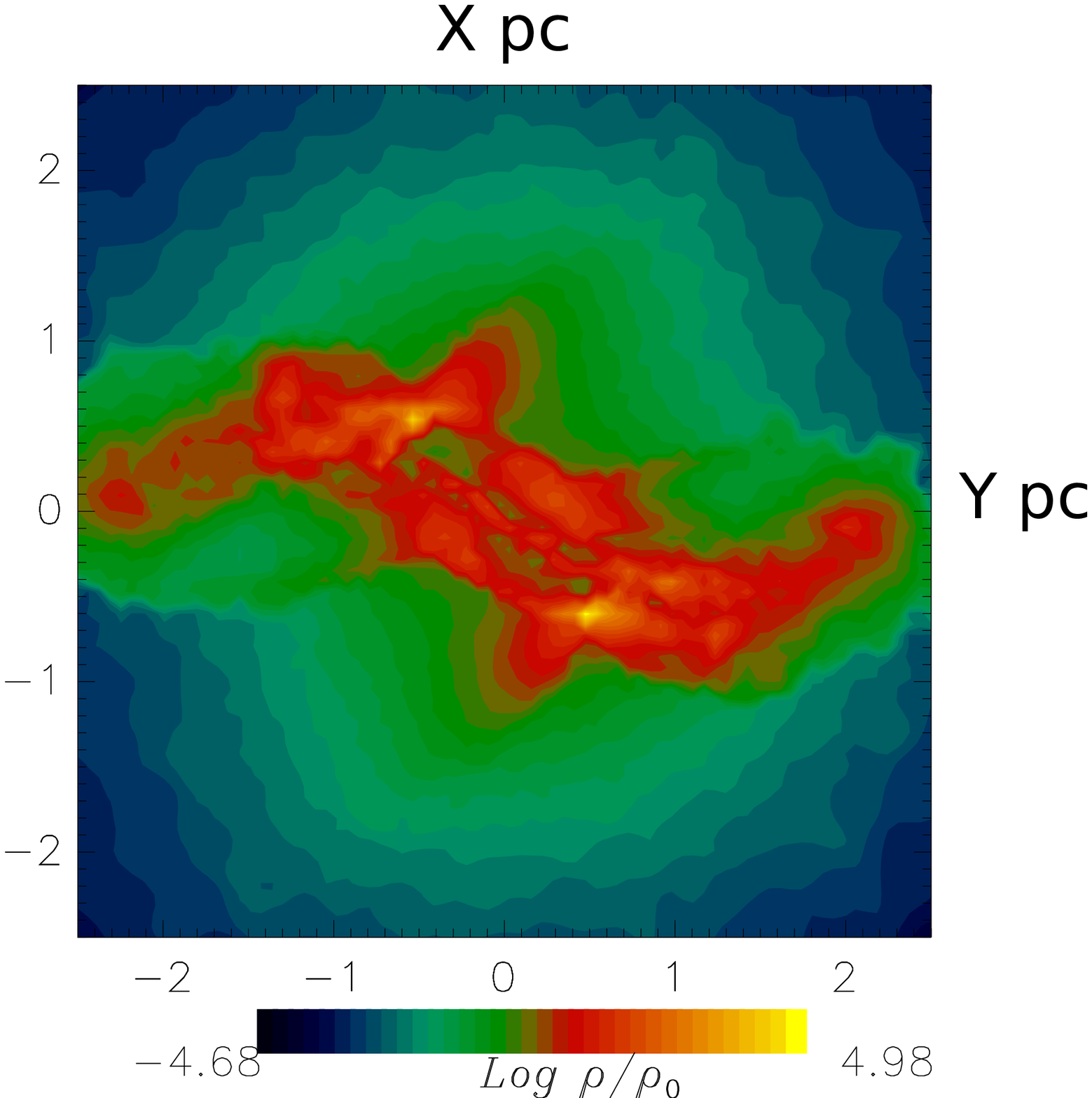}\\
\end{tabular}
\caption{\label{Mosps} Column density plots of the collision models $U$, for a thin slice of gas parallel
to the x-y plane. The unit of length is one parsec. The models are shown in panels as follows:
(top left-hand)
$U5$ (at time $t/t_{ff}=1.65$ and peak density $\log \left( \rho_{\rm max}/\rho_0 \right)=5.0$);
(top right-hand) $U13$ (at time $t/t_{ff}=1.34$ and peak density $\log \left( \rho_{\rm max}/\rho_0 \right)=4.3$);
(bottom left-hand) $U9$ (at time $t/t_{ff}=1.42$ and peak density $\log \left( \rho_{\rm max}/\rho_0 \right)=5.0$);
(bottom right-hand) $U11$ (at time $t/t_{ff}=2.4$ and peak density $\log \left( \rho_{\rm max}/\rho_0 \right)=5.0$).}
\end{center}
\end{figure}
%%%%%%%%%%%%%%%%%%%%%%%%%%%%%%%%%%%%%%%%%%%%%%%%%

In model $U5$, once the head-on collison
between the left-hand and right-hand sub-clouds has taken place, the asymmetry in the original 
sub-clouds in both radius and translational velocity, makes the right-hand sub-cloud 
(the biggest and the fastest cloud) swallow and drag the left-hand sub-cloud 
(the smallest and slowest cloud). The resulting stirred gas oscillates
from the left-hand side to the right-hand side along the x-axis. The spatial symmetry in 
the original configuration of the right-hand sub-cloud is translated to 
the symmetry in the arms developed around the central cloudlet, as can be
seen in the top left-hand panel of Fig.\ref{Mosps}. In Section \ref{subsec:peak}, this phenomenon was
simply referred to as gas dragging, which was a useful way to explain the
time of collapse by means of the density peak curves.

When the asymmetry in the radius is removed, the final result is a central cloudlet, that is 
elongated along the x-axis, with a strong bipolar outflow along the y-axis, as can be seen in the 
top right-hand panel of Fig.\ref{Mosps}, which is the outcome of model $U13$. Because of this result, we note that
the asymmetry in the velocities of the collision model is not as important as the asymmetry in the
radii with respect to the outcome of the simulations.

Different results are obtained when an impact parameter is taken into account. In the case of model 
$U9$, as illustrated in the bottom left-hand panel of Fig.\ref{Mosps}, the
asymmetry in the radii and velocities together with an impact parameter produce a weak binary system, in which the
the left-hand sub-cloud (the smallest and slowest cloud) passes by and is attracted by the right-hand sub-cloud 
(the biggest and fastest cloud), so that part of the mass of the former is pulled out. Nevertheless, an arm
is still visible around the remains of the pre-collision left-hand sub-cloud and a long arm is also developed
around the central cloudlet, which is analogous in origin to the arm formed in model $U5$.

In the case of the model $U11$, when the symmetry in the radii of the pre-collision sub-clouds is restored
but still in the presence of the impact parameter as in model $U9$, a binary system is formed 
as the main outcome, in which several gas bridges are seen to be strongly connecting
strongly the remains of the two colliding clumps, as can be seen in the bottom right-hand panel of Fig.\ref{Mosps}.

%%%%%%%%%%%%%%%%%%%%%%%%%%%%%%%%%%%%%%%%%%%%%%%%%
\begin{figure}
\begin{center}
\begin{tabular}{cc}
\includegraphics[width=2.0 in]{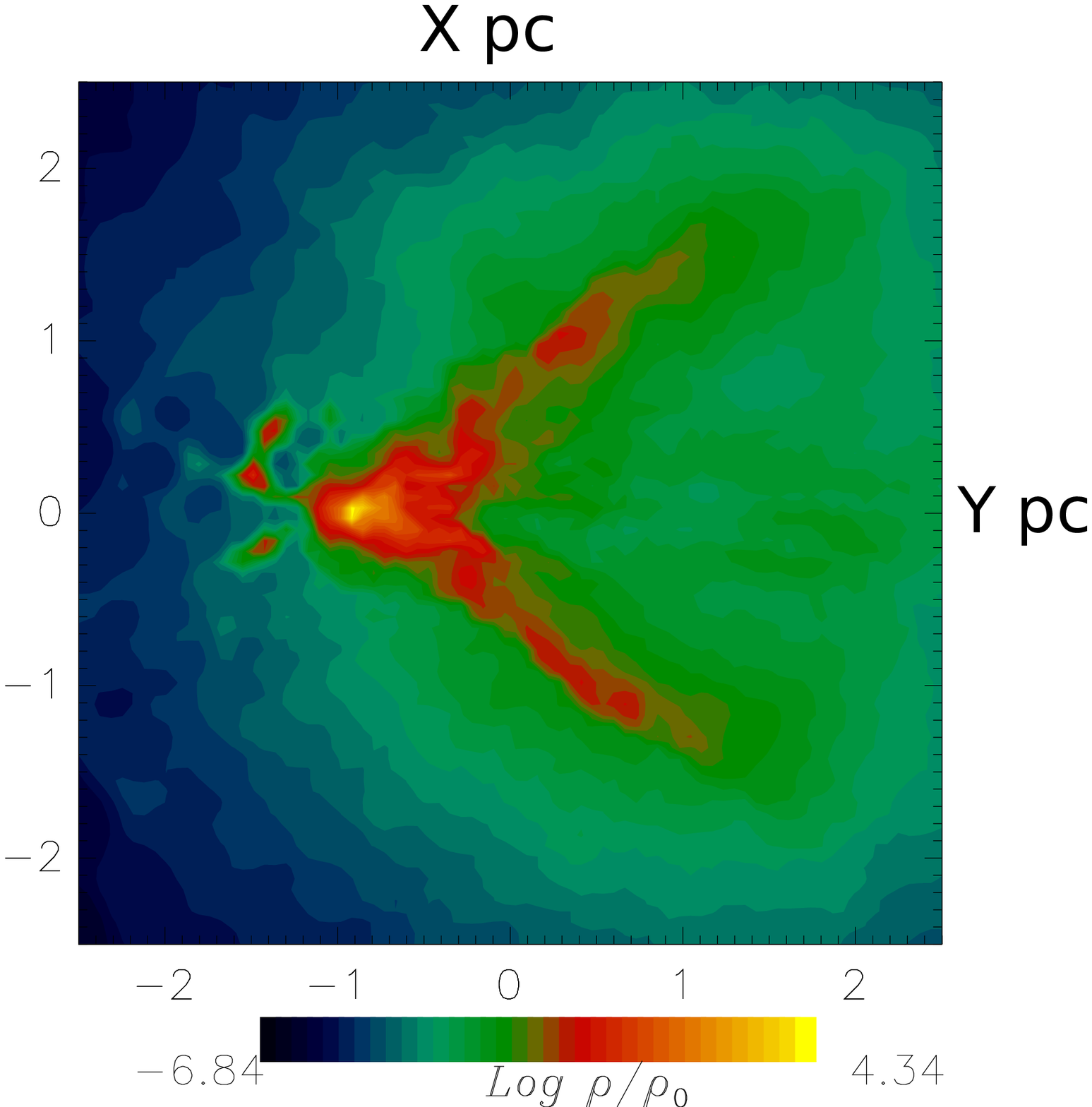} &
\includegraphics[width=2.0 in]{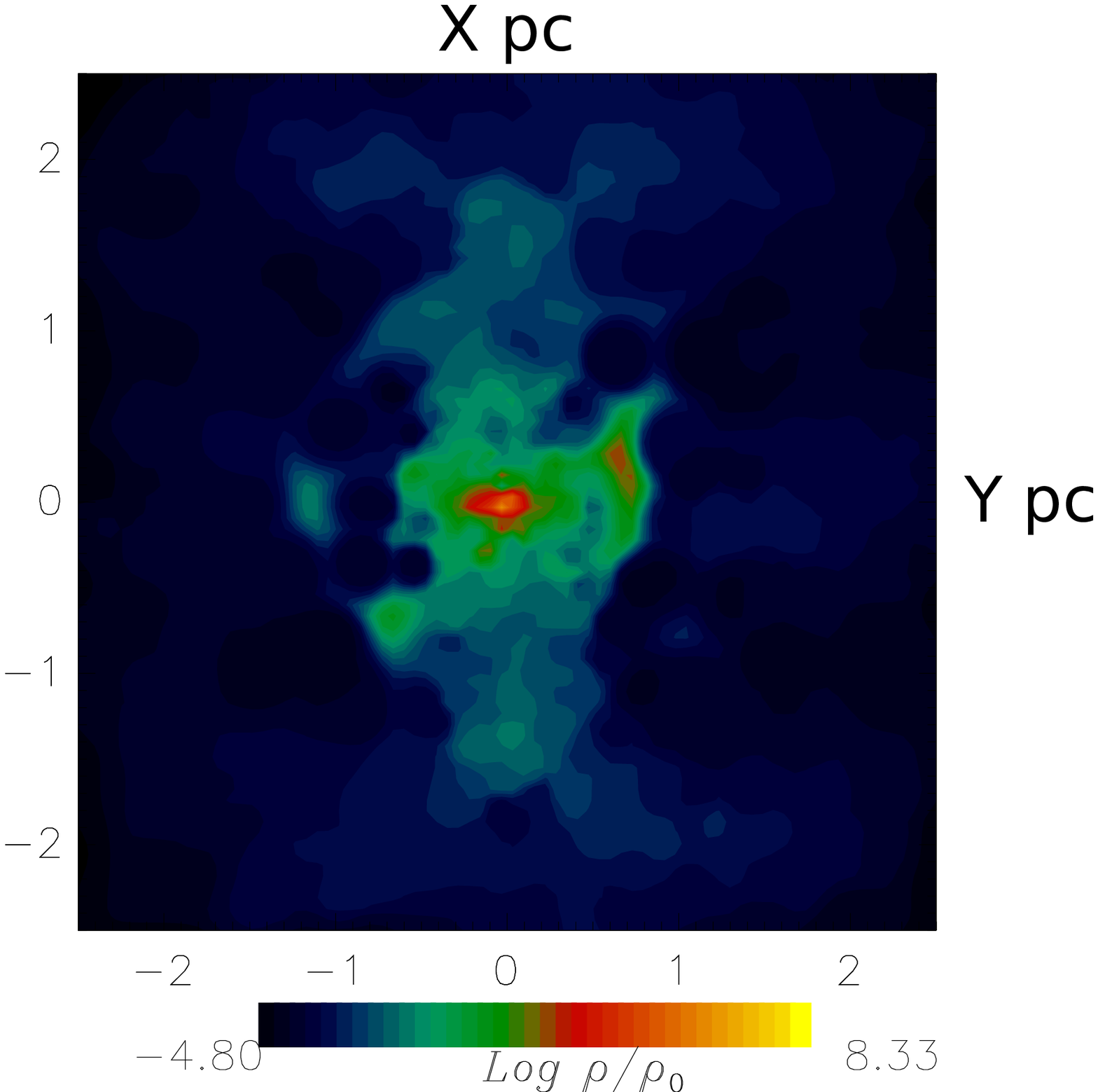}\\
\includegraphics[width=2.0 in]{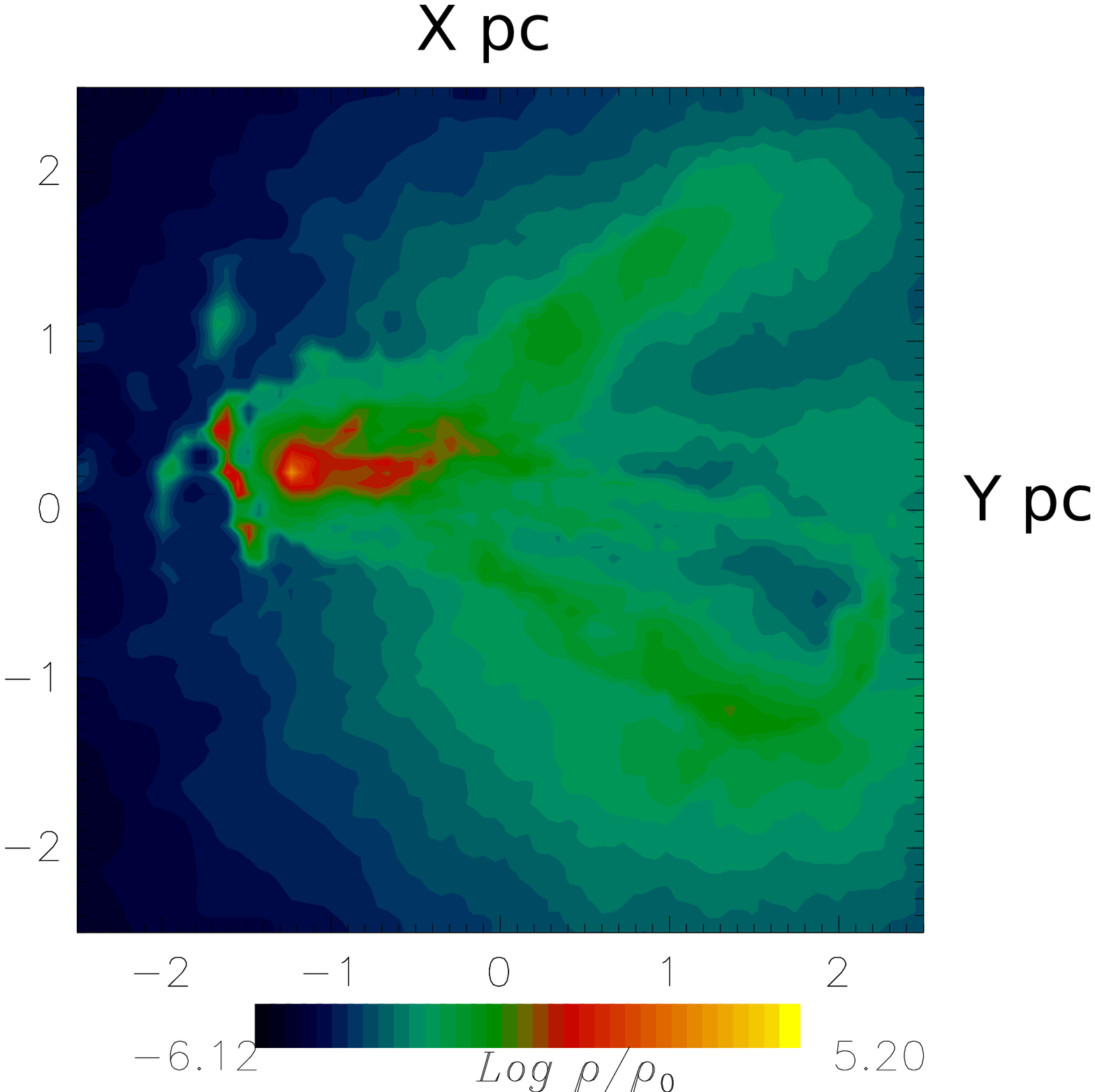} &
\includegraphics[width=2.0 in]{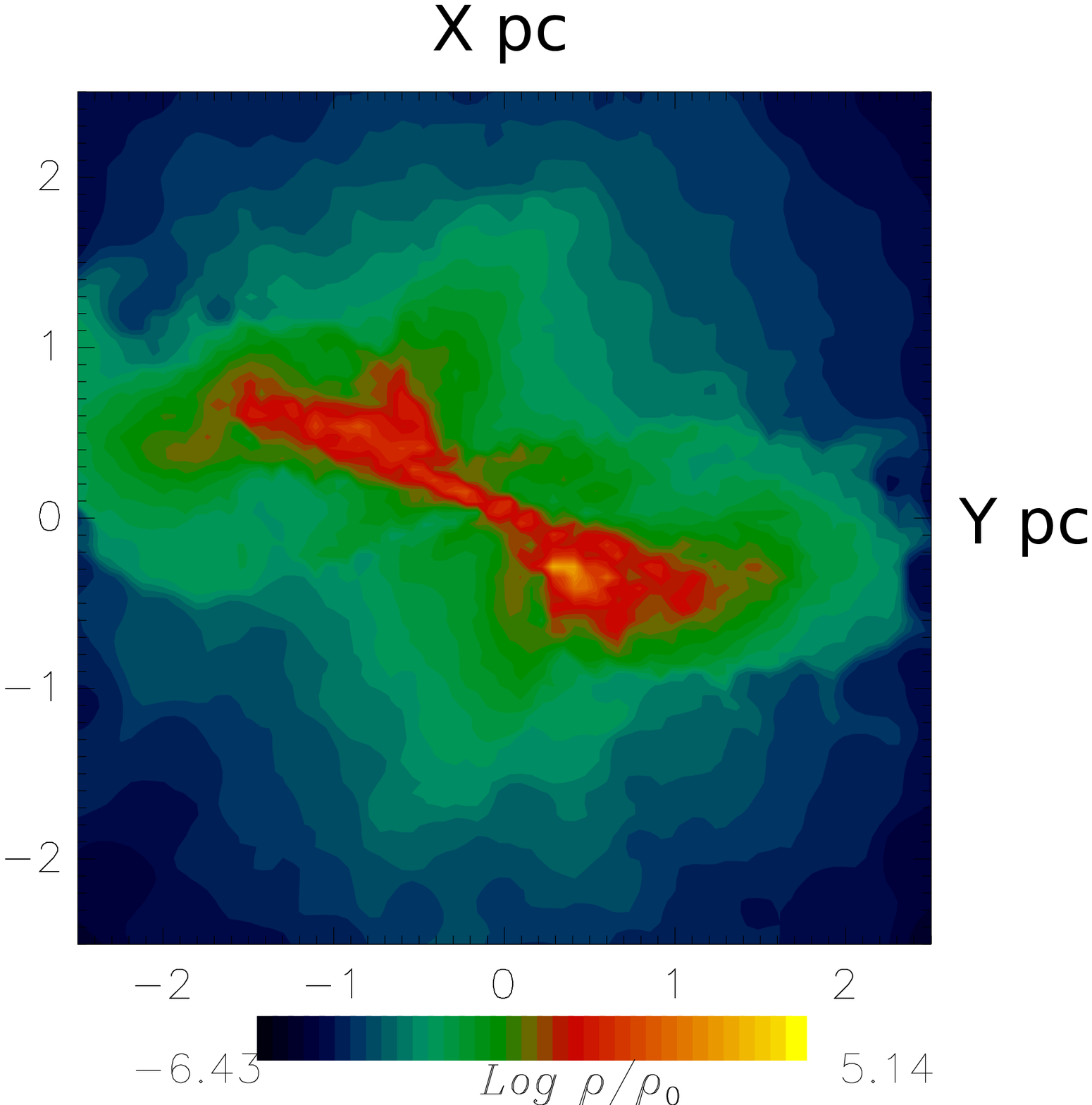}\\
\end{tabular}
\caption{\label{Mospsb2} Column density plots of the collision models $Ub$, for a thin slice of gas parallel
to the x-y plane. The unit of length is one parsec. The models are shown in panels as follows:
(top left-hand)
$U5b$ (at time $t/t_{ff}=2.15$ and peak density $\log \left( \rho_{\rm max}/\rho_0 \right)=4.34$);
(top right-hand) $U13b$ (at time $t/t_{ff}=1.73$ and peak density $\log \left( \rho_{\rm max}/\rho_0 \right)=8.63$);
(bottom left-hand) $U9b$ (at time $t/t_{ff}=2.08$ and peak density $\log \left( \rho_{\rm max}/\rho_0 \right)=5.20$);
(bottom right-hand) $U11b$ (at time $t/t_{ff}=2.42$ and peak density $\log \left( \rho_{\rm max}/\rho_0 \right)=5.14$).}
\end{center}
\end{figure}

In Fig.\ref{Mospsb2} we show the column density plots of the models $Ub$, with a high level of turbulence.
As expected, there is a lot of similarity with the previous models $U$, with a low level of turbulence, because
the initial structure of the velocity spectrum is the same for both models $U$ and $Ub$.
The only difference is the magnitude of the velocity. The larger magnitude of the velocity for models $Urb$ makes
the arms and tails larger and better defined than those in models $U$.

\begin{figure}
\begin{center}
\begin{tabular}{cc}
\includegraphics[width=1.8 in]{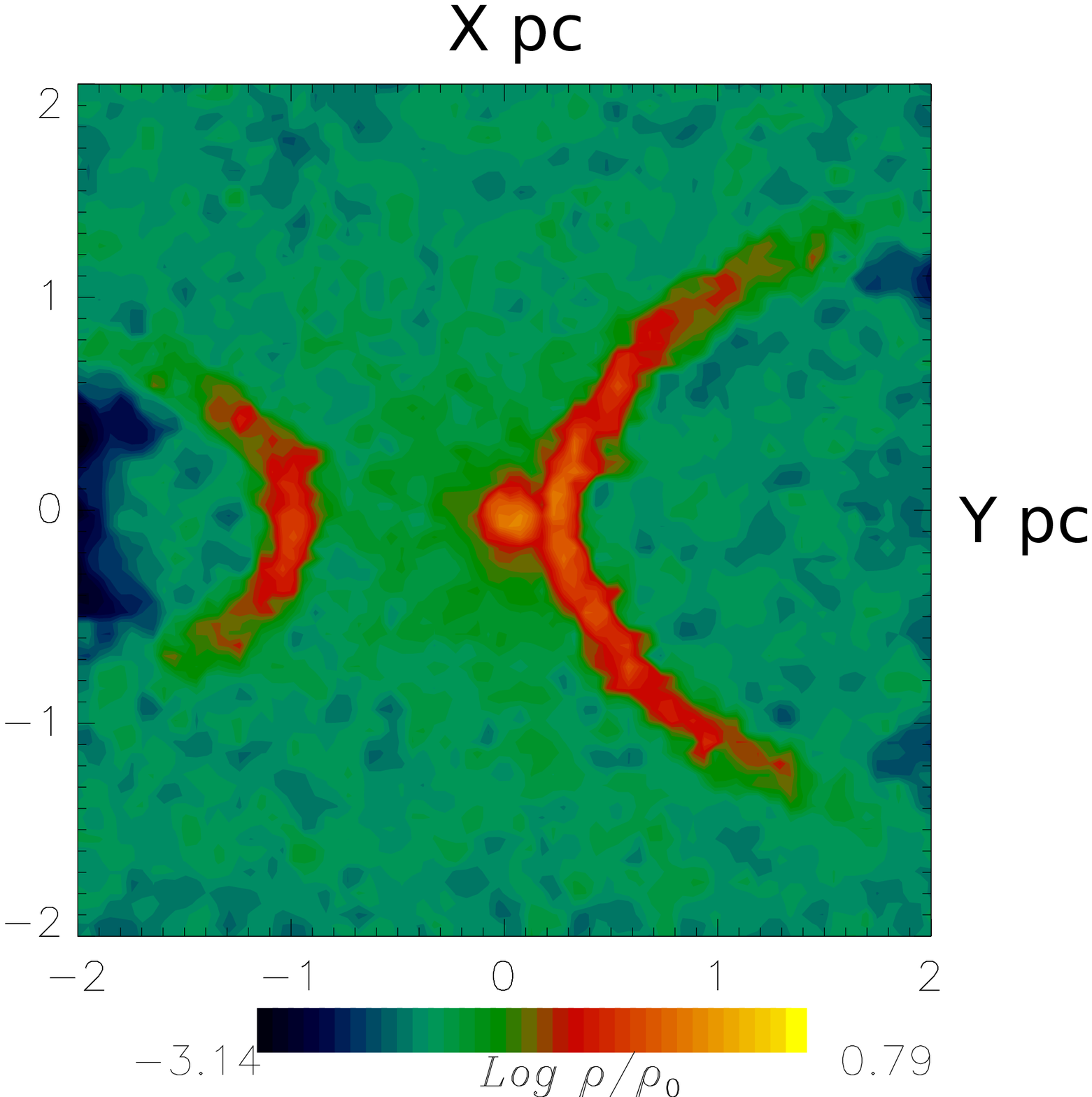} &
\includegraphics[width=1.8 in]{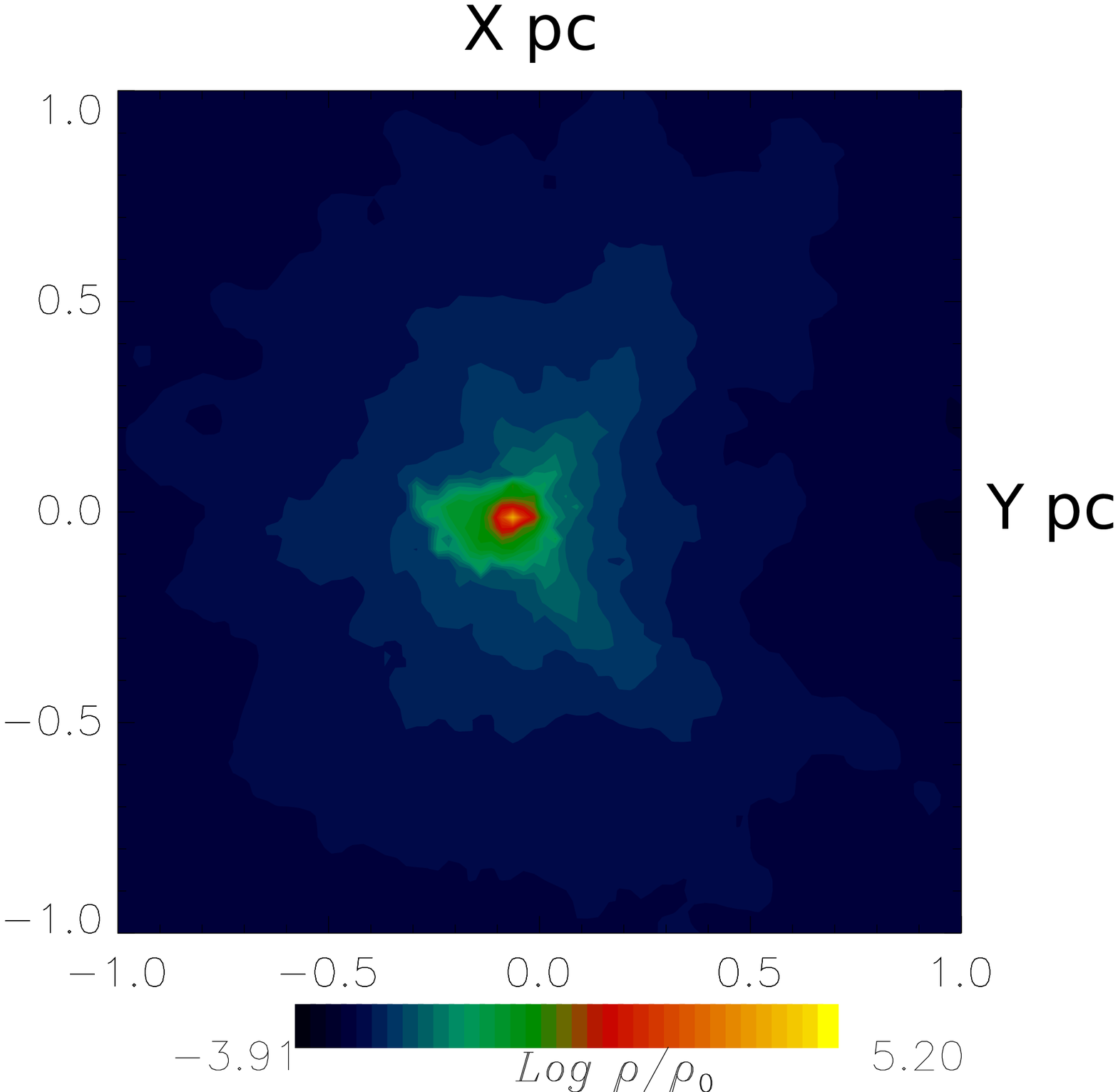}\\
\includegraphics[width=1.8 in]{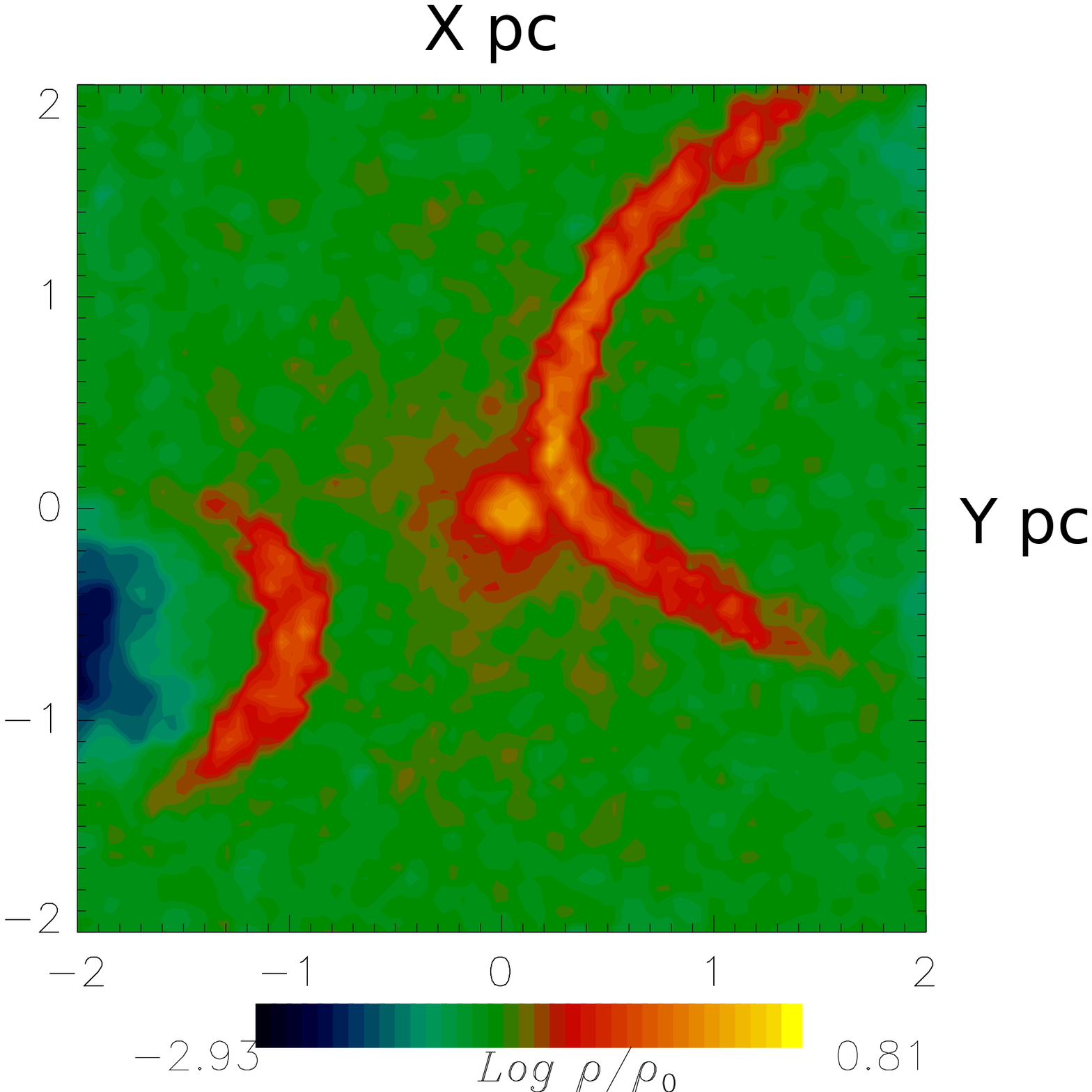} &
\includegraphics[width=1.8 in]{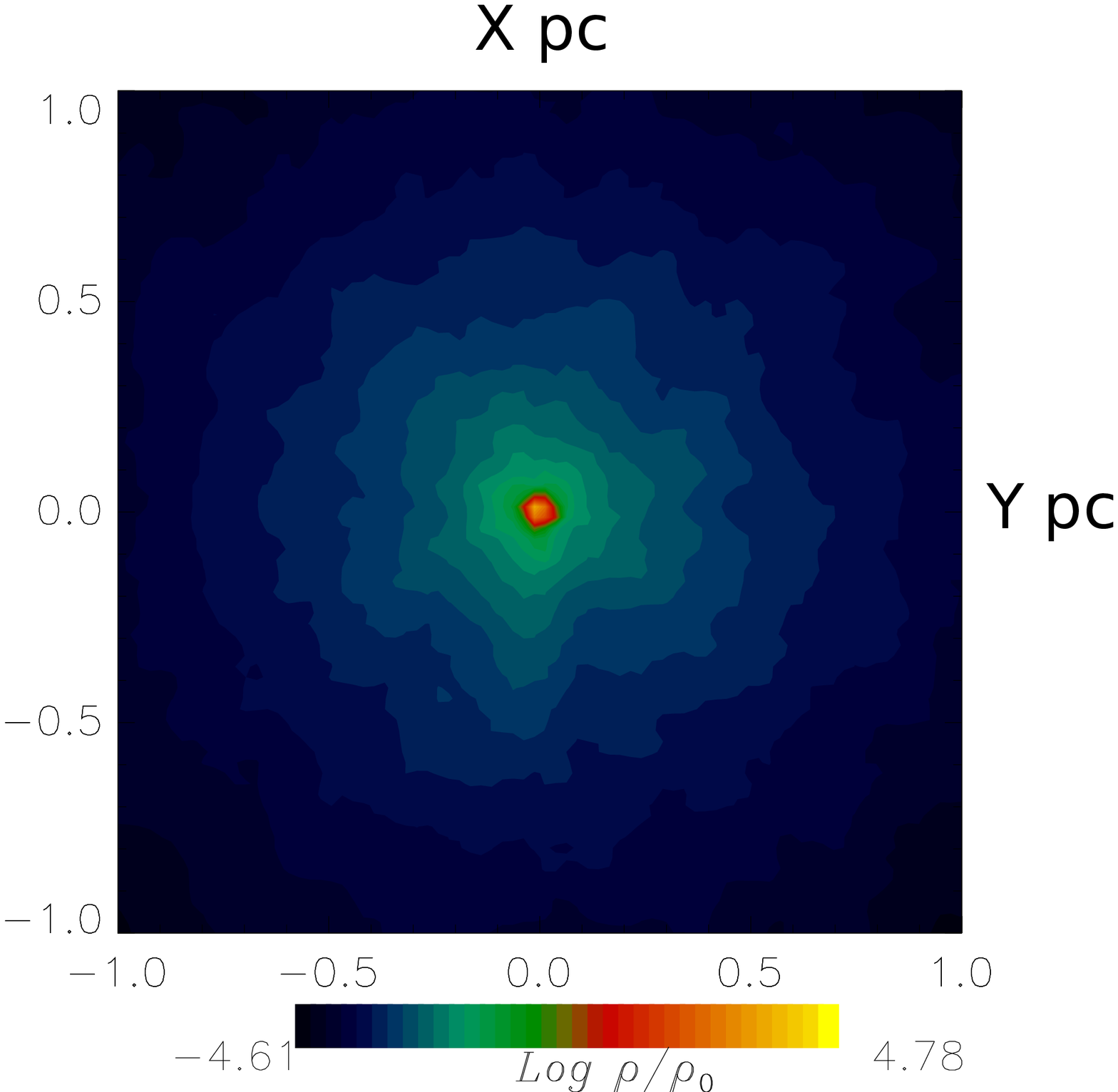}\\
\includegraphics[width=1.8 in]{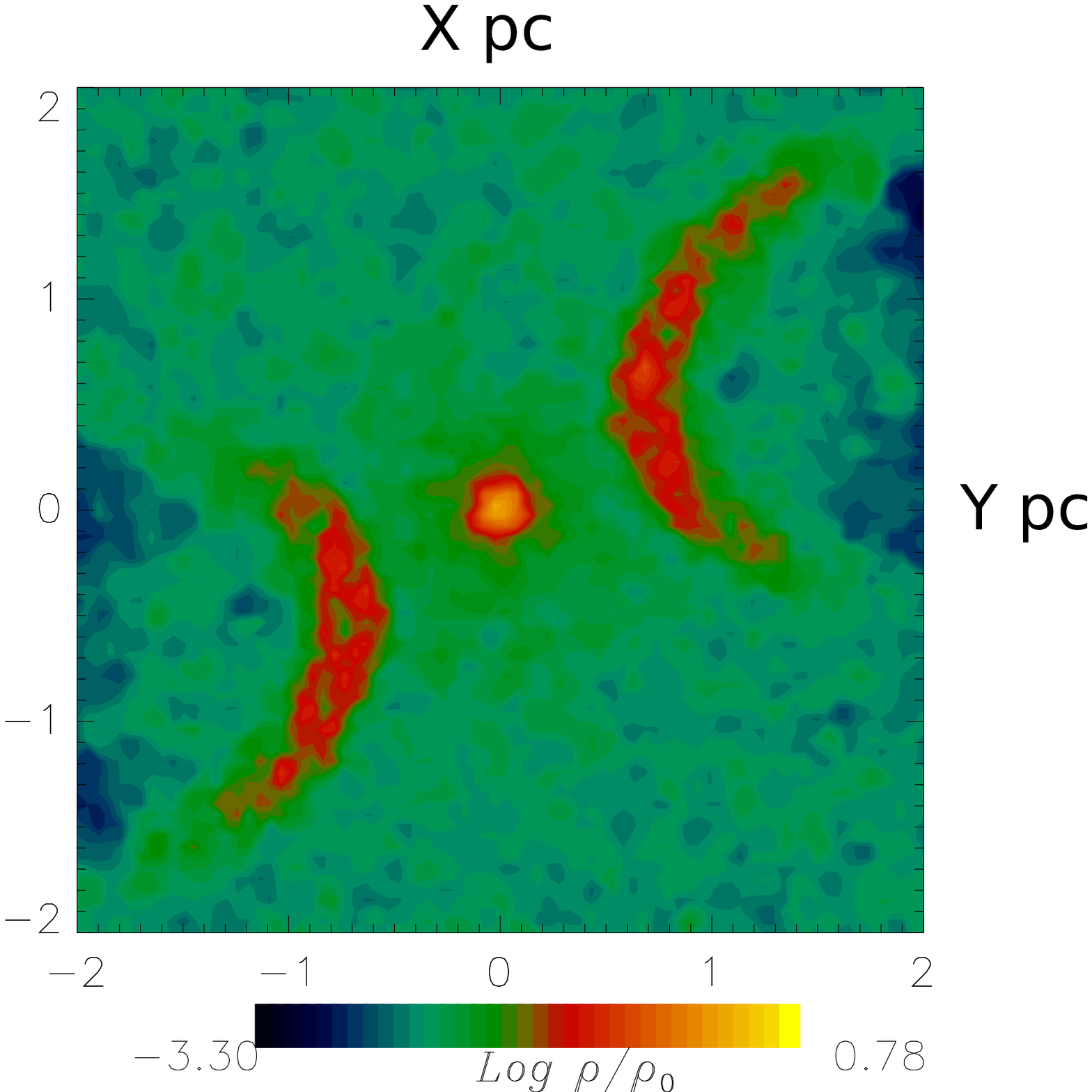} &
\includegraphics[width=1.8 in]{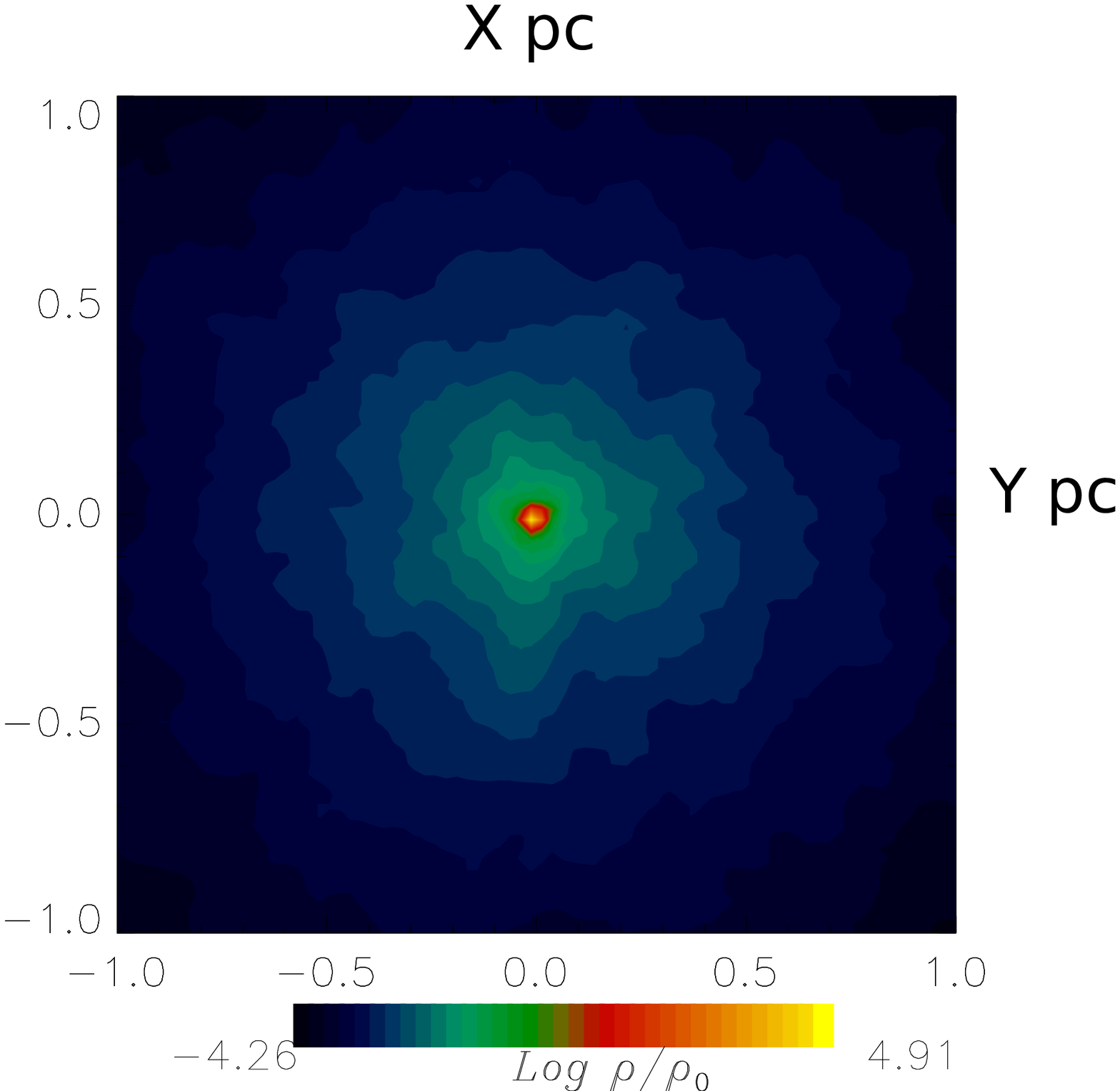}\\
\includegraphics[width=1.8 in]{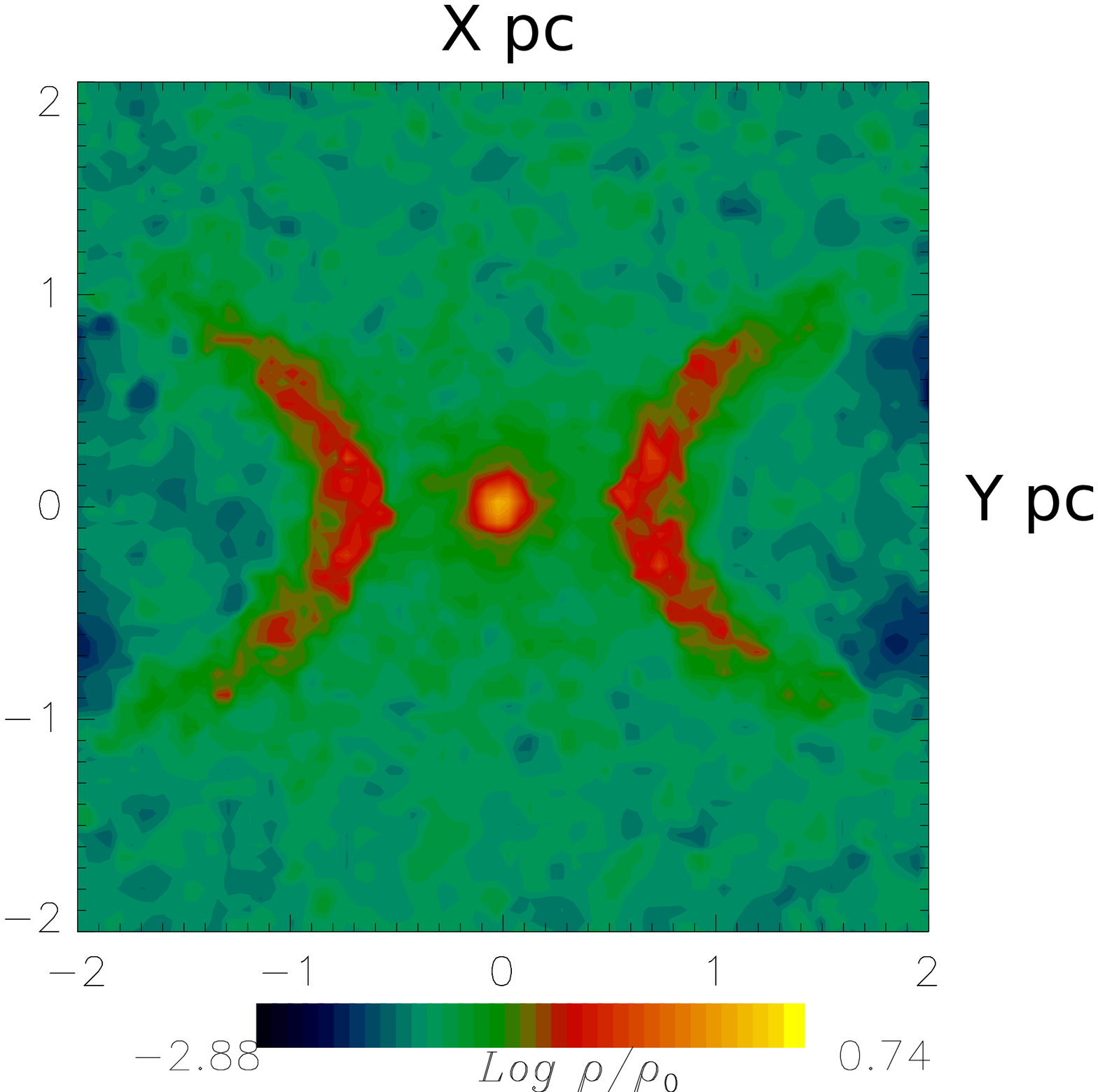} &
\includegraphics[width=1.8 in]{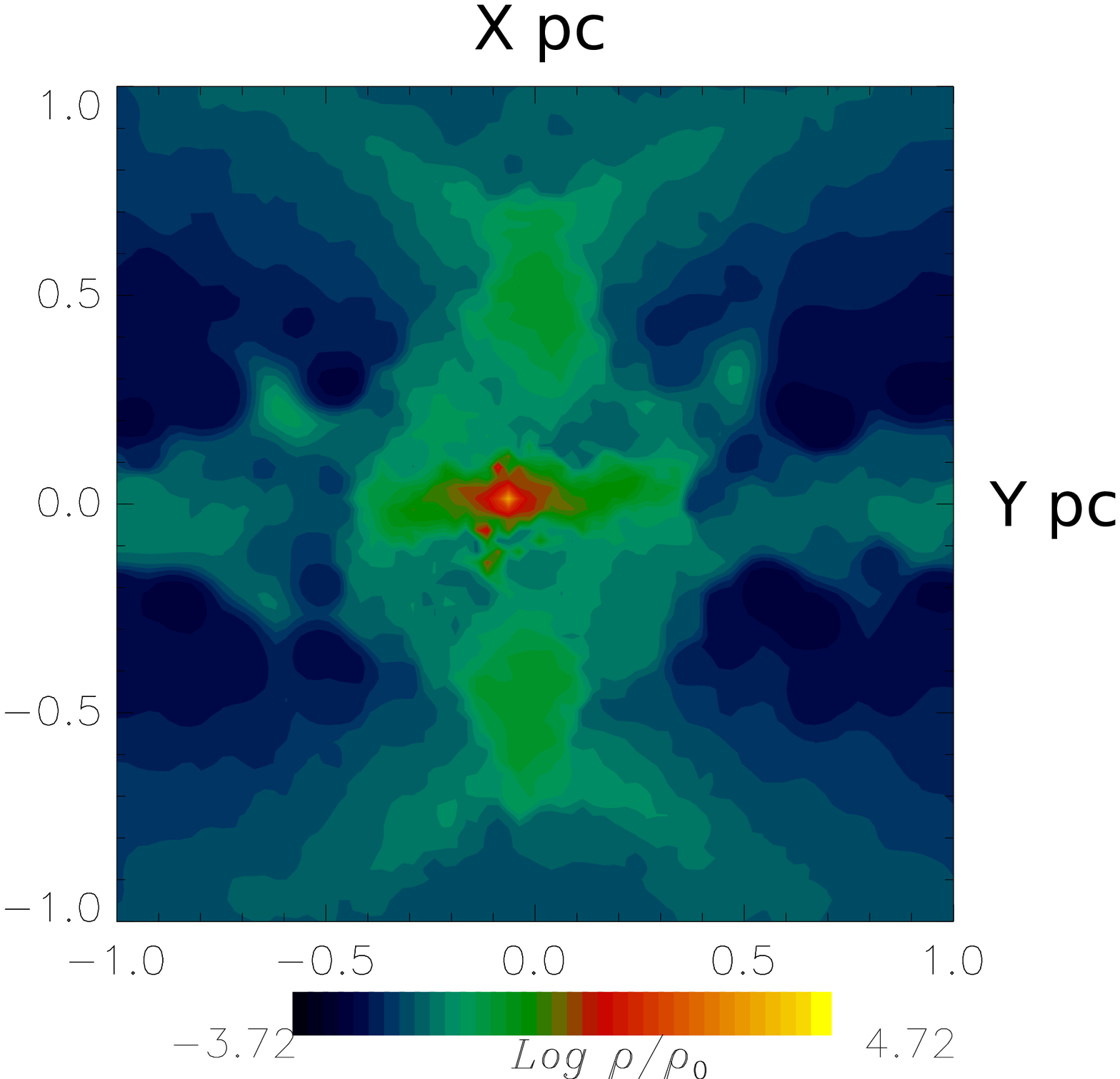}\\
\end{tabular}
\caption{\label{MospsRot} Column density plots of the collision models $Ur$, with an azimuthal velocity included, see
Section\ref{subs:Vcir}. The models are shown in panels (from top to bottom) as follows:
(first line) Model $U5r$,
(left) at time $t/t_{ff}=0.24$, peak density $\log \left( \rho_{\rm max}/\rho_0 \right)=0.79$ and
(right) at time $t/t_{ff}=1.05$, peak density $\log \left( \rho_{\rm max}/\rho_0 \right)=5.2$;
(second top line) Model $U9r$,
(left) at time $t/t_{ff}=0.25$, peak density $\log \left( \rho_{\rm max}/\rho_0 \right)=0.8$ and
(right) at time $t/t_{ff}=1.05$, peak density $\log \left( \rho_{\rm max}/\rho_0 \right)=4.8$;
(third line) Model $U11r$,
(left) at time $t/t_{ff}=0.23$, peak density $\log \left( \rho_{\rm max}/\rho_0 \right)=0.78$ and
(right) at time $t/t_{ff}=1.05$, peak density $\log \left( \rho_{\rm max}/\rho_0 \right)=4.9$;
(fourth line) Model $U13r$,
(left) at time $t/t_{ff}=0.22$, peak density $\log \left( \rho_{\rm max}/\rho_0 \right)=0.74$ and
(right) at time $t/t_{ff}=0.92$, peak density $\log \left( \rho_{\rm max}/\rho_0 \right)=4.7$.}
\end{center}
\end{figure}

Let us now consider the iso-density plots for models $Ur$, which are shown in the Fig.\ref{MospsRot}.
In this case, we show two columns of density plots. In the right-hand column of Fig.\ref{MospsRot}, we show the
density plots for the snapshots with almost the same density peak shown in Fig.\ref{Mosps} to allow comparison with
models $U$ and $Ub$. We only observe an homogeneous and spherical collapse as the final result of simulations
$Ur$.

In the left-hand column of Fig.\ref{MospsRot}, we choose snapshots
of the first stage of evolution, to show the early development of a central
lump of gas, which is a direct consequence of the azimuthal velocity added to the particle velocity, see
Section\ref{subs:Vcir}. This central lump of gas makes the collapse of the cloud faster, as can be seen
in Fig.\ref{fig:DenMax}, because it acts as a centrally located mass attractor.
%%%%%%%%%%%%%%%%%%%%%%%%%%%%%%%%%%%%%%%%%%%%%%%%%%%%%%%%%%%%%%%%%%%%%%%%%%%%%%%%%%%%%%%%%%

\begin{figure}
\begin{center}
\begin{tabular}{cc}
\includegraphics[width=2.0 in]{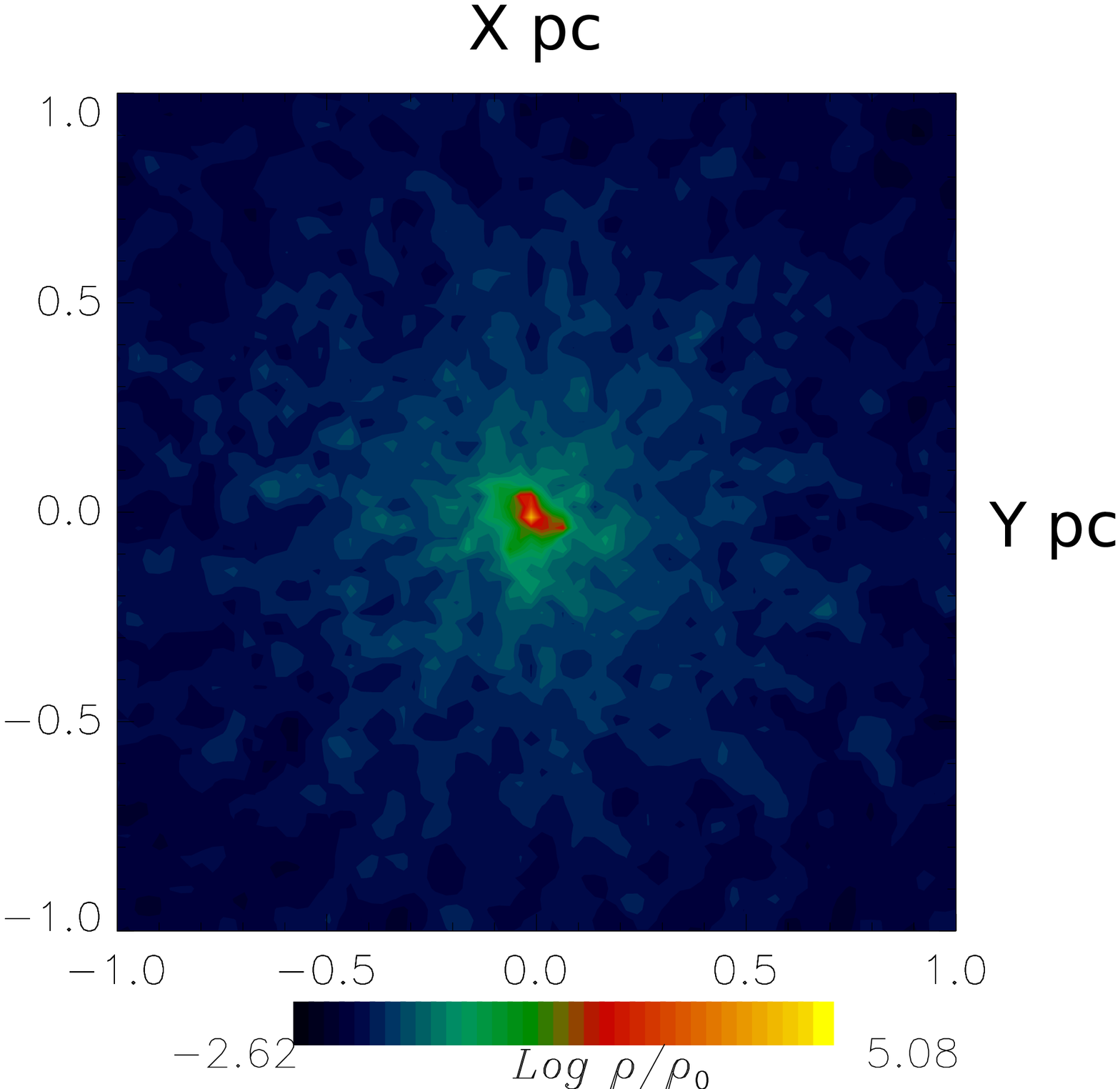} &
\includegraphics[width=2.0 in]{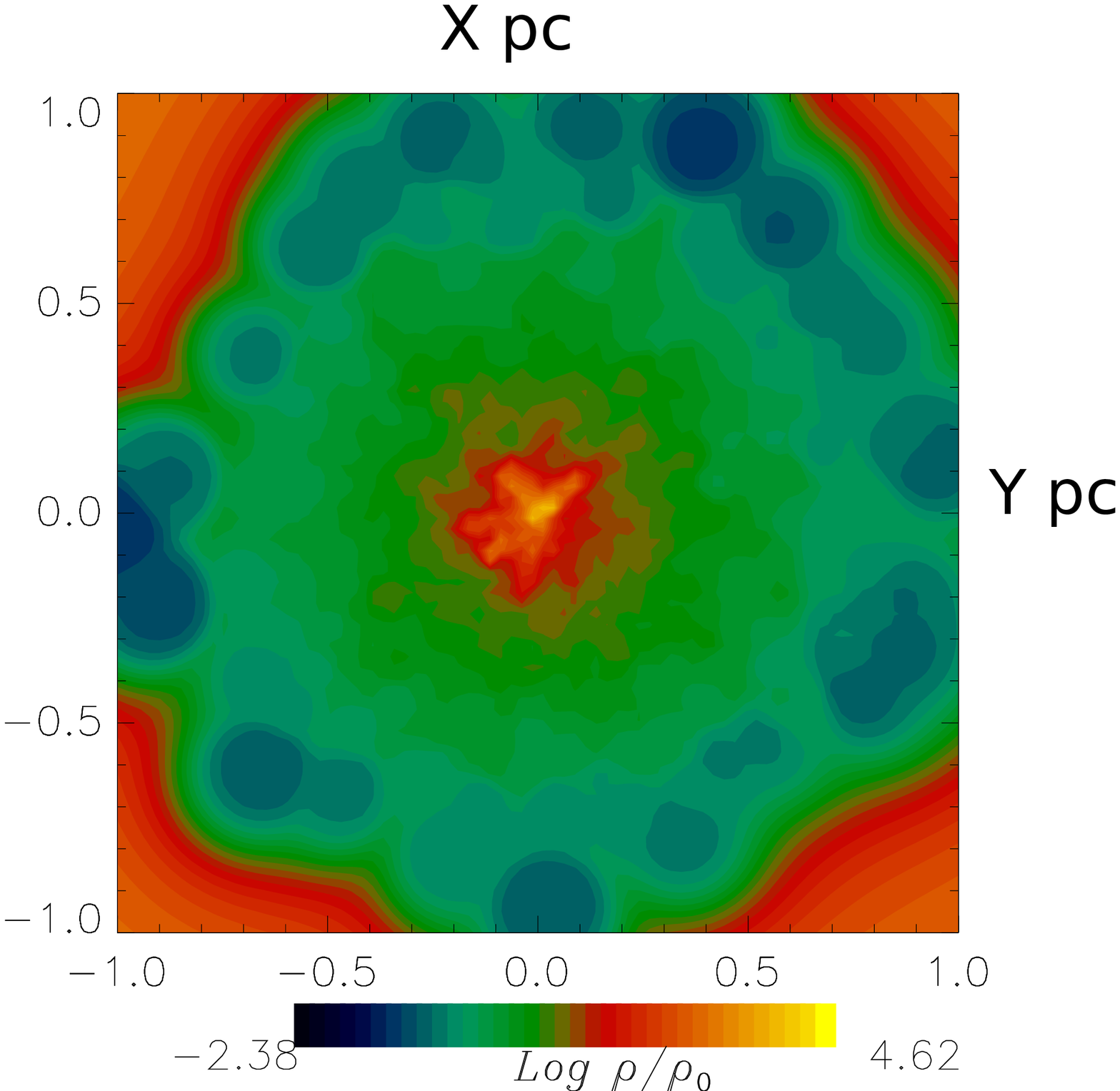}\\
\includegraphics[width=2.0 in]{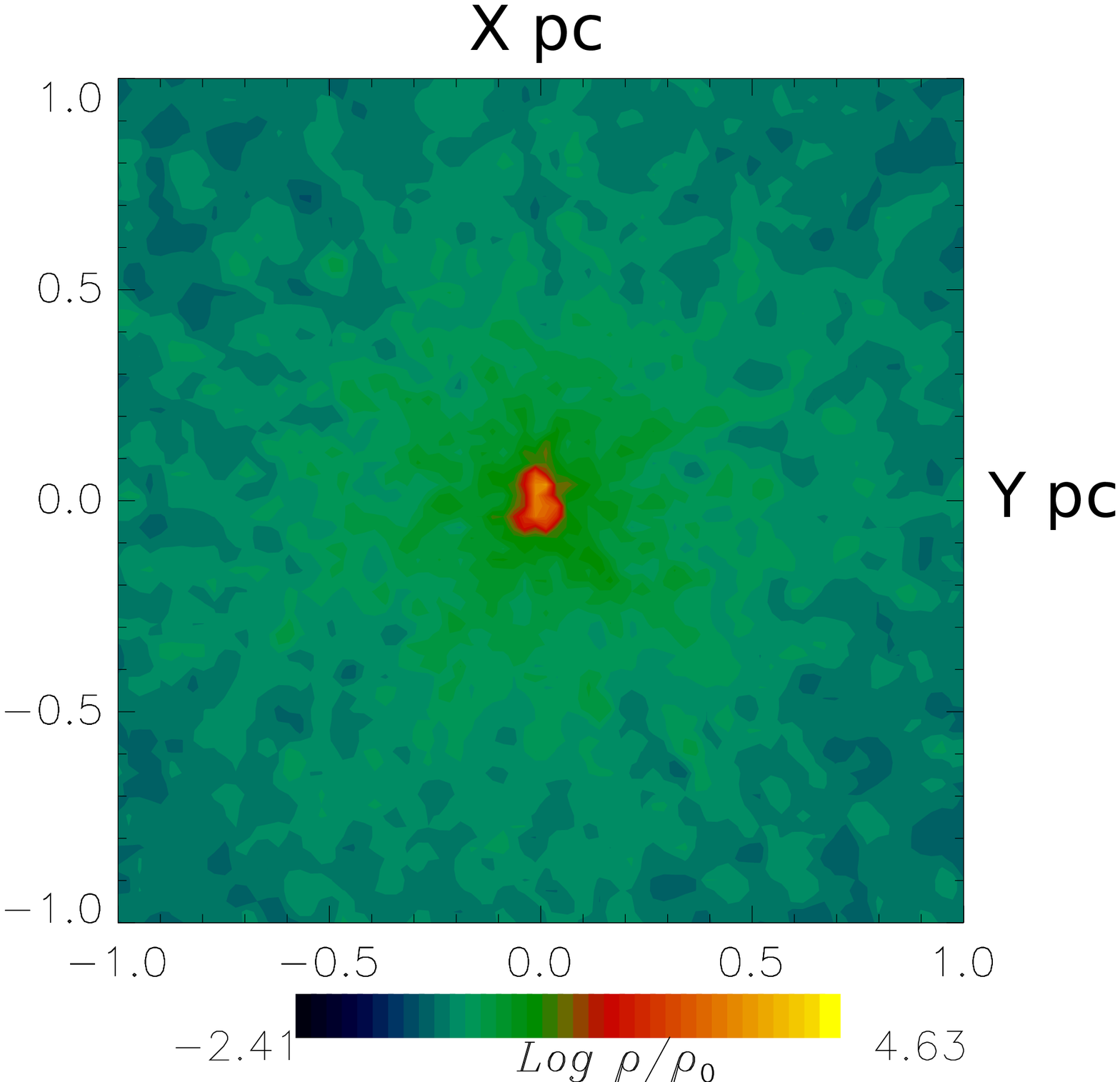} &
\includegraphics[width=2.0 in]{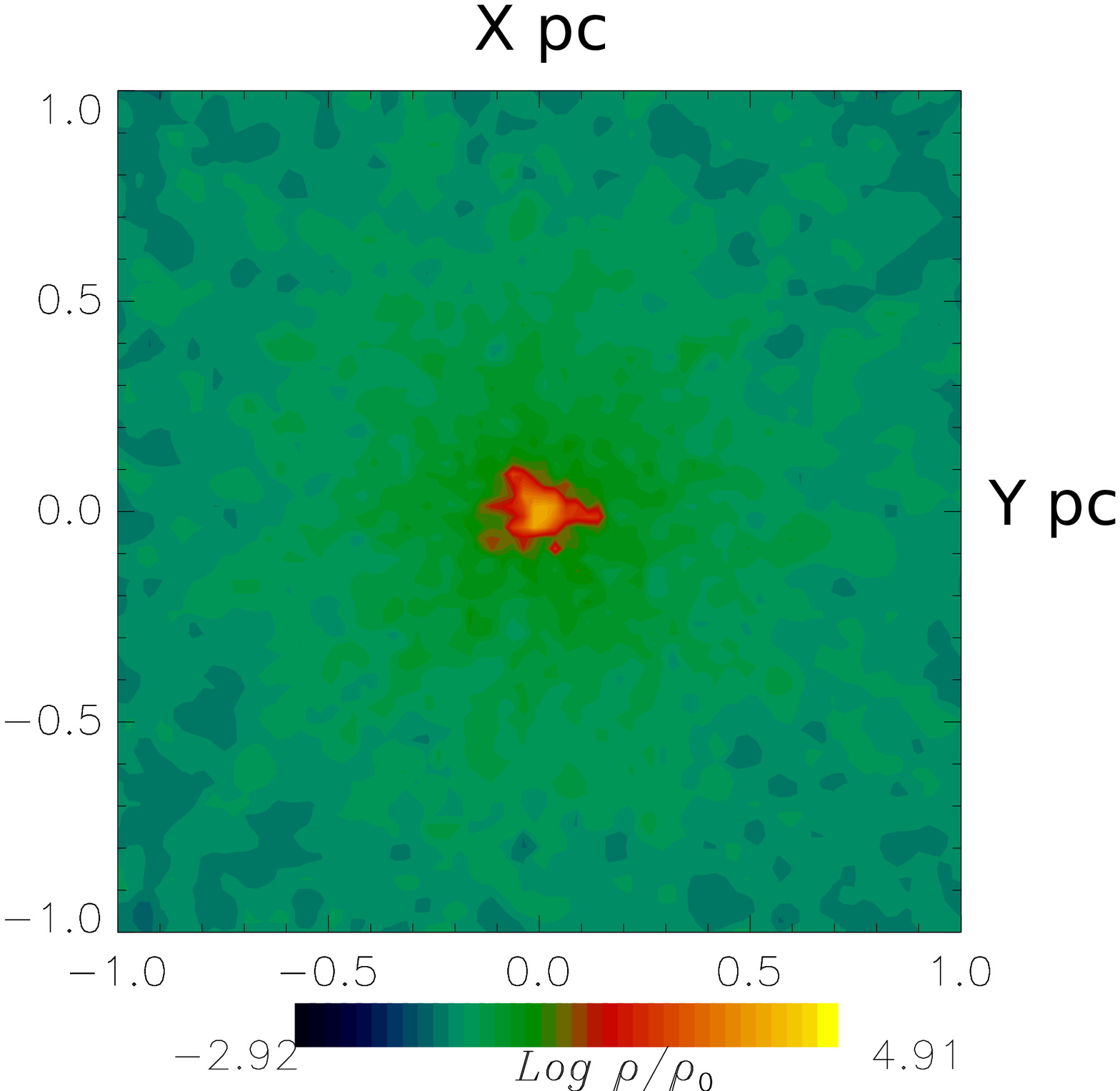}\\
\end{tabular}
\caption{\label{MospsRotb} Column density plots of the collision models $Urb$, for a thin slice of gas parallel
to the x-y plane. The unit of length is one parsec. The models are shown in panels as follows:
(top left-hand)
$U5rb$ (at time $t/t_{ff}=0.12$ and peak density $\log \left( \rho_{\rm max}/\rho_0 \right)=5.0$);
(top right-hand) $U13rb$ (at time $t/t_{ff}=0.12$ and peak density $\log \left( \rho_{\rm max}/\rho_0 \right)=4.62$);
(bottom left-hand) $U9rb$ (at time $t/t_{ff}=0.09$ and peak density $\log \left( \rho_{\rm max}/\rho_0 \right)=4.63$);
(bottom right-hand) $U11rb$ (at time $t/t_{ff}=0.15$ and peak density $\log \left( \rho_{\rm max}/\rho_0 \right)=4.62$).}
\end{center}
\end{figure}

Finally, let us consider the iso-density plots for models $Urb$, which are shown in Fig.\ref{MospsRotb}. We
observe the formation of a massive lump of gas at the cloud center, at the beginning of 
the evolution, similarly to those shown in
the left-hand column of Fig.\ref{MospsRot}. However, for each model $Urb$, the massive lump of gas is quite more bigger
than for models $Ur$. This is a direct consequence of the higher azimuthal velocity added
to the particle velocity, see Section\ref{subs:Vcir}. This massive central lump of gas makes the collapse
quite faster than that observed for models $Ur$, as we notice happens in the bottom right-hand panel
of Fig.\ref{fig:DenMax}.

%%%%%%%%%%%%%%%%%%%%%%%%%%%%%%%%%%%%%%%%%%%%%%%%%%%%%%%%%%%%%%%%%%%%%%%%%%%%%%%%%%%%%%%%%%
%%%%%%%%%%%%%%%%i
%%%%%%%%%%%%%%%%%%%%%%%%%%%%%%%%%%%%%%%%%%%%%%%%%%%%%%%%%%%%%%%%%%%%
\subsection{3D rendered plots}
\label{subsec:3dp}

\begin{figure}
\begin{center}
\begin{tabular}{cc}
\includegraphics[width=2.5 in]{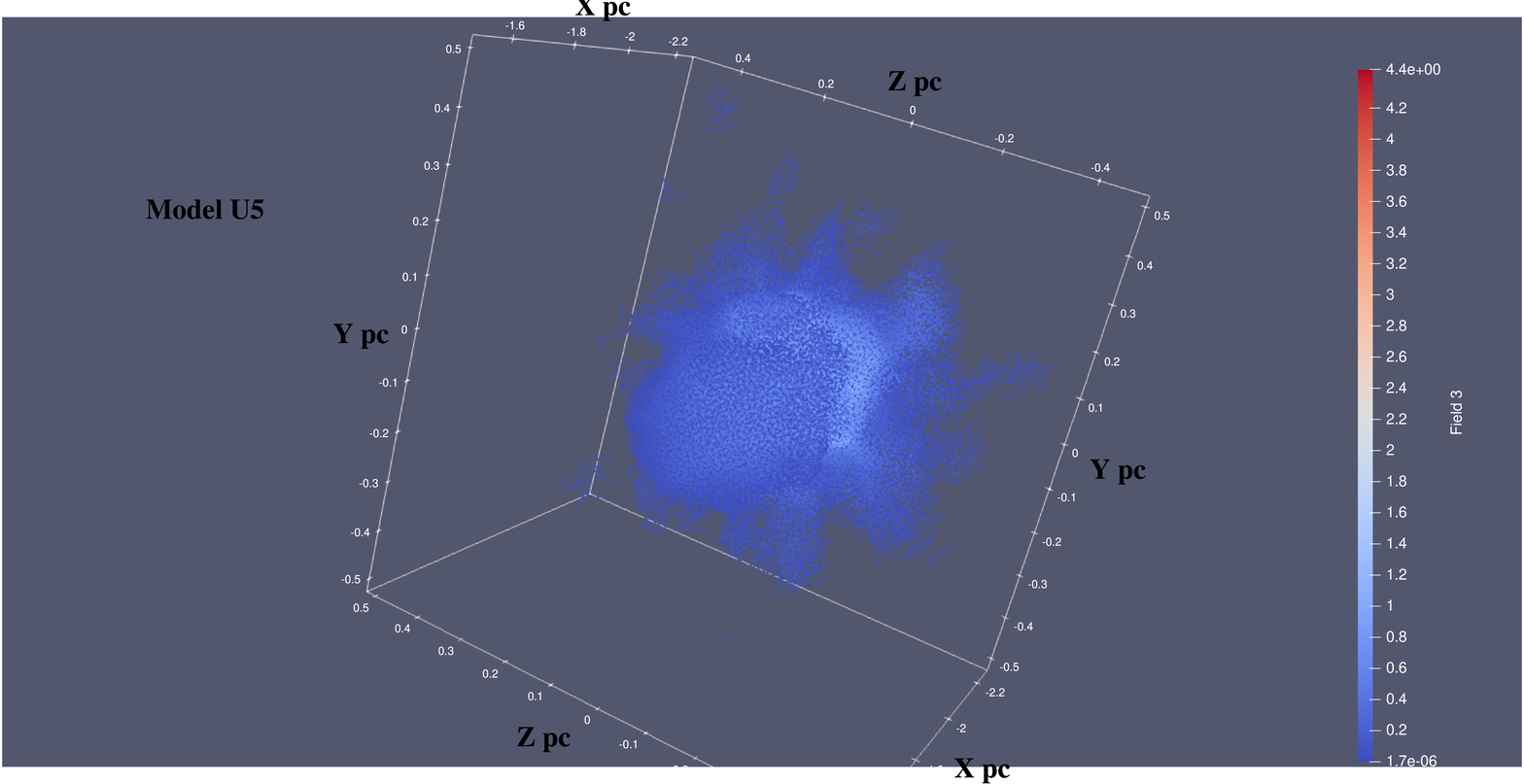} &
\includegraphics[width=2.5 in]{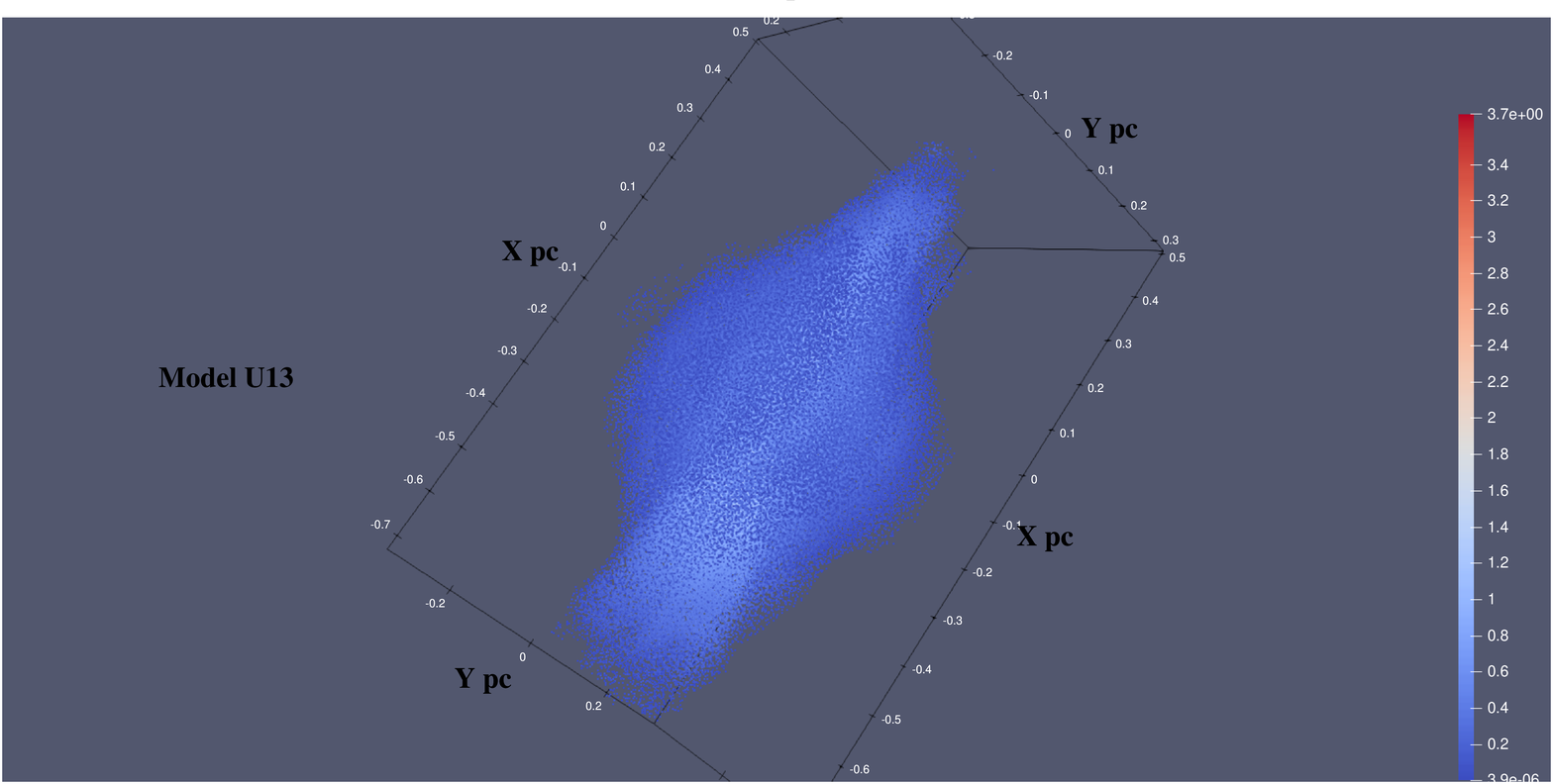}\\
\includegraphics[width=2.5 in]{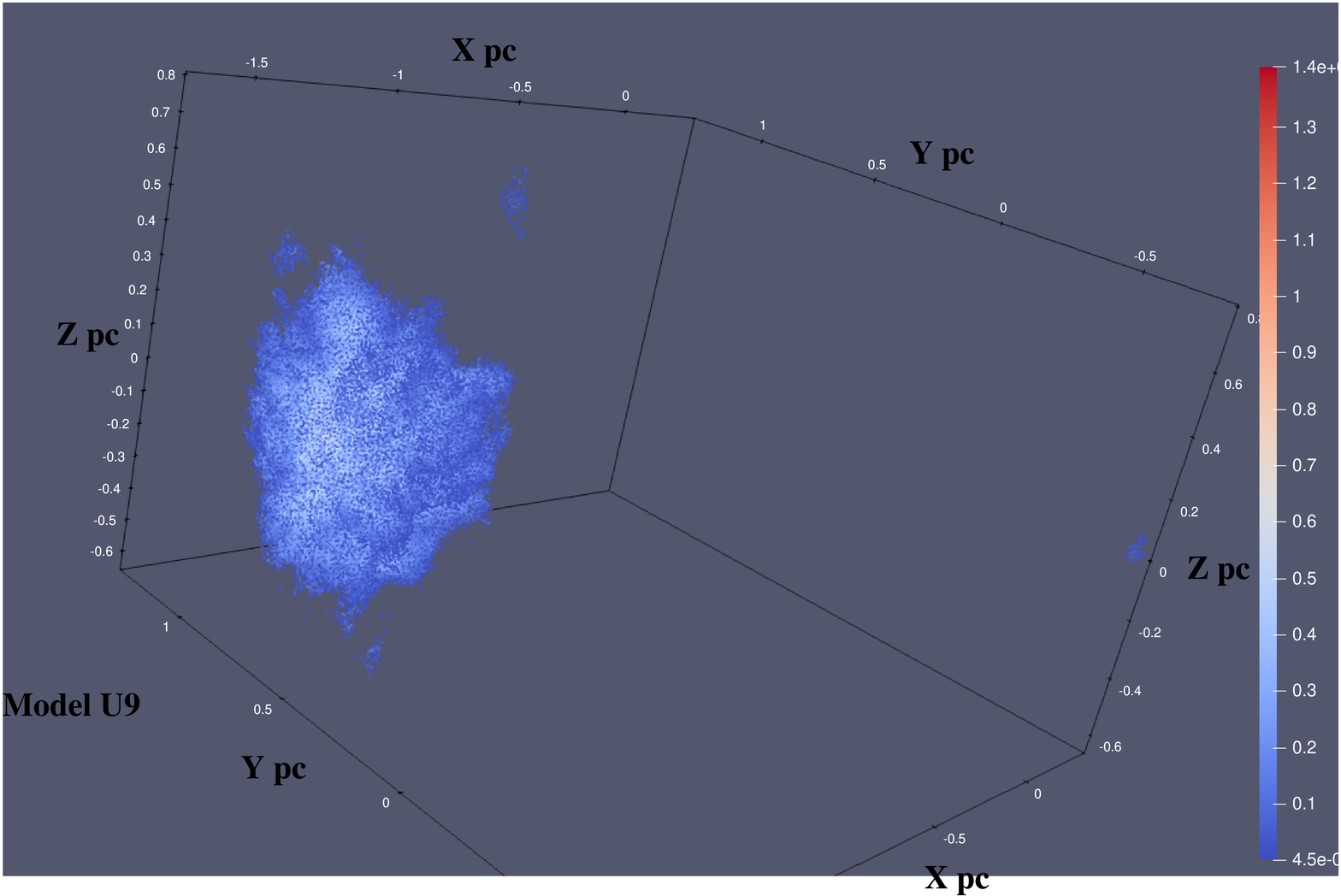} &
\includegraphics[width=2.5 in]{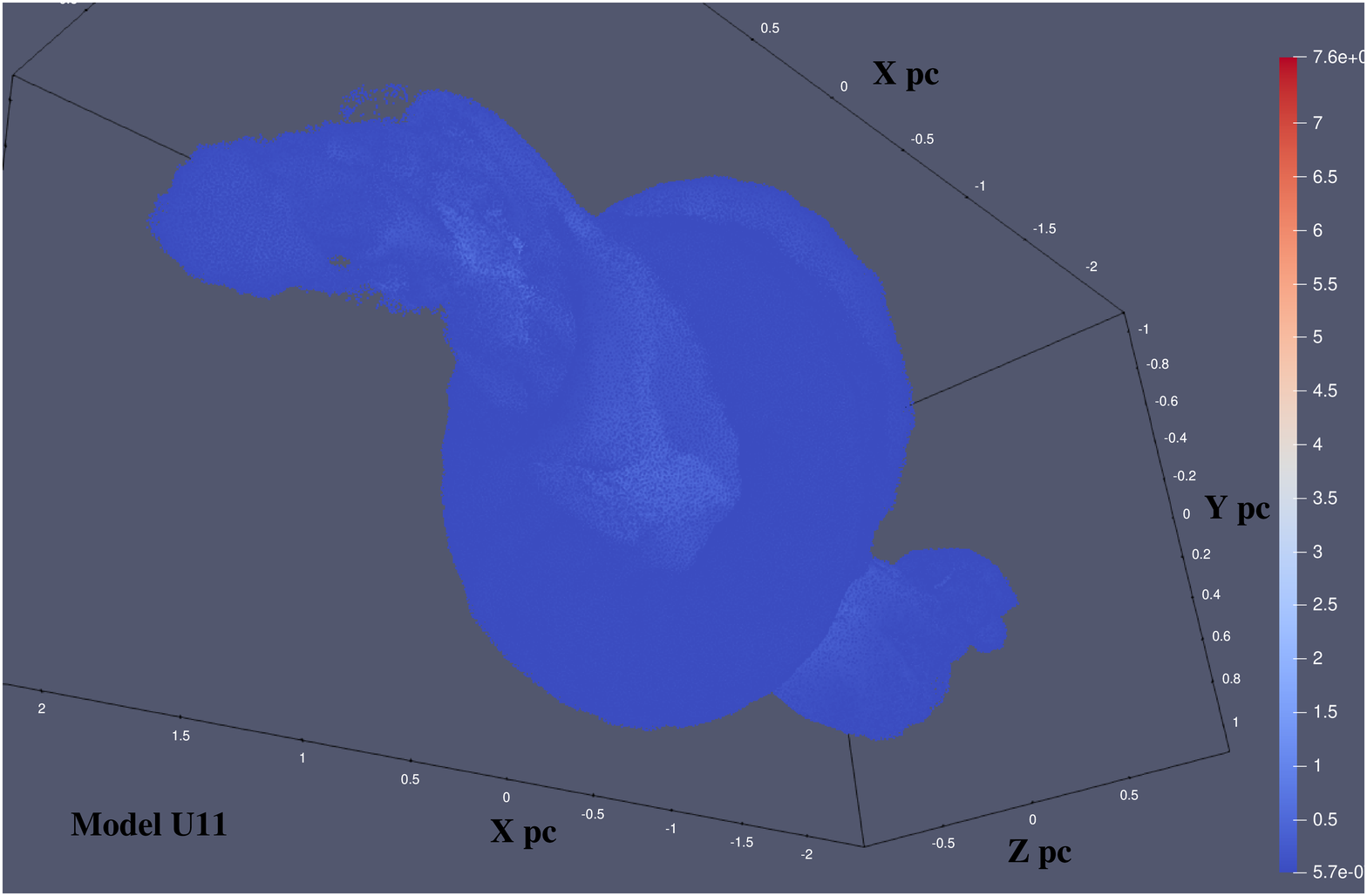}\
\end{tabular}
\caption{\label{VisMoseps} 3D-plots of the collision models $U$ for the same snapshots of
Fig.\ref{Mosps}, rotated arbitrarily to reveal more details of the configurations, in a window
of spatial dimension from -2.5 to 2.5, in each of the axis-XYZ, in which only those
particles with $\log \left( \rho_{\rm max}/\rho_0 \right)> 0.7$ have entered to make the plot. The
panels and the number of particles entered are shown as follows:
(top left-hand) $U5$ with 243455 particles; (top right-hand) $U13$ with 287037 particles;
(bottom left-hand) $U9$ with 181266 particles and (bottom right-hand) $U11$ with 5616884 particles.}
\end{center}
\end{figure}

In Fig.\ref{VisMoseps}, we show the spatial structure of the models $U$ using 3D plots, for the same
time and density chosen for the snapshot shown in Fig.\ref{Mosps}.  Until now, the figures
displayed in Section \ref{subsec:col}, have been cuts parallel to the
equatorial plane of the initial sphere, so that around of 10,000 particles are included in the
slice shown. For the 3D plots, all of the particles with a density greater than
$\log \left( \rho_{\rm max}/\rho_0 \right)  \approx 0.7$
and located within the region [-2.5,2.5] in the three Cartesian coordinates $x,y,z$, entered
in the 3D plots. In this case, the number of particles that are used to make the 3D plots ranges from
181266 to 5616884. The log of the density is rendered in the 3D plots to assign a color
and a vertical bar located in the bottom right-hand corner of each panel. It must be noted that an arbitrary
rotation is done on the Cartesian coordinates to show some of the details of the spatial structure.

%%%%%%%%%%%%%%%%%%%%%%%%%%%%%%%%%%%%%%%%%%%%%%%%%%%%%%%%%%%%%%%%%%%%%

In the top left-hand panel of Fig.\ref{VisMoseps},
we see the remains of the model $U5$, view from the rear (along the positive X-axis), in which one can see
an elongated bulb. In the region where the un-equal sub-cloud collision takes place, on
the negative side of the X-axis, one can see a thick disk of gas surrounding the elongated 
bulb. This structure looks like a mushroom pointing toward the negative X-axis.

In the top right-hand panel of Fig.\ref{VisMoseps}, we see the remains of model $U13$; recall that this collision
is head-on along the X-axis between two equal sized sub-clouds. For this reason, one can see an elongated
solid tube of gas along the X-axis, surrounded by an almost spherical gas region, which is formed by
the particles bounced from the collision, most of which escape away along the Y-axis in both
directions, positive and negative.

In the bottom left-hand panel of Fig.\ref{VisMoseps}, we see the remains of model $U9$, in which two un-equal sized
sub-clouds have an oblique collision. One can see only the remains of each separate sub-clouds, after their close
encounter, so one can notice that the bottom sub-cloud is almost destroyed by the tidal force caused by the
top sub-cloud.

In the bottom right-hand panel of Fig.\ref{VisMoseps}, we see the remains of model $U11$, in which two equal-sized
sub-clouds have an oblique collision. For this reason, the symmetry is evident between the top and bottom gas
tubes, which are formed as the tracks of the original colliding sub-clouds. There is a complex bridge of gas connecting
these top and bottom tubes. A disk of gas is surrounding the bridge.

%%%%%%%%%%%%%%%%%%%%%%%%%%%%%%%%%%%%%%%%%%%%%%%%%%%%%%%%%%%%%%%%%%%%%%%%%%%%%%%
\subsection{Distribution function of the radial component of velocity}
\label{subsec:disvel}

\begin{figure}
\begin{center}
\begin{tabular}{cc}
\includegraphics[width=3.0 in]{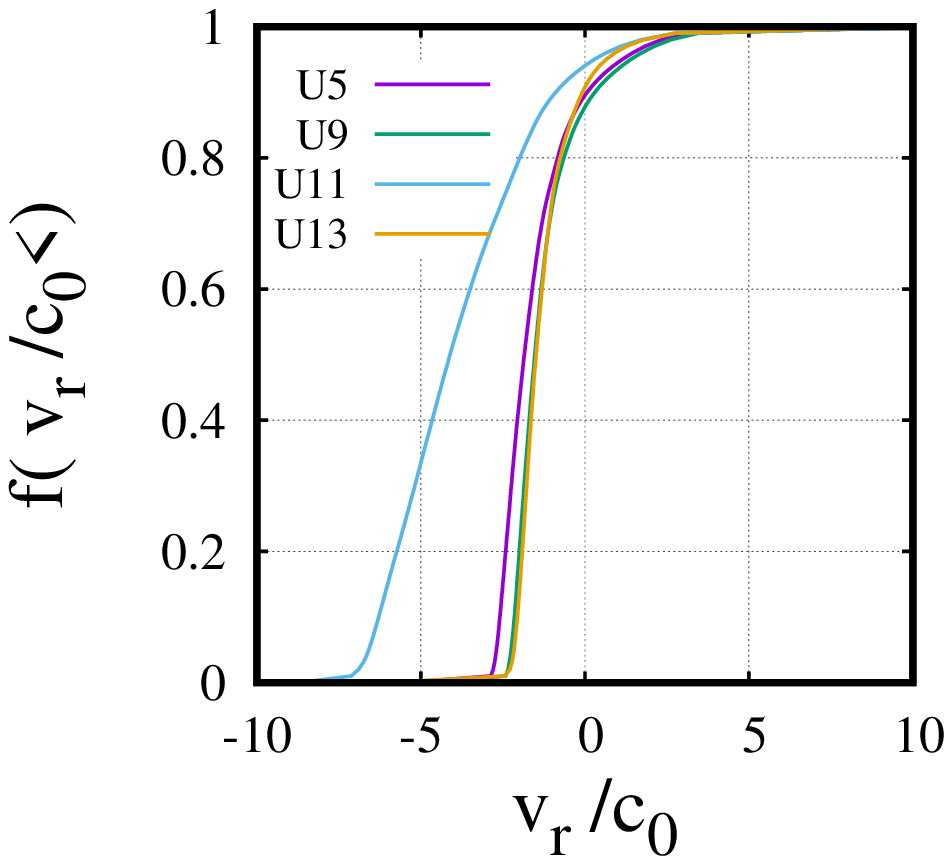}  & \includegraphics[width=3.0 in]{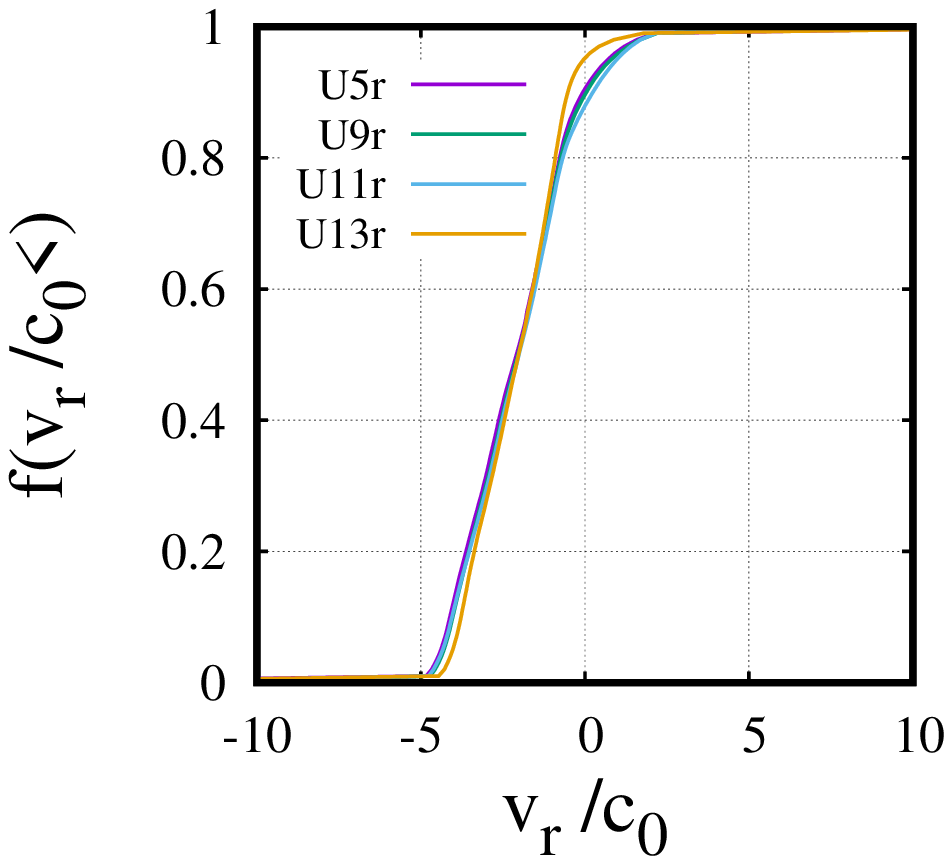} \\
\includegraphics[width=3.0 in]{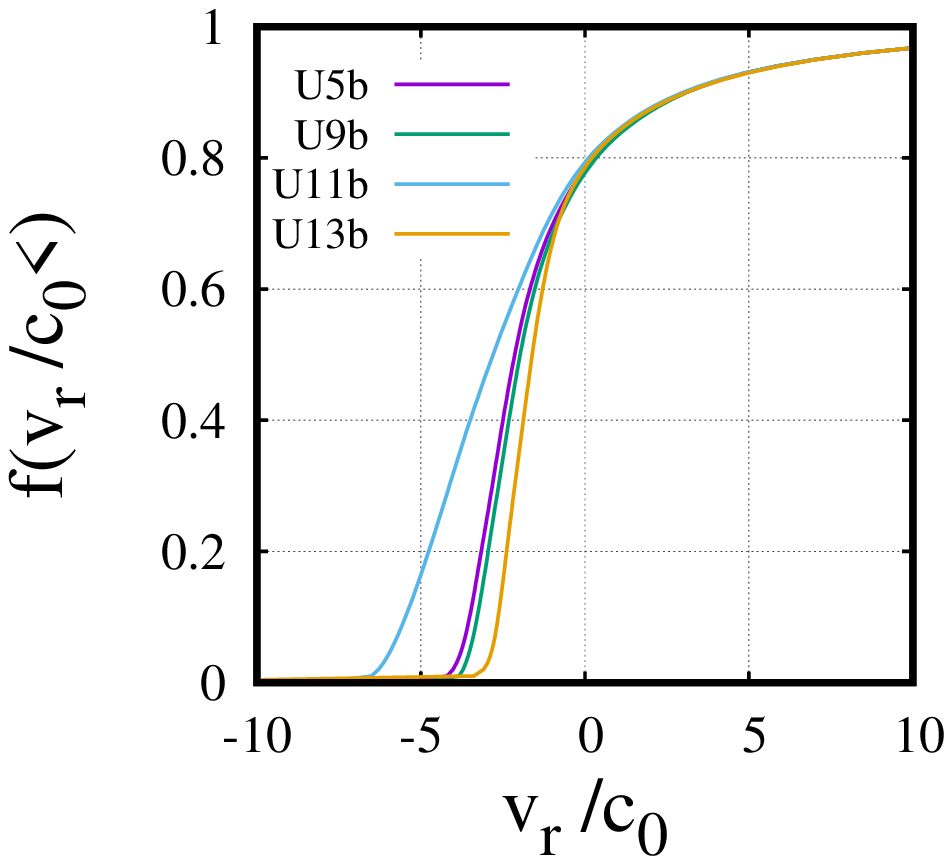} &  \includegraphics[width=3.0 in]{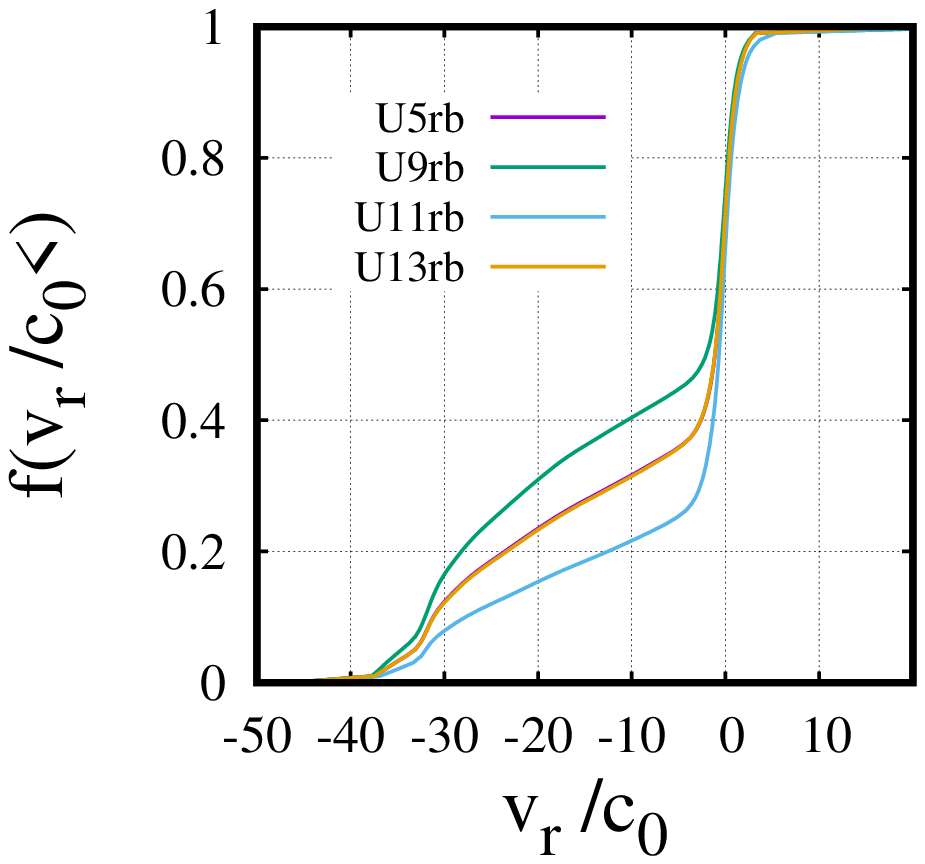} \\
 \\
\end{tabular}
\caption{\label{VelDistMos} Distribution function of the
the radial component of the velocity, calculated with respect to the center of the cloud.
The snapshot are taken at the same time and density of the snapshots shown in Fig.\ref{Mosps},
\ref{Mospsb2}, \ref{MospsRot} and Fig.\ref{MospsRotb}.
(top left-hand) models $U$ with a low level of turbulence;
(top right)-hand $Ur$ with a low azimuthal velocity
(bottom left-hand) $Ub$ with high level of turbulence and
(bottom right-hand) $Urb$ with a high azimuthal velocity.}
\end{center}
\end{figure}

In Fig.\ref{VelDistMos} we show in the vertical axis the fraction of particles with a velocity smaller
than that shown in the horizontal axis. This distribution function of the velocity is taken at the same time
and density as the snapshots shown in Figs.\ref{Mosps}, \ref{Mospsb2}, \ref{MospsRot} and \ref{MospsRotb}. We consider
only the radial component of the velocity, which is calculated with respect to the center of the cloud.

By comparing with the panels of Fig.\ref{VelDist} in Section \ref{subs:velocities}, one can see that
the fraction of the simulation particles with a negative component of the radial velocity has increased
from 0.5 at time $t/t_{ff}=0$ for all the models, to 0.9 for models $U$, at time of the
Fig.\ref{Mosps}; to 0.8 for models $Ub$ and to 0.9 for models $Ur$. This means that a high fraction of the
simulation particles have likely reached already (models $U5$, $U9$ and $U13$) or are flowing towards
(model $U11$) an accretion center.

%%%%%%%%%%%%%%%%%%%%%%%%%%%%%%%%%%%%%%%%%%%%%%%%%%%%%%%%%%%%%%%%%%%%

\section{Dynamic characterization of the simulations outcome}
\label{subsec:charac}

\begin{figure}
\begin{center}
\begin{tabular}{cc}
\includegraphics[width=2.5 in]{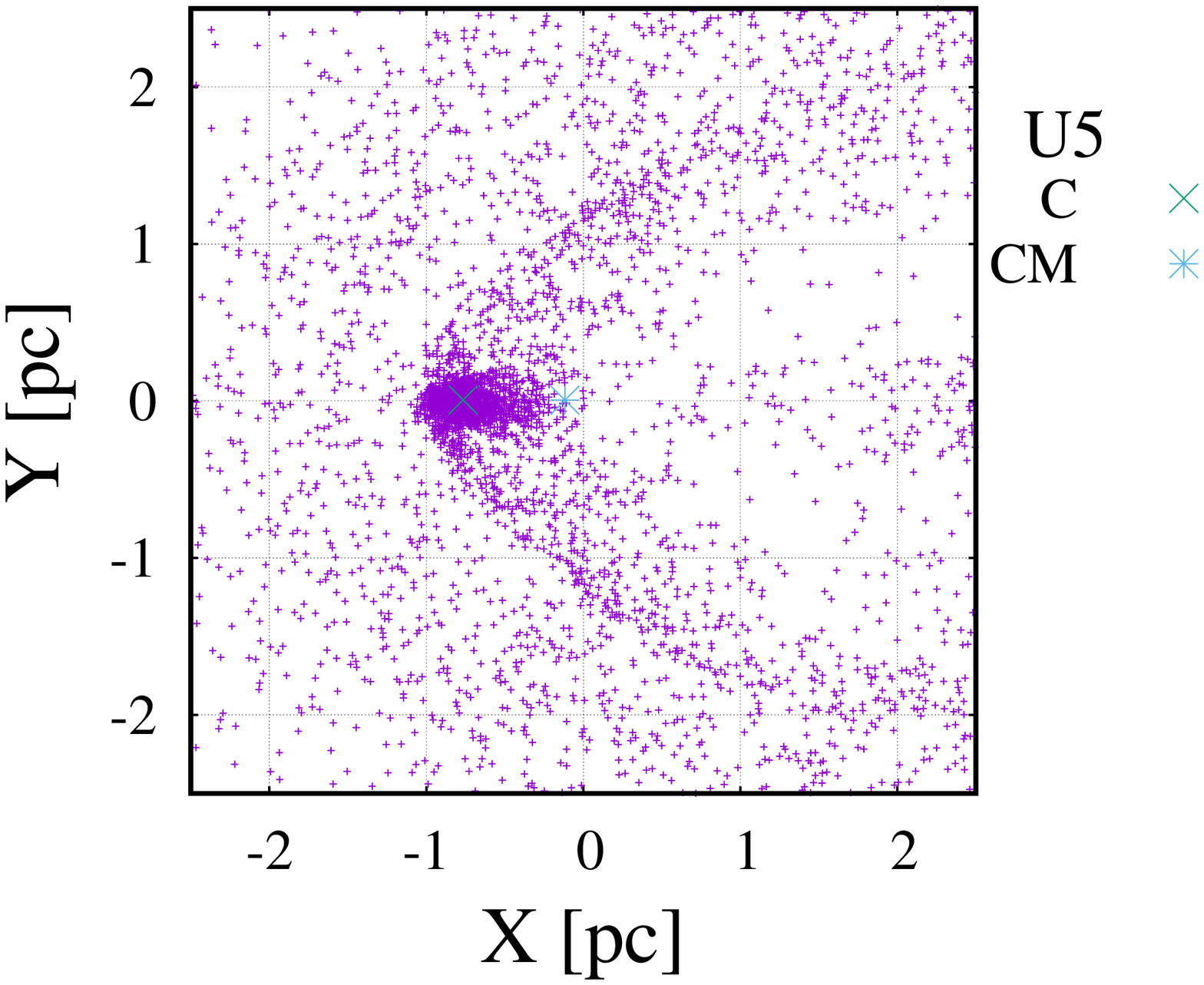} &
\includegraphics[width=2.5 in]{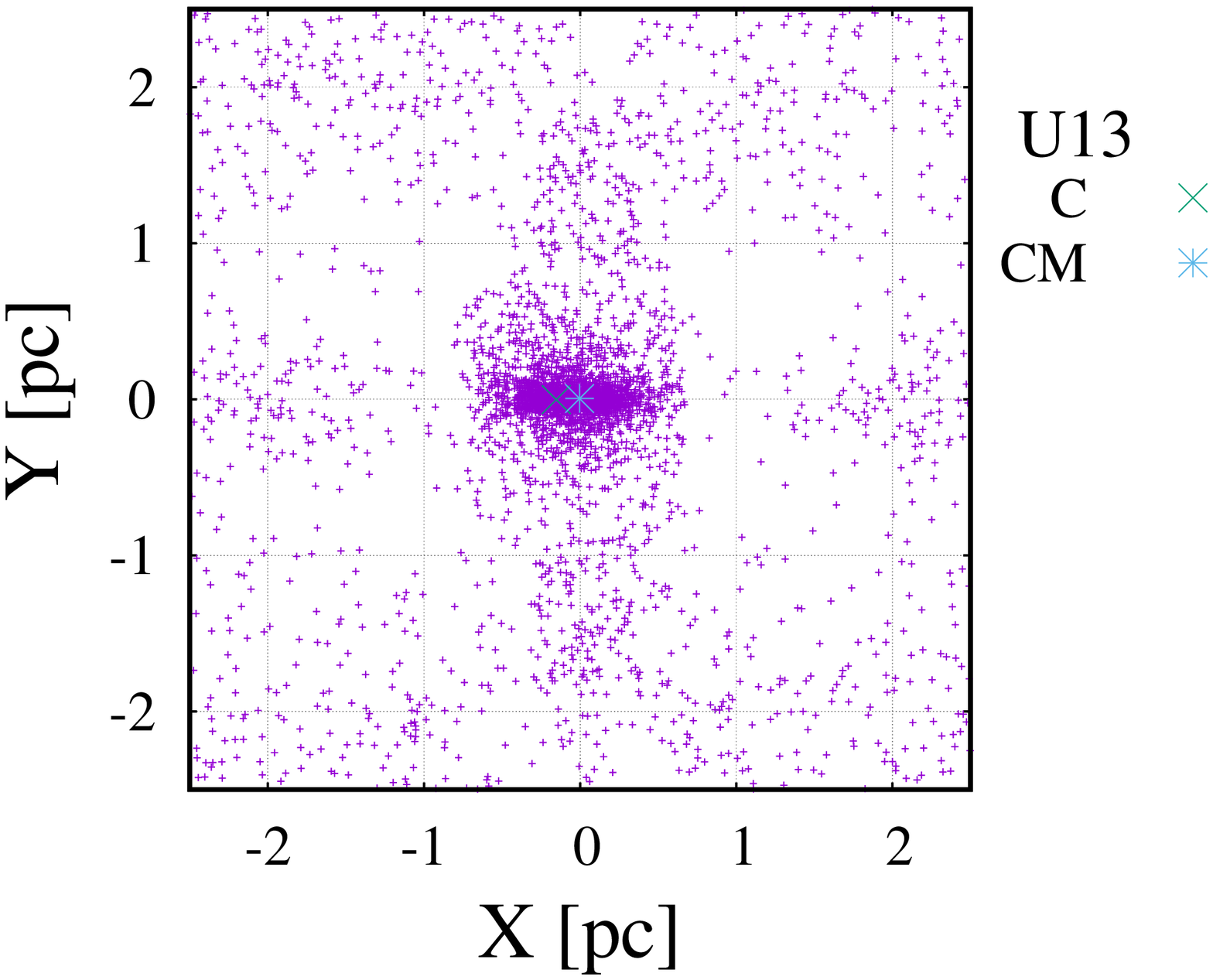}\\
\includegraphics[width=2.5 in]{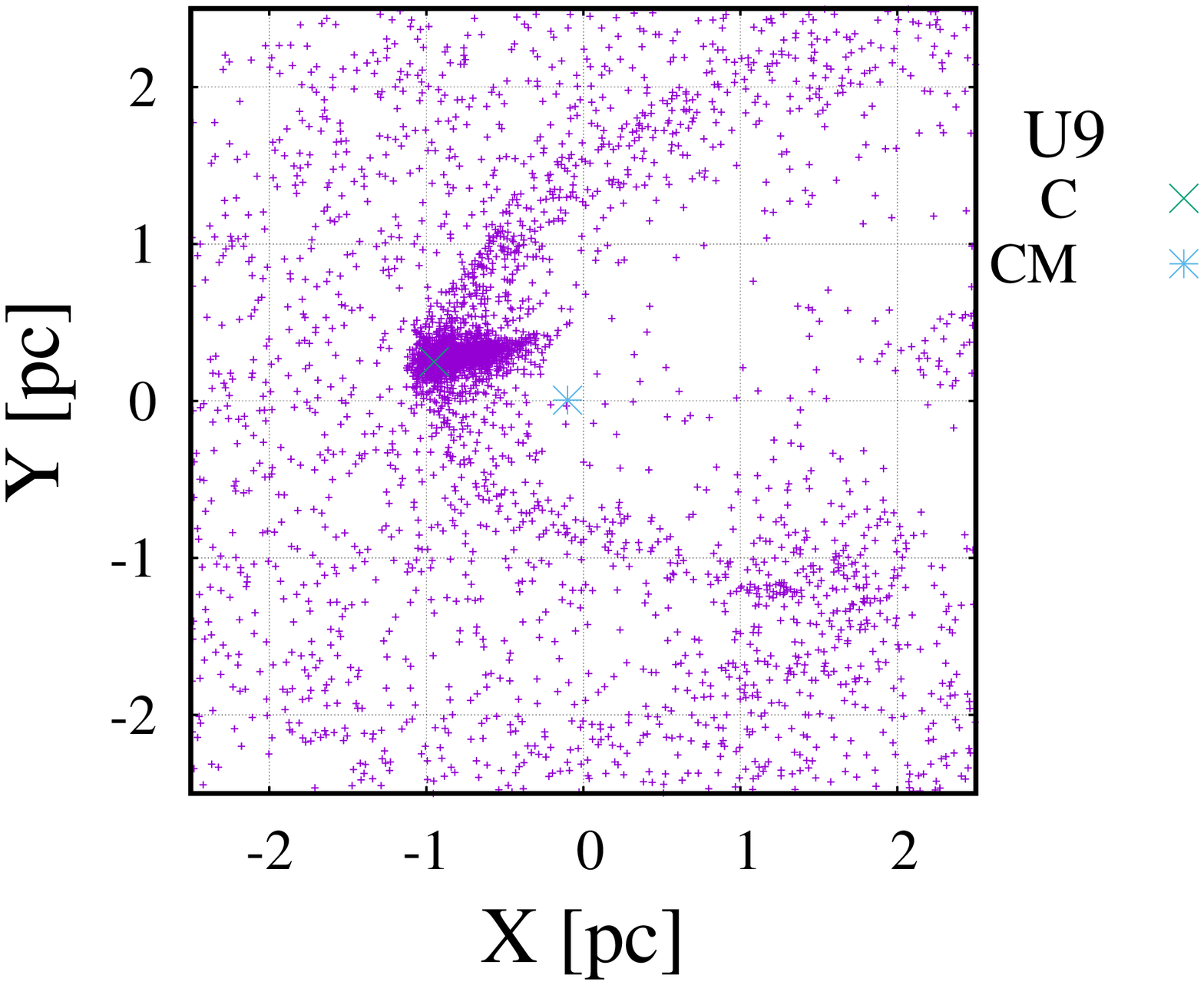} &
\includegraphics[width=2.5 in]{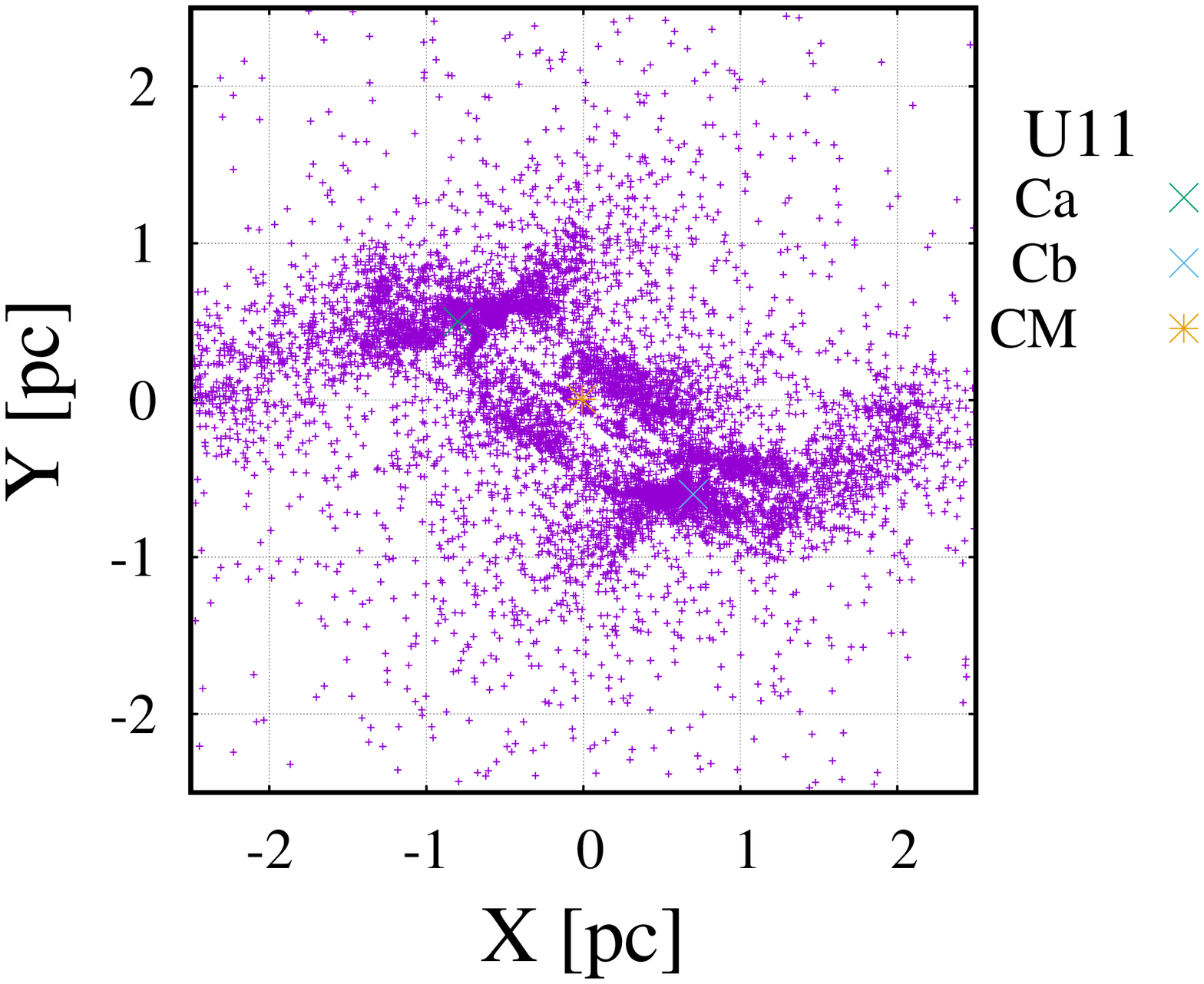}\\
\end{tabular}
\caption{\label{fig:CentrosFragF_XY} Cloud centers are marked with
a symbol "x"; the center of mass is marked with the symbol "*"; both of them are located in a thin slice
parallel to the XY plane, in which the symbol "+" represents a
gas particle. The models are as follows: (top left-hand) $U5$; (top right-hand) $U13$;
(bottom left-hand) $U9$ and (bottom right-hand) $U11$.}
\end{center}
\end{figure}

Let us define a cloudlet as the densest region of a simulation outcome, whose physical properties
must be determined. The center of the cloudlet and a radius are the main parameters to delimit the cloudlet region 
and calculate its physical properties. These centers do not coincide in general the center of mass of each
simulation, although both kind of centers are close to each other, as can be seen in
Fig.\ref{fig:CentrosFragF_XY}, in which we show the center of each cloudlet in models $U$.
%%%%%%%%%%%%%%%%%%%%%%%%%%%%%%%%%%%%%%%%%%%%%%%%%%%%%%%%%%%%%%%%%%%%%%%%%%%%%%%%

%%%%%%%%%%%%%%%%%%%%%%%%%%%%%%%%%%%%%%%%%%%%%%%%%
\subsection{Radial profile of the density and mass}
\label{subsec:radialprofile}

\begin{figure}
\begin{center}
\begin{tabular}{cc}
\includegraphics[width=2.5 in]{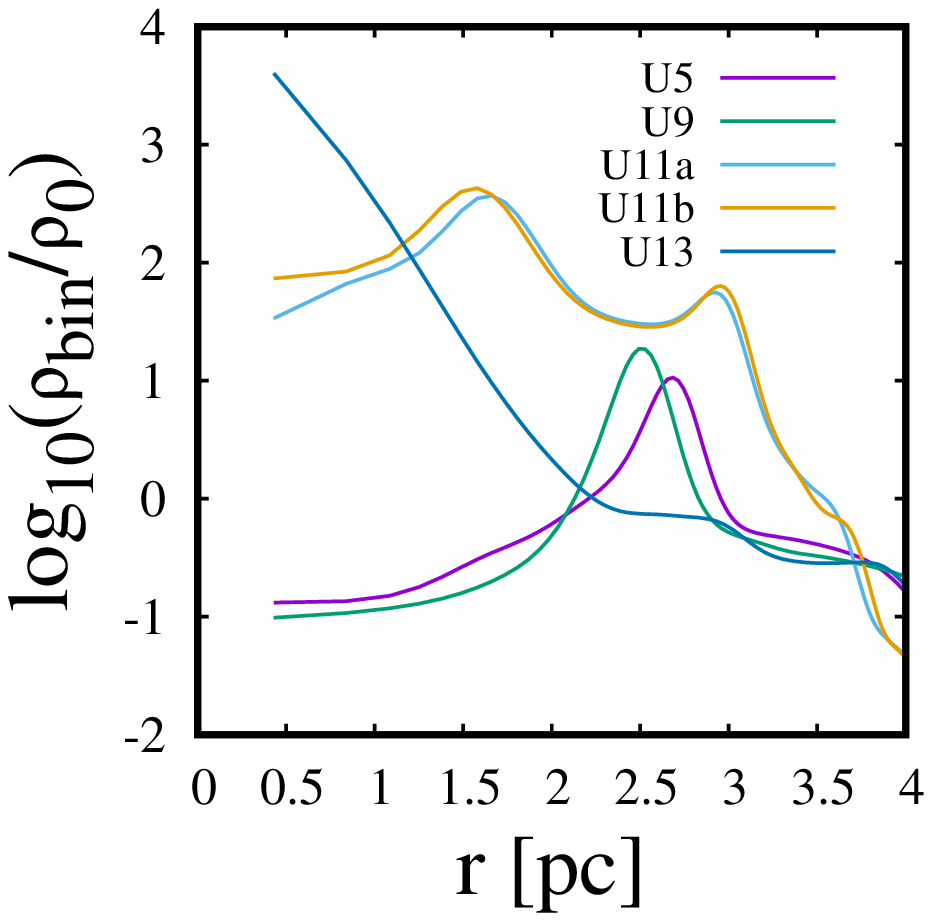} & \includegraphics[width=2.5 in]{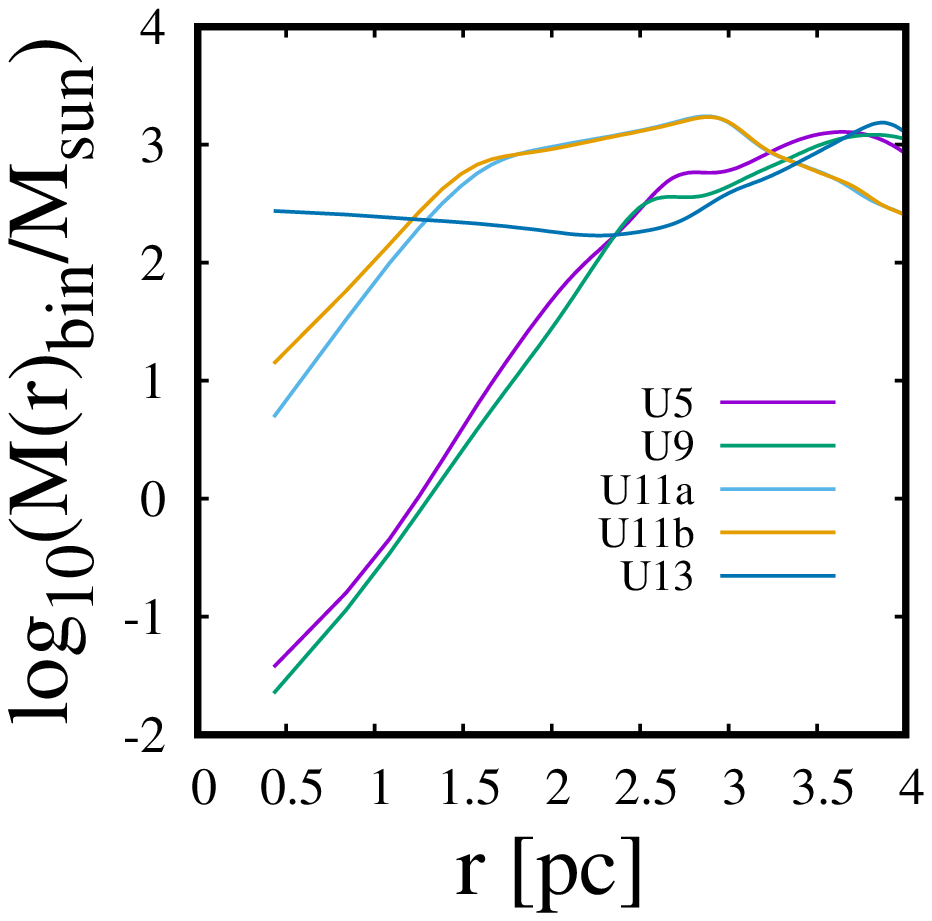}\\
\includegraphics[width=2.5 in]{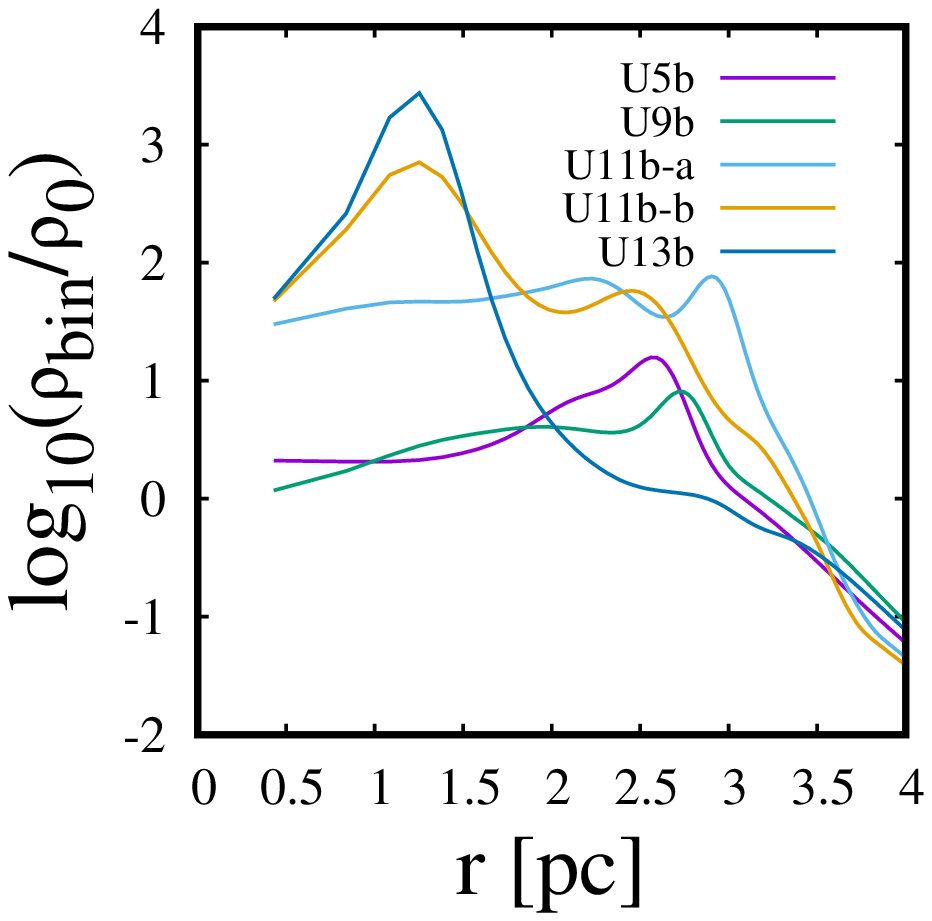} & \includegraphics[width=2.5 in]{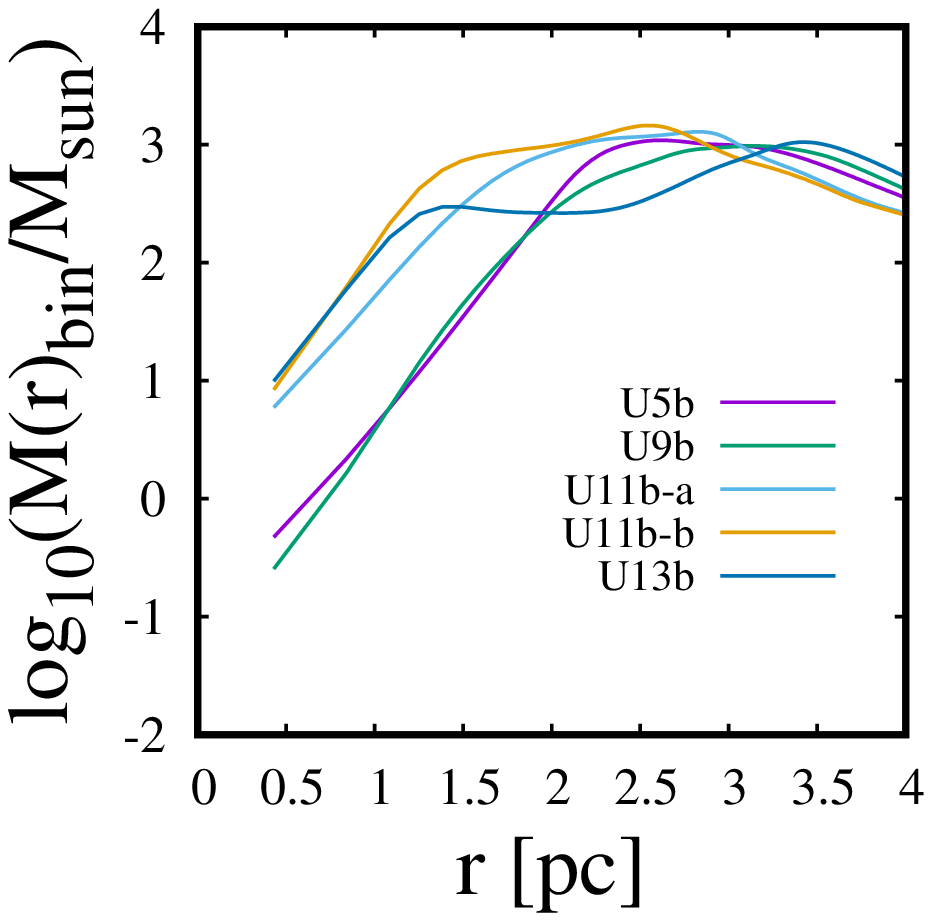}\\
\includegraphics[width=2.5 in]{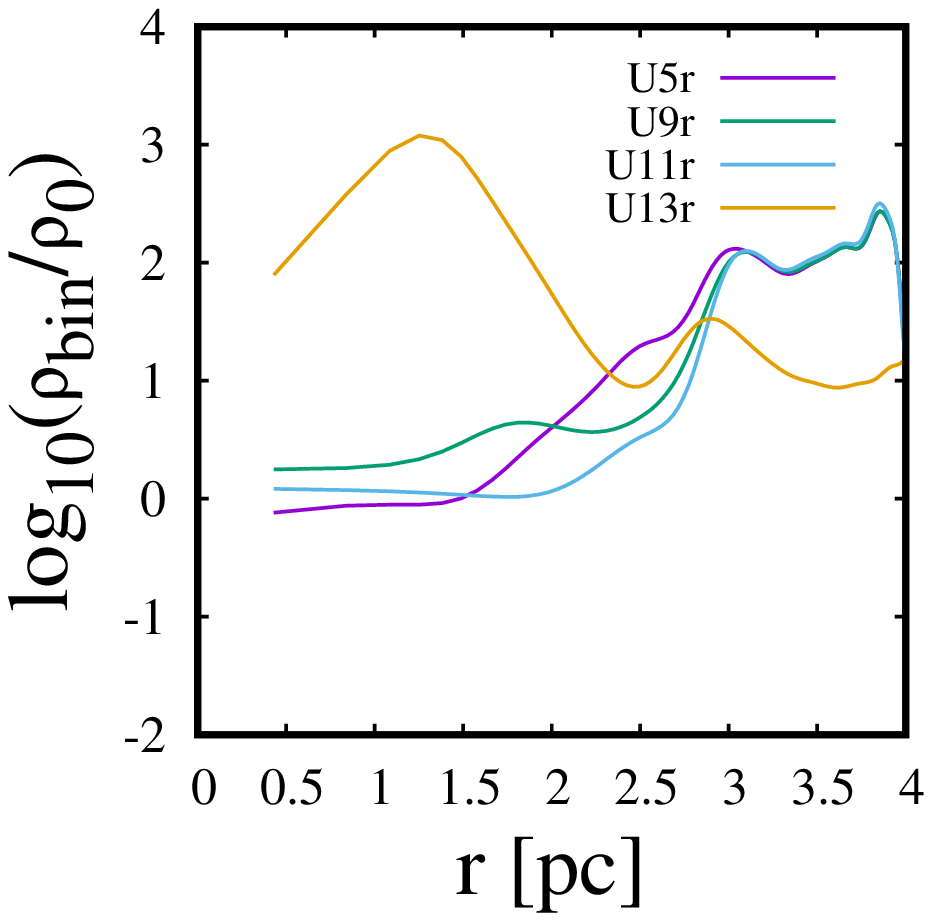} & \includegraphics[width=2.5 in]{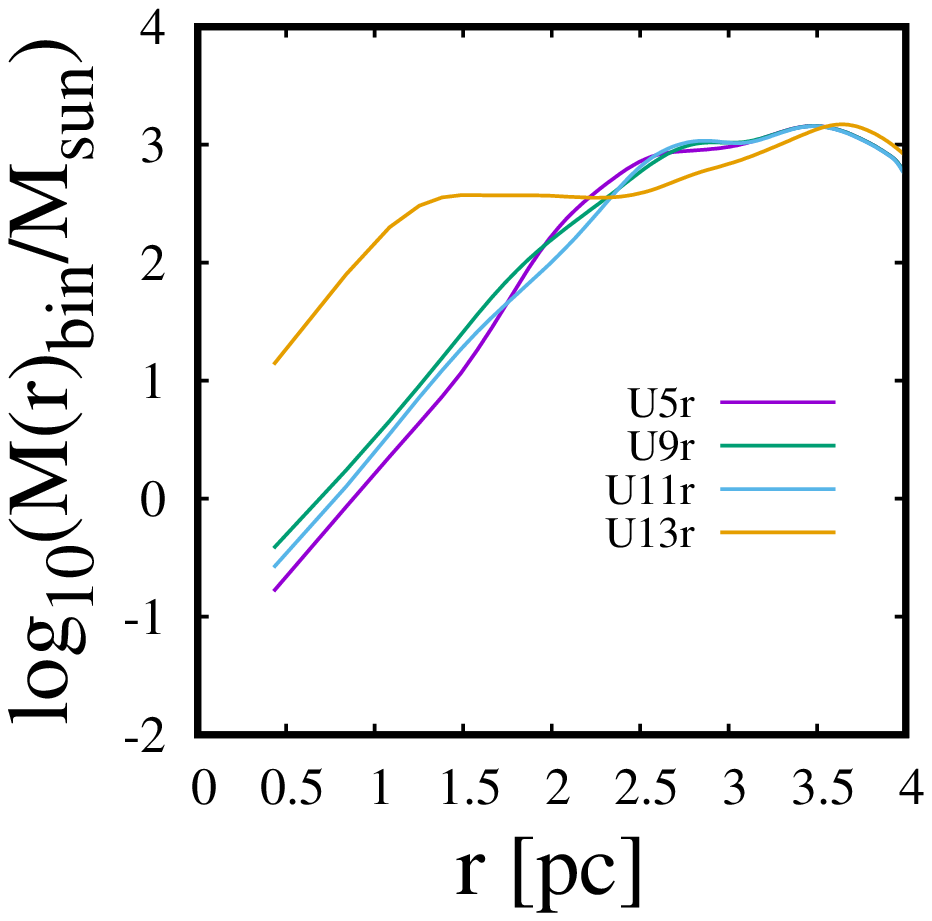}\\
\includegraphics[width=2.5 in]{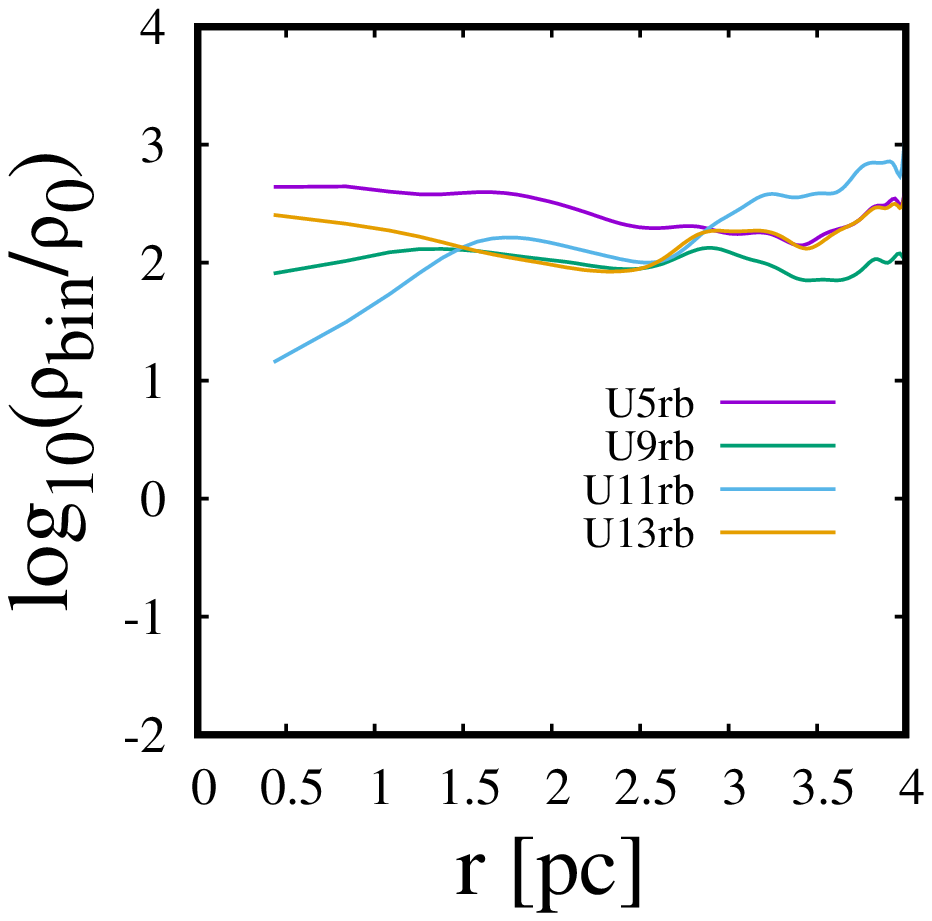} & \includegraphics[width=2.5 in]{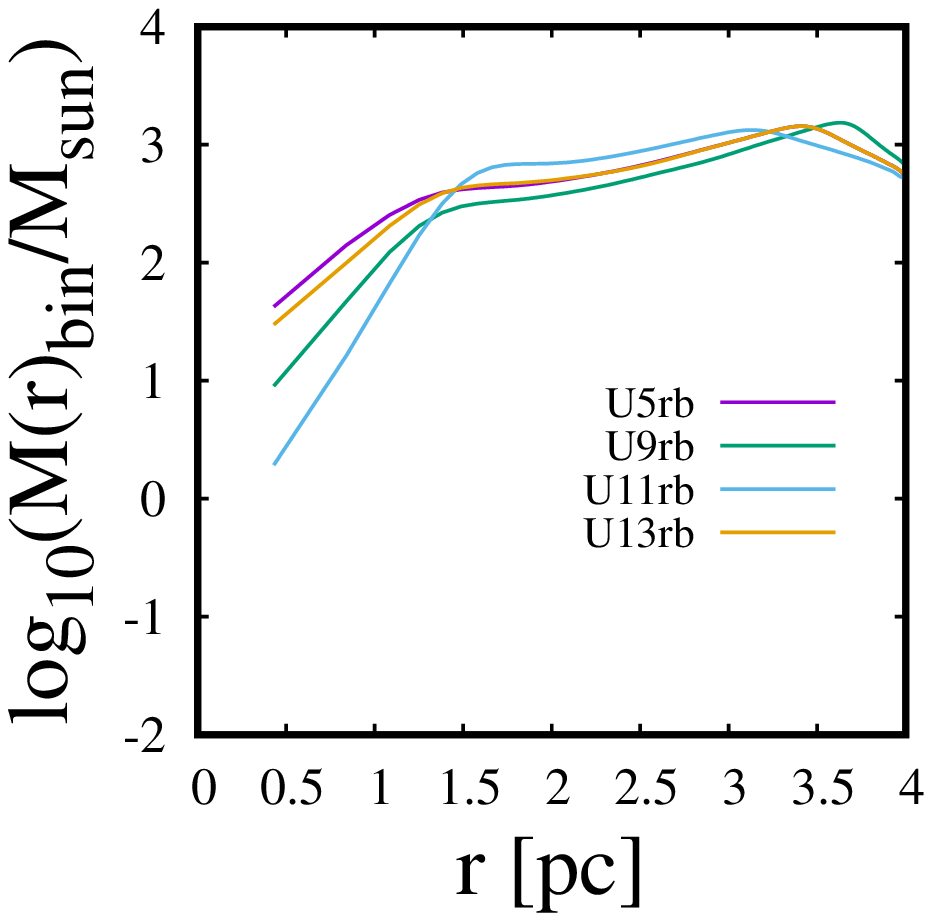}\\
\end{tabular}
\caption{\label{MasayRhor} (left-hand panel) Radial profile of the density and
(right-hand panel) radial profile of the mass. In the vertical axis we
show the mass and density averaged over those particles located within the radial
spherical shell defined by $r$, so that these functions are not cumulative. In the horizontal
axis, $r$ starts at 0, the center of each cloudlet, as illustrated in Fig.\ref{fig:CentrosFragF_XY} for models $U$
and analogously calculated for all the other models.  The panels on the top line are for 
models $U$; the panels on the middle line are for models $Ur$ and the panels on the
bottom line are for models $Urb$. The snapshots are taken at the same time and peak
density shown in Fig.\ref{Mosps}, Fig. \ref{Mospsb2}, Fig.\ref{MospsRot} and Fig.\ref{MospsRotb}, respectively.}
\end{center}
\end{figure}

In Fig.\ref{MasayRhor} we show the radial profile of the density (in the left-hand column) and the mass (in the
right-hand column), calculated with respect to the center of each cloudlet as defined in Fig.\ref{fig:CentrosFragF_XY}.

It must be clarified that the density $\rho_{\rm bin}(r)$ and mass $M(r)_{\rm bin}$, shown in the vertical
axis of Fig.\ref{MasayRhor}, are determined by taking into account only those particles located within
the radii $r$ and $\delta r$, where $\delta r$ is given by $4/500$ pc per bin, and the $r$ goes
from $r=0$ (the cloudlet center) to $r_{\rm max}=4$ pc (even further than the edge of the
cloudlet, because we want to study the environment of the cloudlets as well). 

%From this definition, it must
%be noted that the  density $\rho_{\rm bin}(r)$ and mass $M(r)_{\rm bin}(r)$ and are not cumulative
%functions up to $r$, but the "averaged" functions at the radial shell $r$.

Let us consider the top line of Fig.\ref{MasayRhor} for models $U$. For model $U11$, there are
two cloudlets. Consequently, we label the cloudlet located to the left
of the vertical axis and above the horizontal axis, as cloudlet "a" (in the upper left-hand region). We label
the cloudlet located to the right of the y-axis and below the x-axis as cloudlet "b", as can be seen
in the bottom right-hand panel of Fig.\ref{fig:CentrosFragF_XY}.

The curves of $\rho_{\rm bin}$ for the
models $U5$, $U9$, $U11a$ and $U11b$ are not steep, as opposed 
to the curve for model $U13$. This means that the cloudlets
of these models must have a mass increasing with radius, as can be seen in the right-hand panel of the top line of
Fig.\ref{MasayRhor}.

Model $U13$ is the only one that shows a density curve $\rho_{\rm bin}(r)$ decreasing
significantly with radius $r$. For this behavior, the mass $M(r)_{\rm bin}(r)$ is almost kept
constant for an wide range of radii. This would be the standard behavior of a dense cloudlet
formed by gravitational attraction in a simulation.

In the second and third lines of Fig.\ref{MasayRhor}, from top to bottom, we show the curves for models
$Ub$ and $Ur$, respectively. The behavior observed here is quite similar to the one described earlier
for model $U$. In the bottom line, we show the curves for models $Urb$, which include a high azimuthal
velocity and we have seen in Section \ref{subsec:col}, the simulation outcome changed
significantly. In this case, the density curves are kept constant and the mass curves are
slightly increasing functions of the radius, above all in the range of radius 0-3 pc.
	
\citet{rathborne} found curves for the radial profile 
of the mass and density of the Brick, such that the mass curve is always an increasing function 
of their effective radius, while the density curve
is always a decreasing function, both curves extended up to a effective radius of $2$ pc. For 
instance, the mass contained within its central one pc is approximately $6\, 10^3 \, M_{\odot}$, while 
the density curve follows a power-law over radii $r^{-1.2}$.

The increasing mass curves shown in the right-hand column of Fig.\ref{MasayRhor} are of the same order 
of magnitude as those reported by \citet{rathborne}, taking into account that the curves of 
Fig.\ref{MasayRhor} are not cumulative (as we
mentioned earlier). Therefore, the exterior layers (outside of the densest central region) of the cloud, contain
a substantial amount of mass.

%%%%%%%%%%%%%%%%%%%%%%%%%%%%%%%%%%%%%%%%%%%%%%%%%%%%%%%%%%%%%%%%%%%%%%%%%%%%%%%%
%%%%%%%%%%%%%%%%%%%%%%%%%%%%%%%%%%%%%%%%%%%%%%%%%%%%%%%%%%%%%%%%%%%%%%%%%%%%%%%%

\subsection{Radial profile of the radial and tangential components of the velocity}
\label{subsec:radialprofilevel}

\begin{figure}
\begin{center}
\begin{tabular}{cc}
\includegraphics[width=2.5 in]{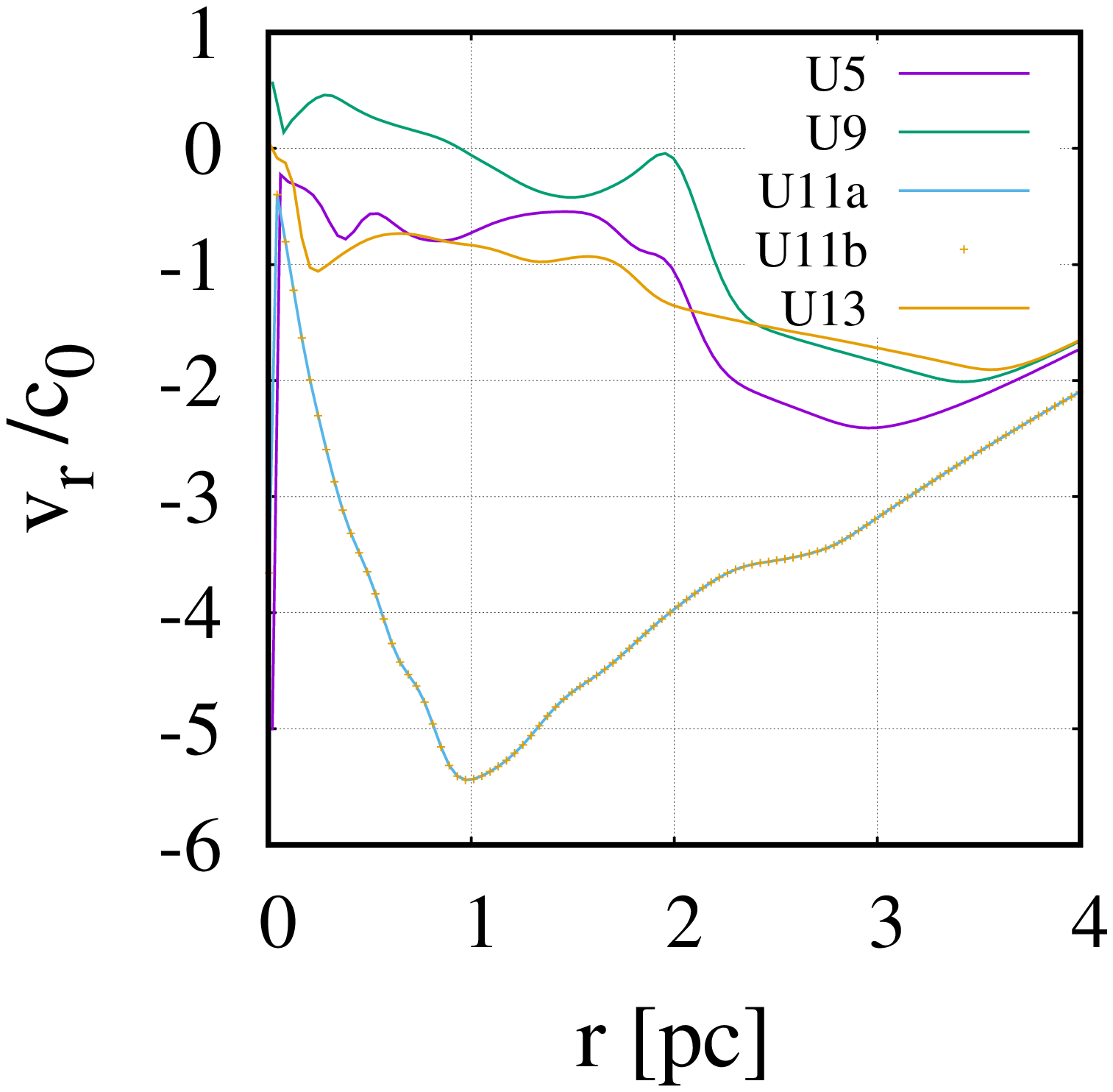} & \includegraphics[width=2.5 in]{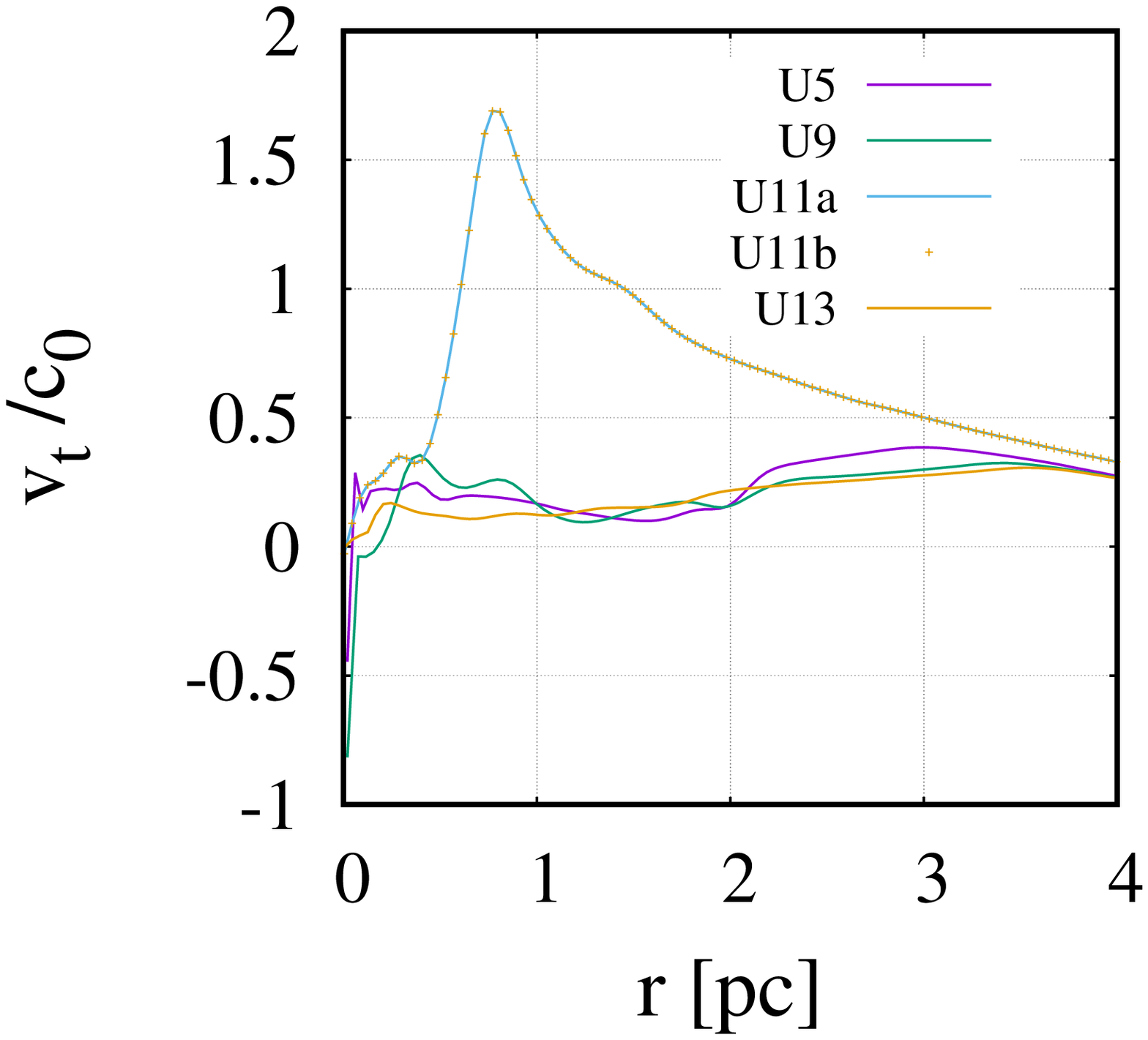} \\
\includegraphics[width=2.5 in]{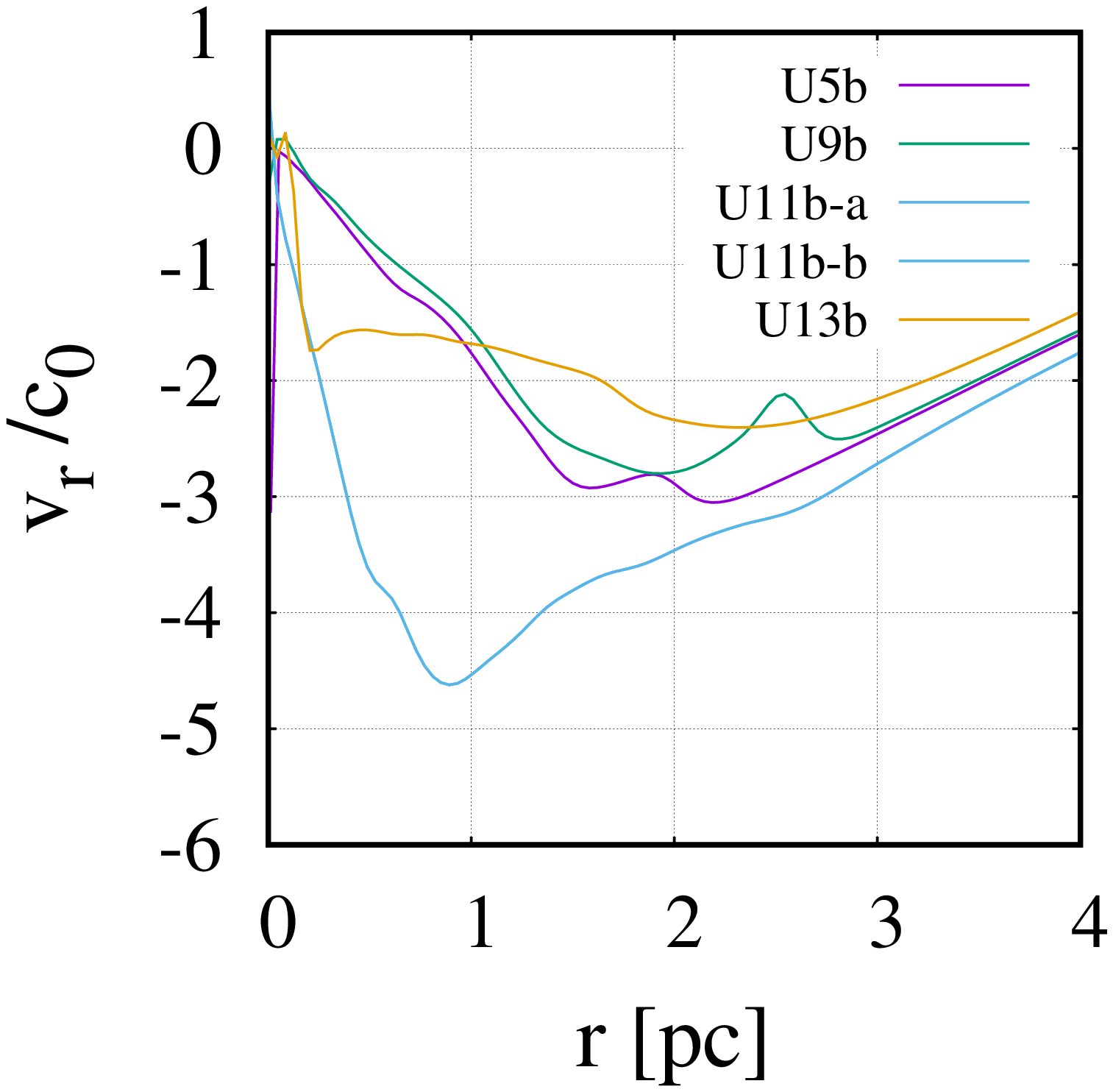} & \includegraphics[width=2.5 in]{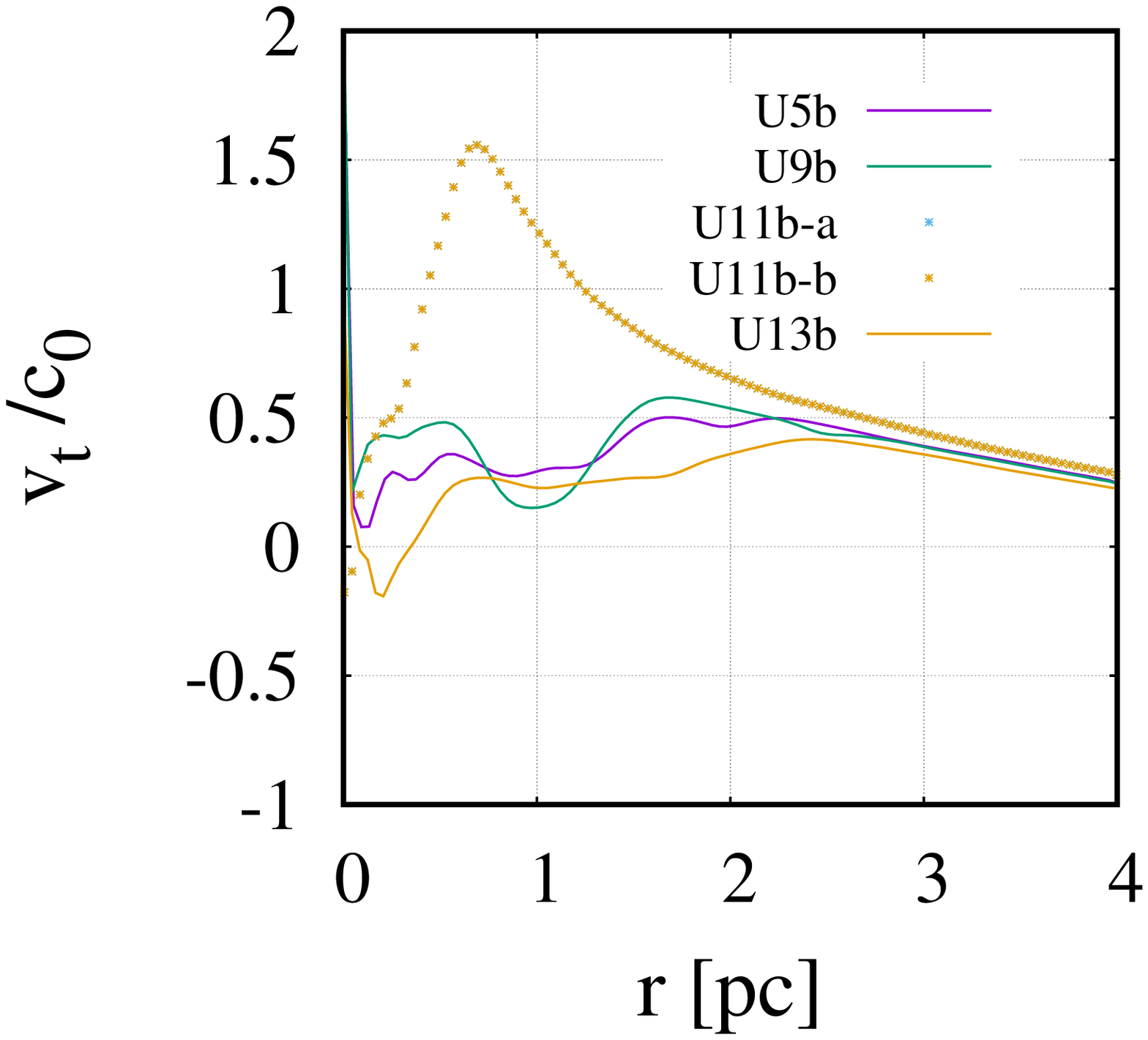}\\
\includegraphics[width=2.5 in]{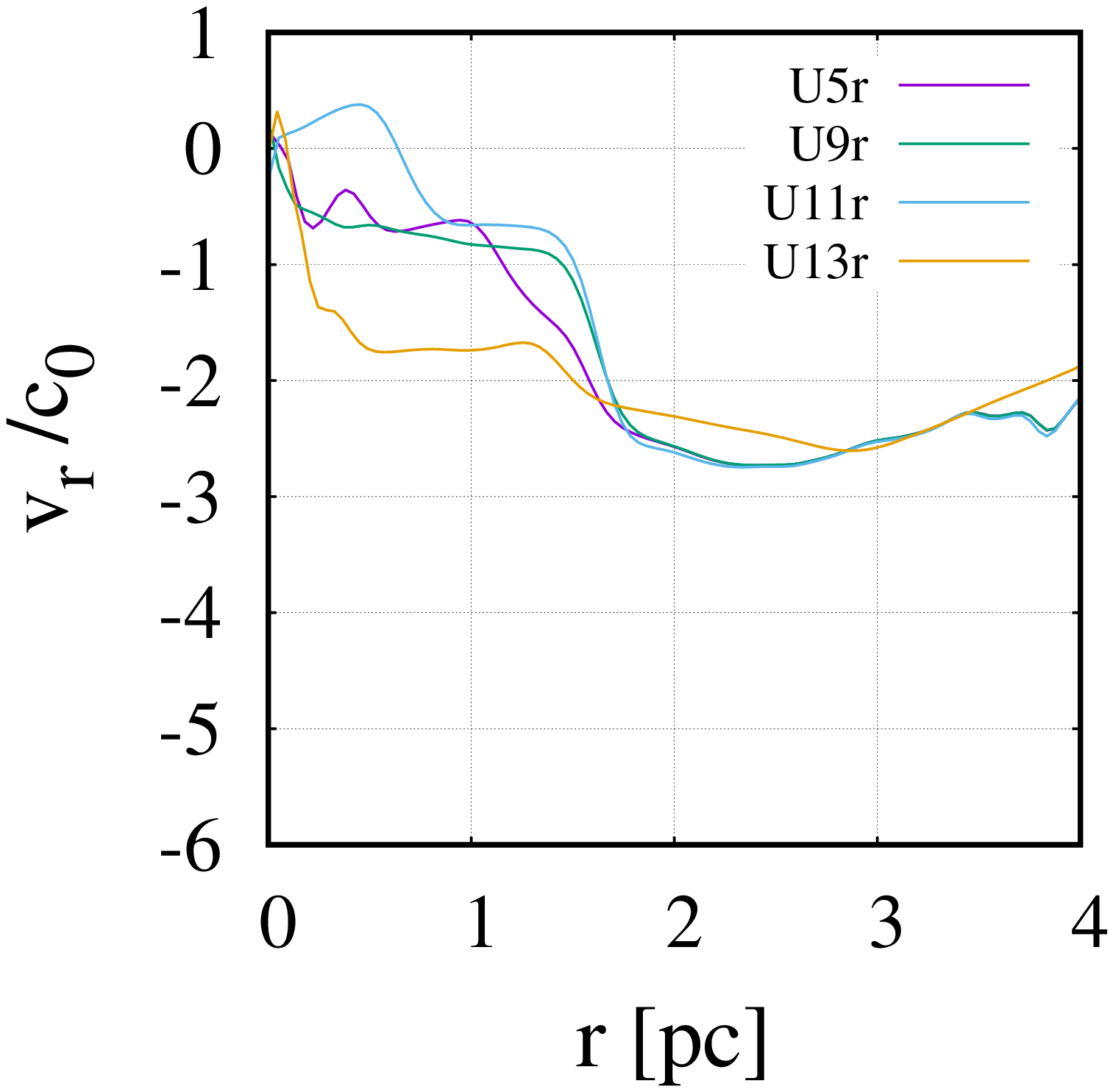} & \includegraphics[width=2.5 in]{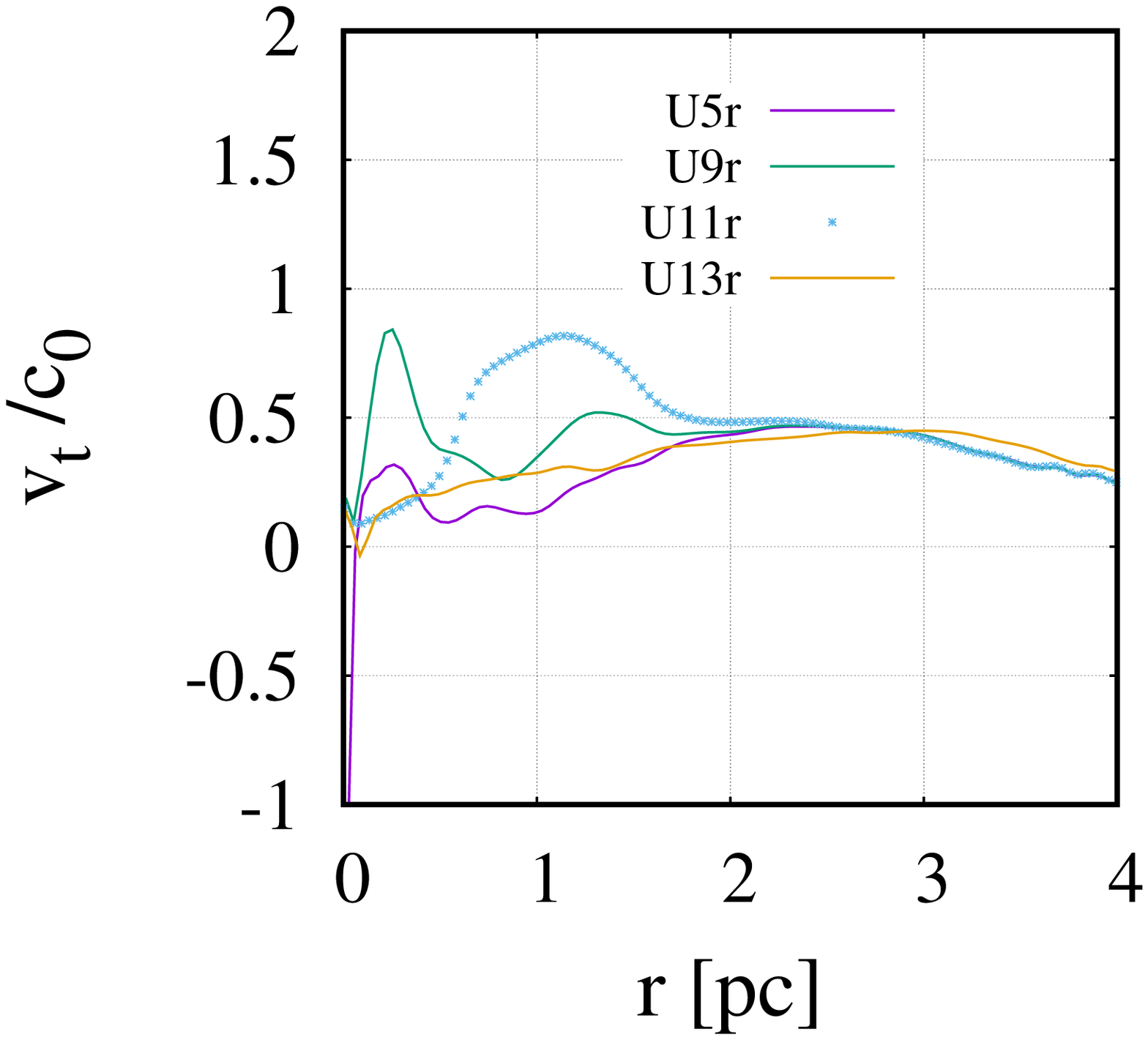}\\
\includegraphics[width=2.5 in]{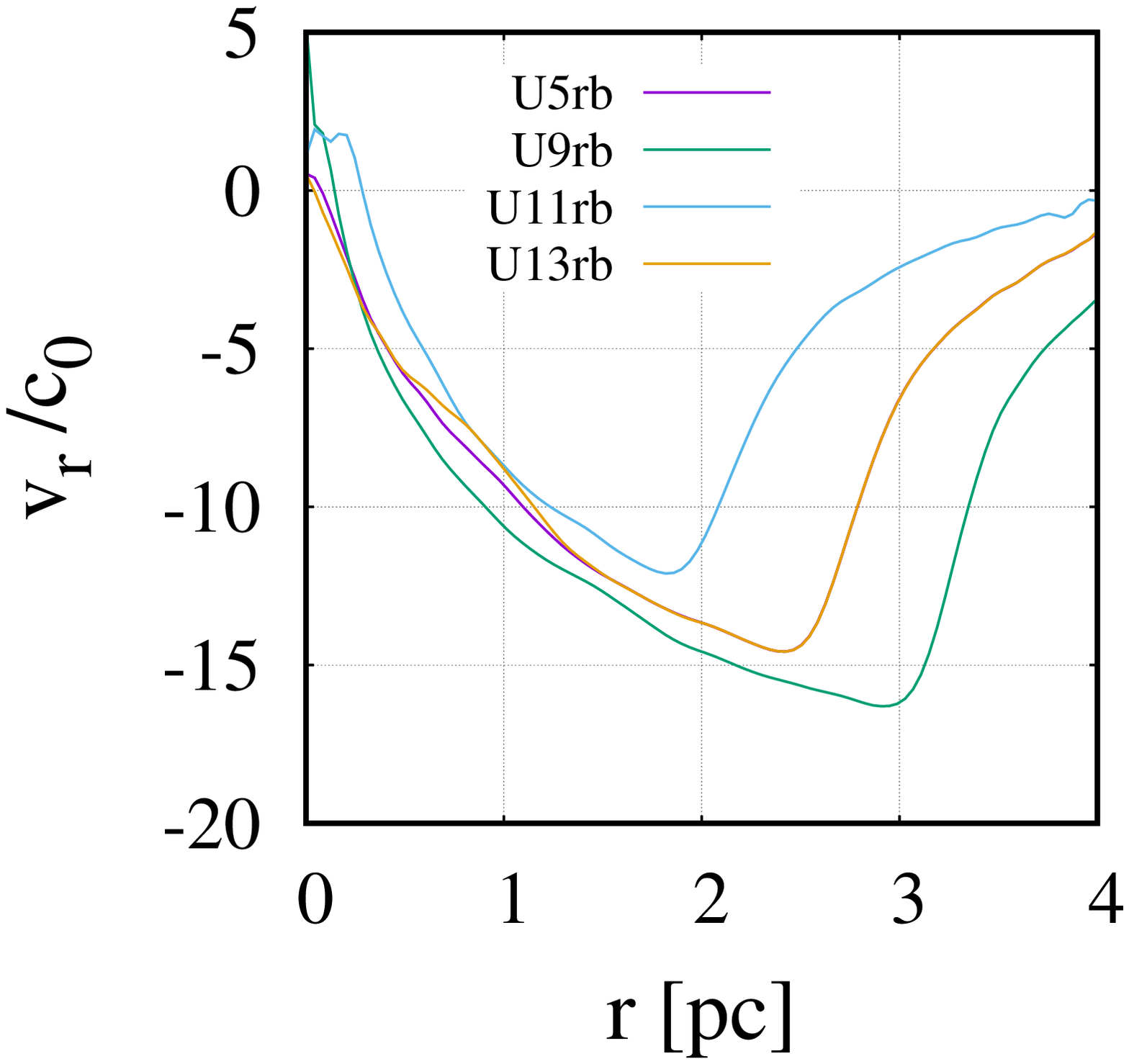} & \includegraphics[width=2.5 in]{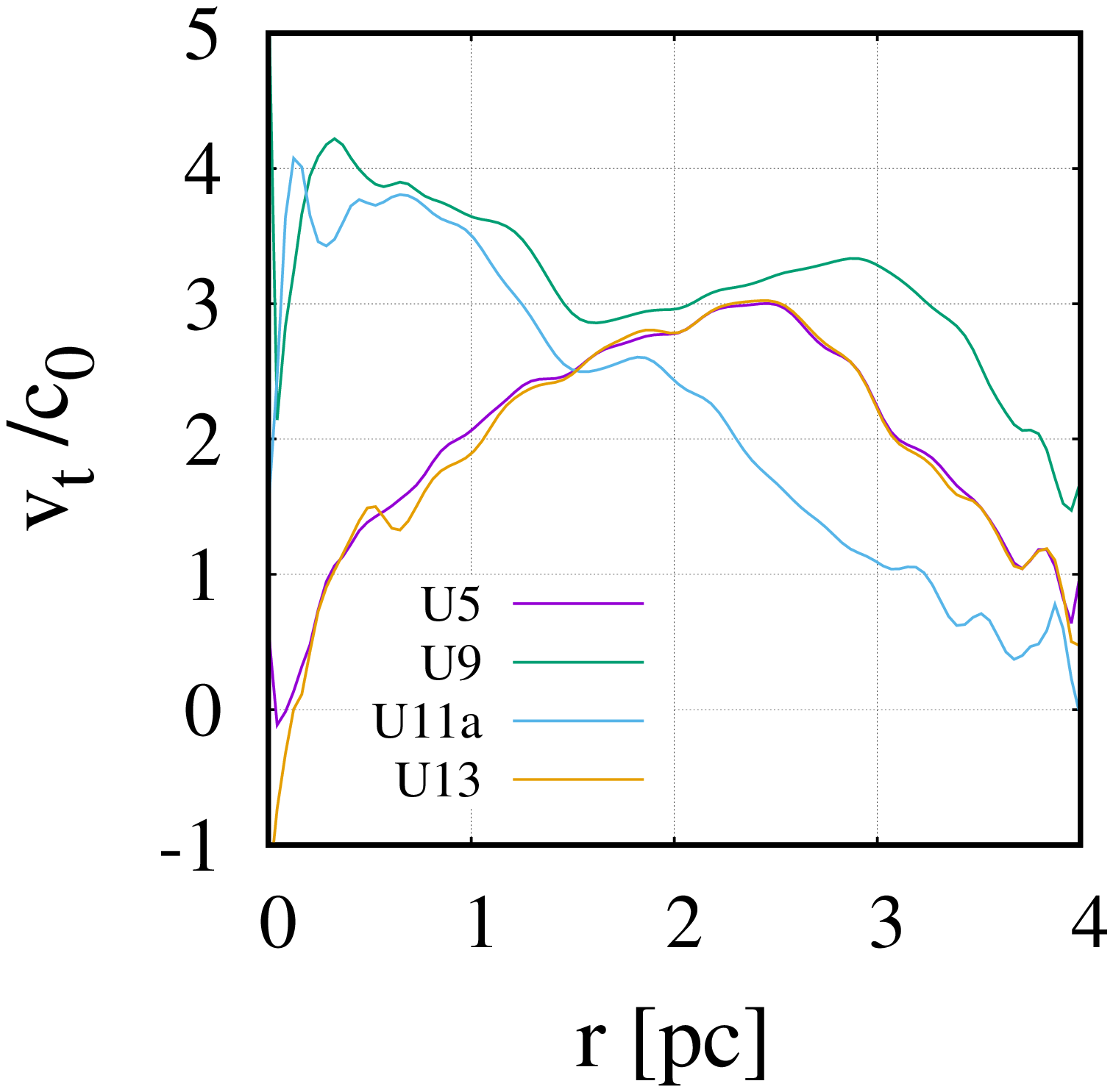}
\end{tabular}
\caption{\label{fig:VelrMosCen2} (left-hand column) Radial profile of the radial component
of the particle velocity and (right-hand column) the radial profile of the tangential component
of the particle velocity, both calculated with respect the center of each cloudlet illustrated in
Fig.\ref{fig:CentrosFragF_XY} and normalized with the speed of sound.
From top to bottom, the first line of panels is for the models $U$; the second line is for models $Ub$ and
the third and four lines are for models $Ur$ and $Urb$, respectively. The snapshot considered for these calculations
are those at the same time and peak density shown in Fig.\ref{Mosps}, Fig. \ref{Mospsb2}, Fig.\ref{MospsRot} and Fig.\ref{MospsRotb},
respectively. }
\end{center}
\end{figure}

In Fig.\ref{fig:VelrMosCen2}, we show the radial profile of the
radial (left-hand column) and tangential (right-hand column) components of
the velocity. We apply here the same radial partition described in
Section \ref{subsec:radialprofile}; that is, from the cloudlet center up to
4 pc.

Let us clarify the meaning of the tangential component of the velocity. In spherical
coordinates $(r,\theta, \phi)$, a gas particle has a magnitude of the velocity vector $v_p$ with the
components $v_r$,$v_{\theta}$ and $v_{\phi}$. Then, we
split the components of the velocity in radial $v_r$ and tangential $v_t=(v_{\theta}+v_{\phi})/2$, so that we
can follow both components separately. We do this separation because the radial component
can be associated with a collapse trend while the tangential component can be considered
as a manifestation of turbulence, see \citet{guerrero}.

The left-hand column panels Fig.\ref{fig:VelrMosCen2} indicates that many particles move to the cloudlet center,
mostly from the innermost region of the cloud. The right-hand column panels Fig.\ref{fig:VelrMosCen2} indicate that
there is a non-zero, almost constant, tangential component of the velocity. These observations
indicate that a lot of particles are falling towards the cloudlet center in trajectories
that are slightly curved (i.e., not from a purely radial direction, such as in a free-fall).

Let us recall the behavior of a test particle and let its velocity magnitude
be given by $v_g$. This $v_g$ is determined by $v_g=\sqrt{2\,G\,M(r)/r}$, where $M(r)$ is the mass contained up to
radius $r$ and $G$ is the Newton's gravitational constant. This $v_g$ can be considered as the velocity
when a test particle arrives at distance $r$ from the central mass $M$, having started from rest
at infinity, where its gravitational potential is zero. As is well-known, for a spatially bounded
mass of radius $R_g$, the velocity $v_g$ must increase with the radius $r$, such that $0<r<R_g$. Once
the radial coordinate $r$ is outside the bounded mass- that is, for $r>R_g$ the velocity $v_g$
simply decreases.

It must be noted that a significant fraction of the total velocity magnitude $v_p$ comes
from its radial component $v_r$, though a minor fraction of $v_p$ comes from the tangential components grouped in
$v_t$. Then, we can think about a curve of $v_p$ if we see a curve of $v_r$, because we only
need to transform from $v_r$ to $v_p$ by changing the velocity sign, from negative to
positive values, so that the behavior of both curves $-v_r$ and $v_p$ would have a lot of similarity.

Let us consider the panels of the top line of Fig.\ref{fig:VelrMosCen2}, which are for models $U$.
In terms of the "imagined curves" of $v_p(r)$, the cloudlets $U11a$ and $U11b$ follow the behavior expected for the
test particle velocity, indicating that the cloudlet radius $R$ (the analog of the bonded mass) is around $1$ pc
for both cloudlets of model $U11$. The curves for models $U9$ and $U13$ indicate that the bounded mass
has a very small radius $R$, which is slightly smaller than $0.5$ pc. For model $U5$, the resolution of the
radial partition is not fine enough to indicate a radius $R$ of the cloudlet found. This bounded mass can
be identified with the size of the region from the center of the cloud, which is a particle reservoir, out of which
the particles flow towards the cloud center. The core of the collapsing cloud, which is formed by
the particles with higher density of the simulation, is located in these cloud centers.

The panels of the second line of Fig.\ref{fig:VelrMosCen2}, from top to bottom, show the curves for 
the models $Ub$, which are very similar to those already described for models $U$.

The panels of the third line of Fig.\ref{fig:VelrMosCen2}, which are for models $Ur$, with a
low azimuthal velocity, indicate a behavior very similar to that observed for models $U9$ and $U13$, that is,
a region of particle reservoir is about 1 pc in radius from the center of the cloud. These behaviors
can be better seen in the panels on the bottom line of Fig.\ref{fig:VelrMosCen2}, which are for
models $Urb$. These models $Urb$ have a high azimuthal velocity, whose effect is more clearly
seen because the in-fall velocity is quite higher than in the previous models $U$, $Ub$ and $Ur$. Instead of
a bounded mass, the size of the region of strong in-fall gas is determined by the size of the centrally located
lump of gas induced by the azimuthal velocity, see the left-hand column of Fig.\ref{MospsRot}, for an illustration.
%%%%%%%%%%%%%%%%%%%%%%%%%%%%%%%%%%%%%%%%%%%%%%%%%%%%%%%%%%%%%%%%%%%%%%%%%%%%%%%%

%%%%%%%%%%%%%%%%%%%%%%%%%%%%%%%%%%%%%%%%%%%%%%%%%%%%%%%%%%%%%%%%%%%%%%%%%%%%%%%%%%%%%%%%%%%%%%%%%%
\subsection{Integral properties of the cloudlets}
\label{subsec:prop}

\begin{figure}
\begin{center}
\begin{tabular}{cc}
\includegraphics[width=2.5 in]{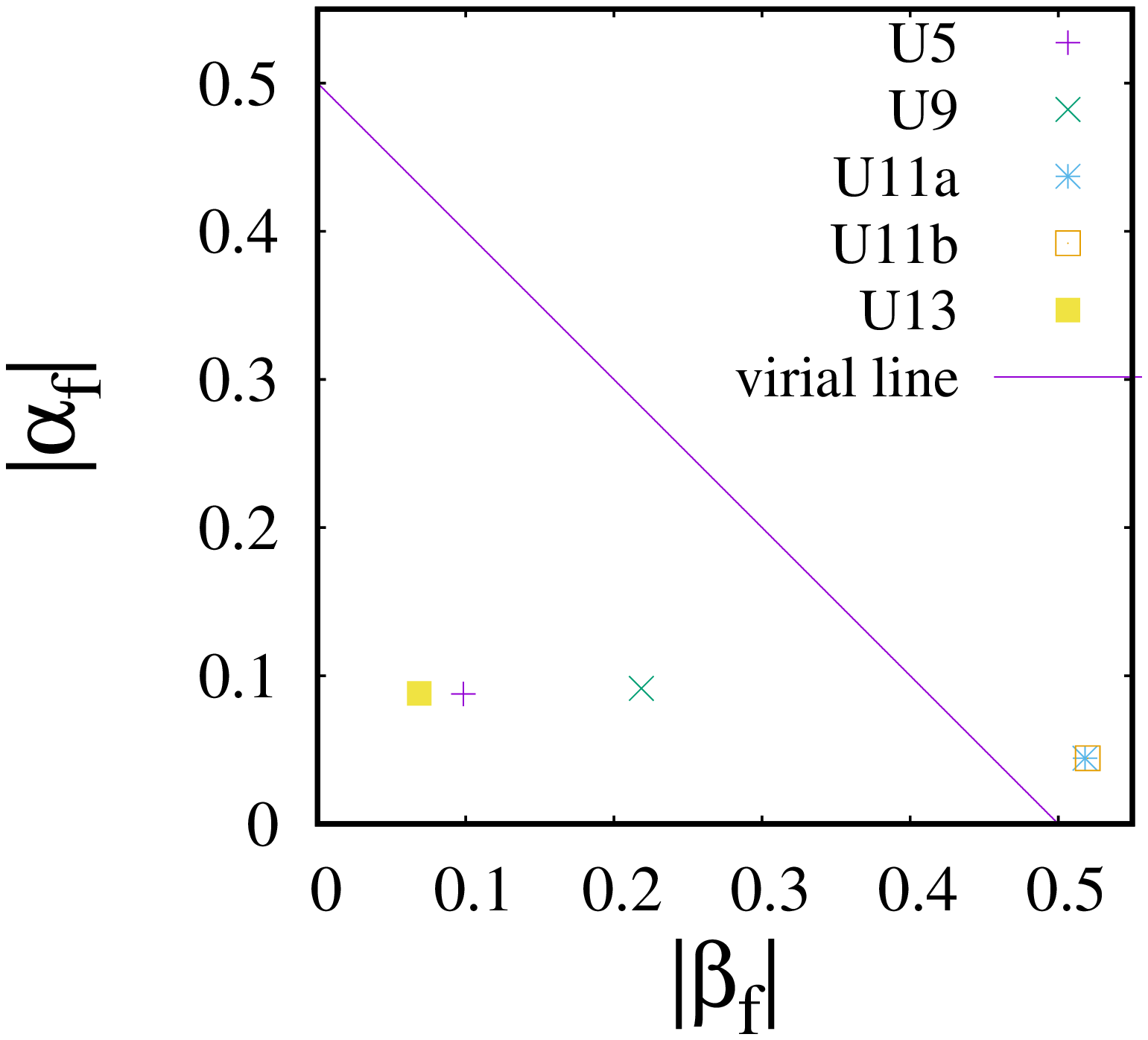} &
\includegraphics[width=2.5 in]{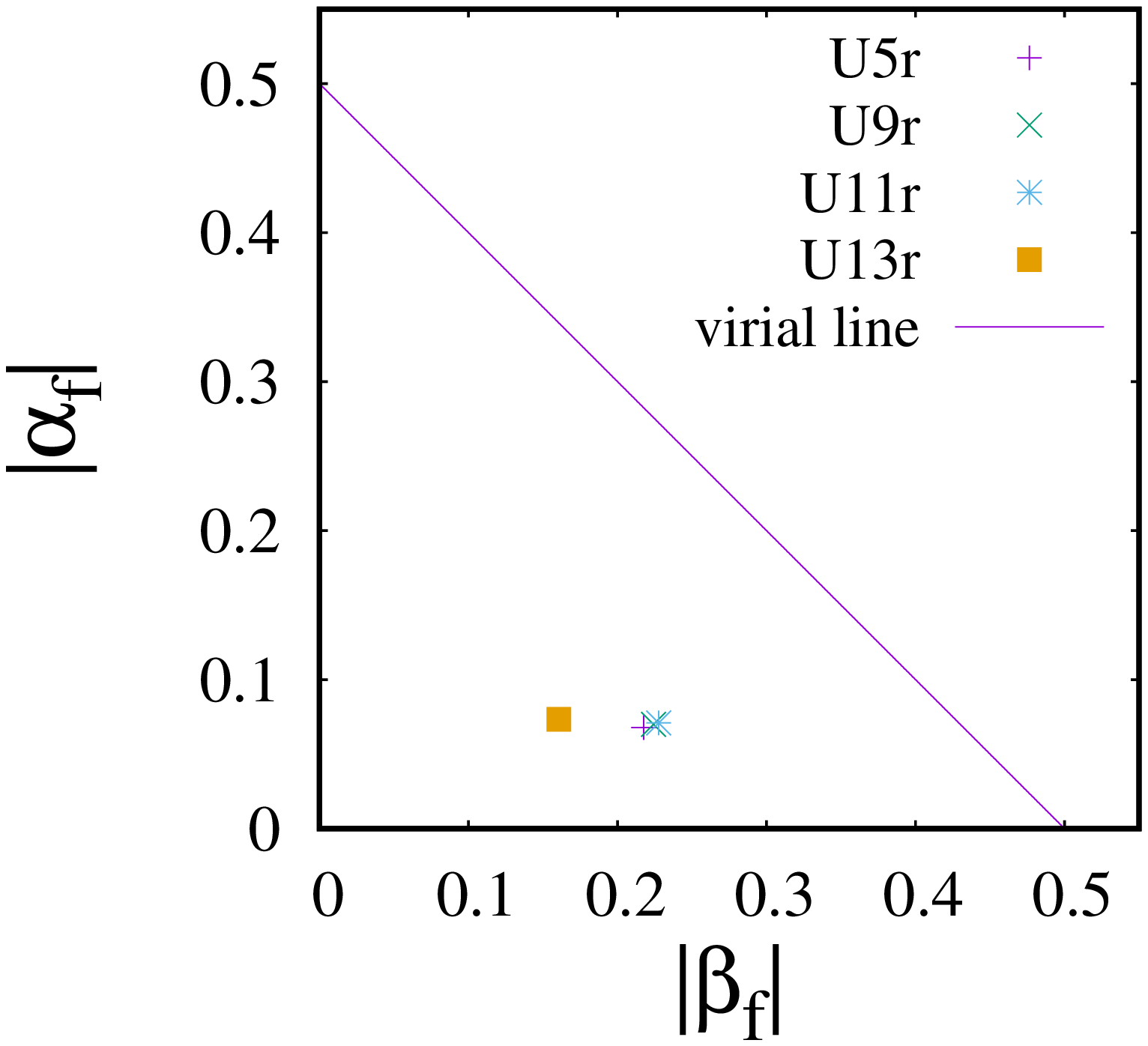} \\
\includegraphics[width=2.5 in]{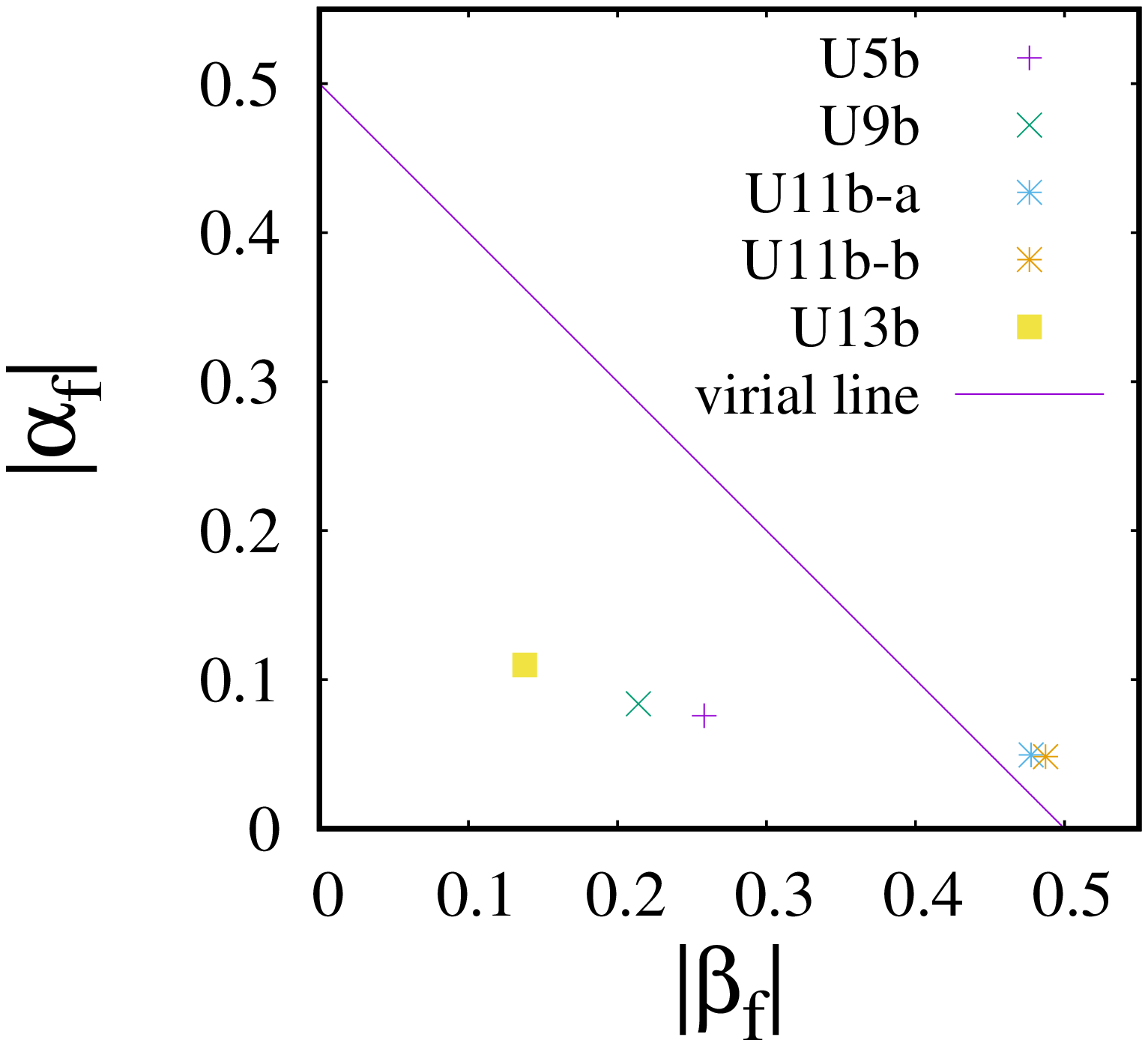} &
\includegraphics[width=2.5 in]{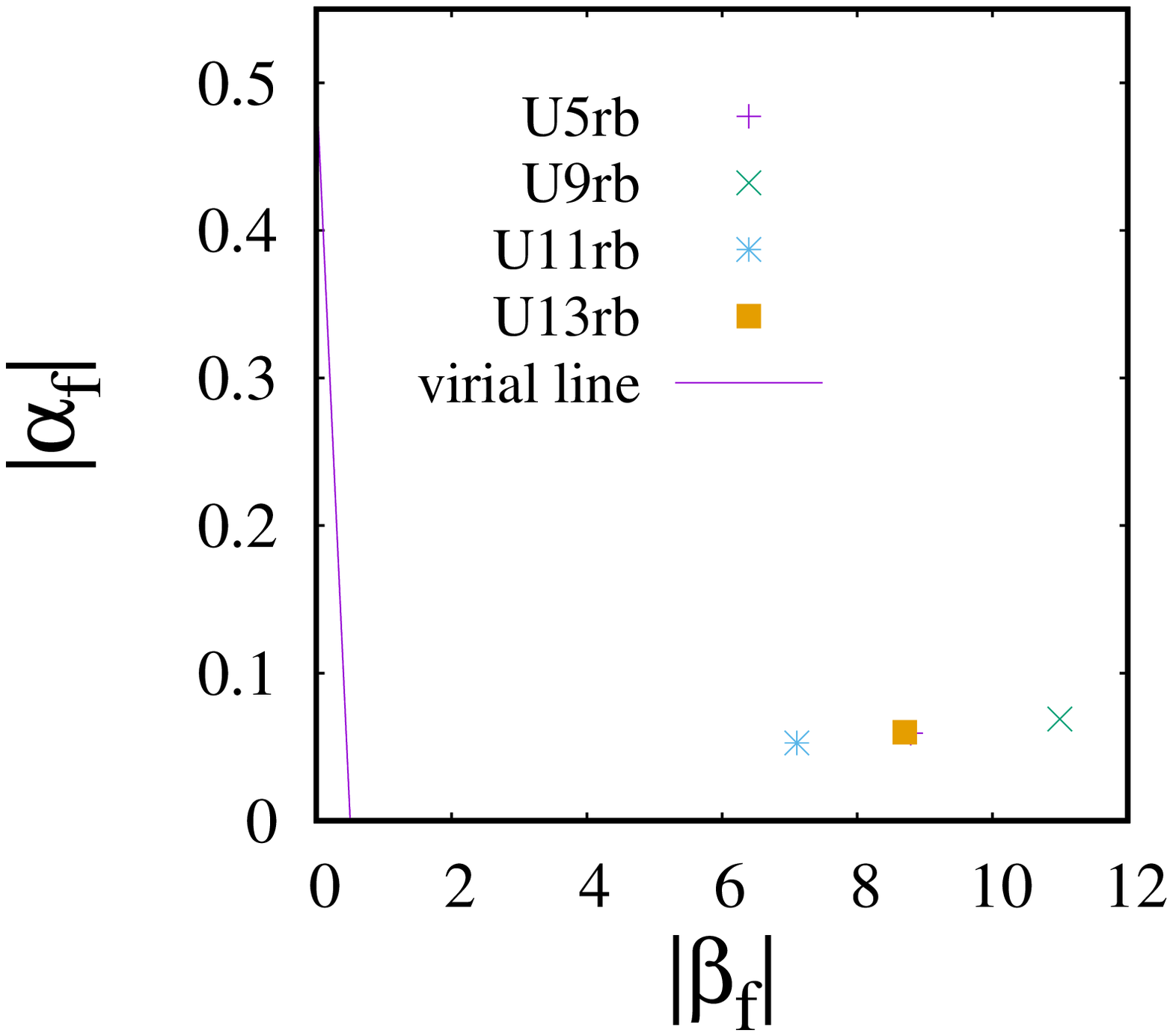}
\end{tabular}
\caption{\label{fig:AlphavsBetaFragsCen2MosCen2} The ratio of the thermal energy to the
gravitational energy $\alpha$ {\it versus} the ratio of the kinetic energy to the gravitational
energy $\beta$, of the cloudlets defined in Section \ref{subsec:prop}, at the same time that
the snapshots shown in Fig.\ref{Mosps}, Fig.\ref{Mospsb2}, Fig.\ref{MospsRot} and Fig.\ref{MospsRotb}, respectively.
(top left-hand) $U$ with a low level of turbulence;
(top right-hand) $Ur$ with a low azimuthal velocity;
(bottom left-hand) $Ub$ with a high level of turbulence and
(bottom right-hand) $Urb$ with a high azimuthal velocity.}
\end{center}
\end{figure}

In Fig.\ref{fig:AlphavsBetaFragsCen2MosCen2} we show the values of the dimensionless
ratios $\alpha$ and $\beta$, respectively, calculated only for a cloud region 
that includes the cloudlets and their surroundings. Two parameters are used: the first parameter 
is $\log \left( \rho_{\rm min} \right) $, which is a lower bound for density; and the
second parameter $r_{\rm max}$ is a maximum radius, which 
is taken with respect to the cloudlet's center. To calculate 
the ratios $\alpha_{f}$ and $\beta_{f}$ for the cloudlets,
we consider only those particles that have a density greater than $\log \left( \rho_{\rm min} \right)$
and are located at a radius smaller than $r_{\rm max}$. To make a comparison 
between all the models, to calculate 
the properties of all the models shown in Fig.\ref{fig:AlphavsBetaFragsCen2MosCen2}, we have used 
the following values $\log \left( \rho_{\rm min} \right)=0.0$ and $r_{\rm max}=1.5$ pc.

One can see in the top left-hand panel of Fig.\ref{fig:AlphavsBetaFragsCen2MosCen2}
that the models $U5$,$U9$ and $U13$ have a similar value for the
ratio $\alpha_{f}$, which is around 0.1. For the cloudlets "a" and "b" of model $U11$ (i.e., $U11a$
and $U11b$) $\alpha_{f}$ is around 0.05. The values of the ratio $\beta_{f}$ ranges from 0.1
for model $U13$, around 0.13 for model $U5$, and a little higher than 0.2 for model $U9$. Cloudlets
$U11a$ and $U11b$ have the highest values of $\beta_{f}\, \approx \, 0.55$. Cloudlets $U11a$ and
$U11b$ are the only ones over-virialized, because their sum $\alpha_{f} + \beta_{f} >1/2$; while
the models $U5$,$U9$ and $U13$ are sub-virialized, because their sum $\alpha_{f} + \beta_{f} <1/2$. The
same behavior is observed for models $Ub$, with a high turbulence, as can be seen in the bottom
left-hand panel of Fig.\ref{fig:AlphavsBetaFragsCen2MosCen2}.

In both the top right-hand panel and the bottom left-hand panel of Fig.\ref{fig:AlphavsBetaFragsCen2MosCen2}, we see
that curves for the models $Ur$ and $Ub$, respectively, show a behavior that is very similar to
that already observed for the curves of models $U$. For all models
$Ur$ and $Ub$ the ratio $\alpha_{f}$ is around the value 0.09, with a clear tendency to
lower values. The ratio $\beta_{f}$ for these models ranges from 0.1 to 0.5
Models $Ur$, $Ur5$, $Ur9$ and $Ur13$ are sub-virialized, because their
sum $\alpha_{f} + \beta_{f} <1/2$. Meanwhile, for the two cloudlets of model $Ur11$ are over-virialized, because
their sum $\alpha_{f} + \beta_{f} > 1/2$. The same behavior is observed for models $Ub$.

On the opposite side, all models $Urb$, as shown in
bottom right-hand panel of Fig.\ref{fig:AlphavsBetaFragsCen2MosCen2}, are over-virialized, because
the high initial azimuthal velocity is manifested in the excess of kinetic energy, such that the
$\beta_{f}$ ratio for these models is in the range 7-11.  In spite of this excess of kinetic energy, all
models $Urb$ have collapsed.

%%%%%%%%%%%%%%%%%%%%%%%%%%%%%%%%%%%%%%%%%%%%%%%%%%%%%%%%%%%%%%%%%%%%%%%%%%%%%%%%%%%%%%%%%%%%
%%%%%%%%%%%%%%%%%%%%%%%%%%%%%%%%%%%%%%%%%%%%%%%%%%%%%%%%%%%%%%%%%%%%%%%%%%%%%%%%%%%%%%%%%%%%

\begin{figure}
\begin{center}
\begin{tabular}{cc}
\includegraphics[width=2.5 in]{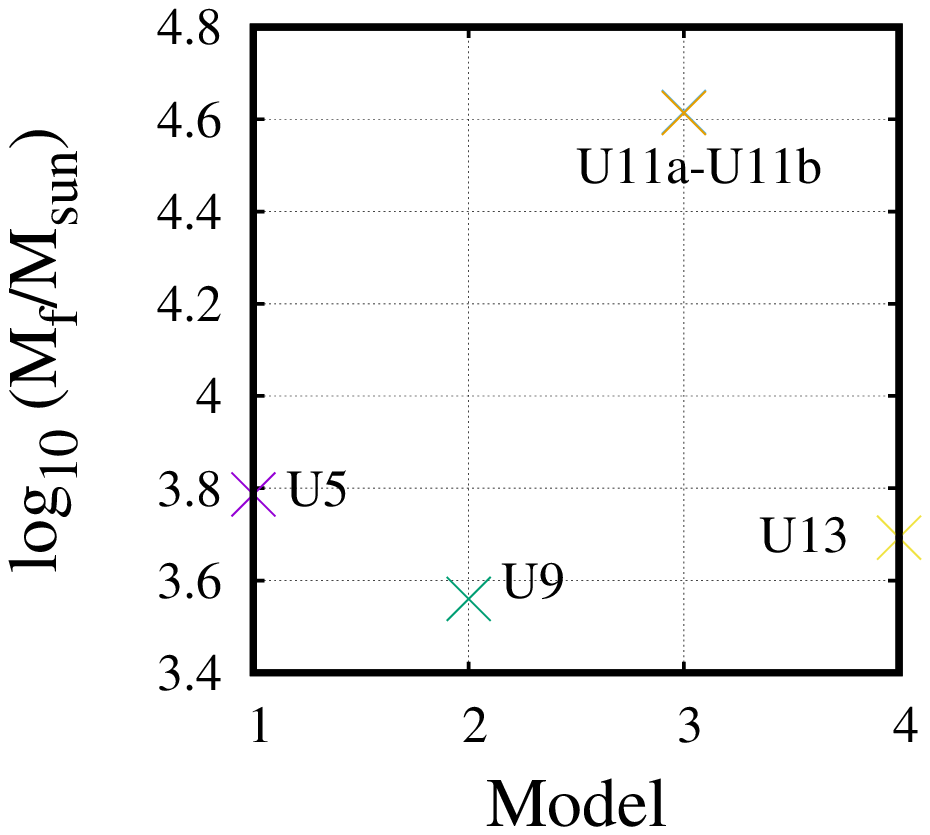} &
\includegraphics[width=2.5 in]{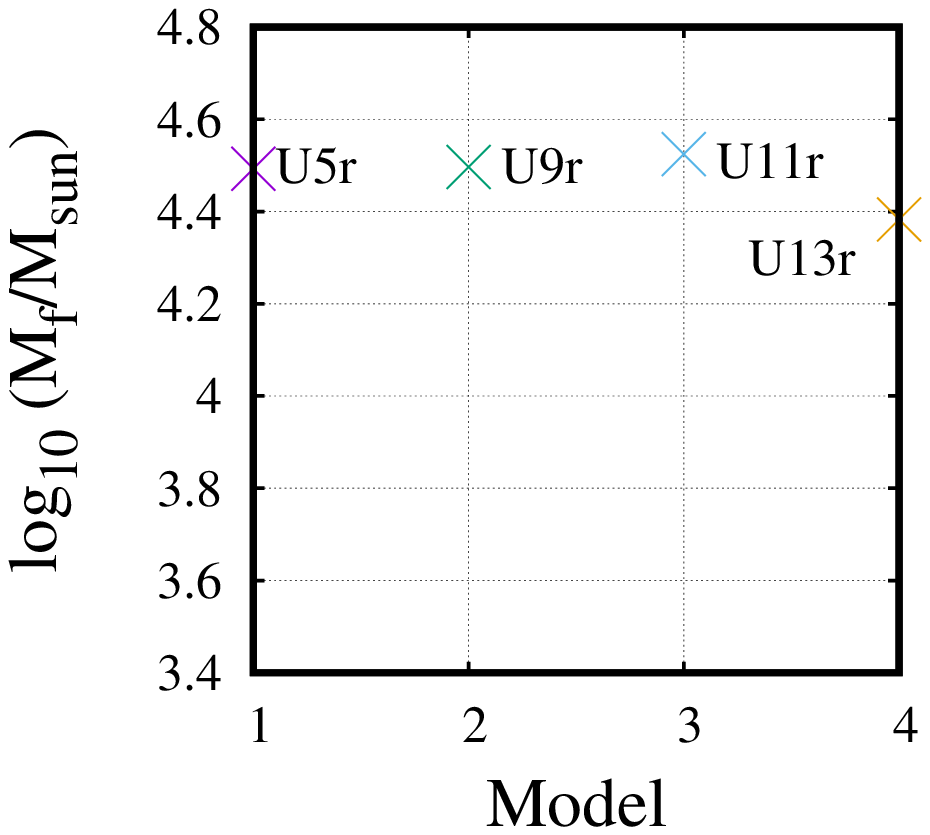}\\
\includegraphics[width=2.5 in]{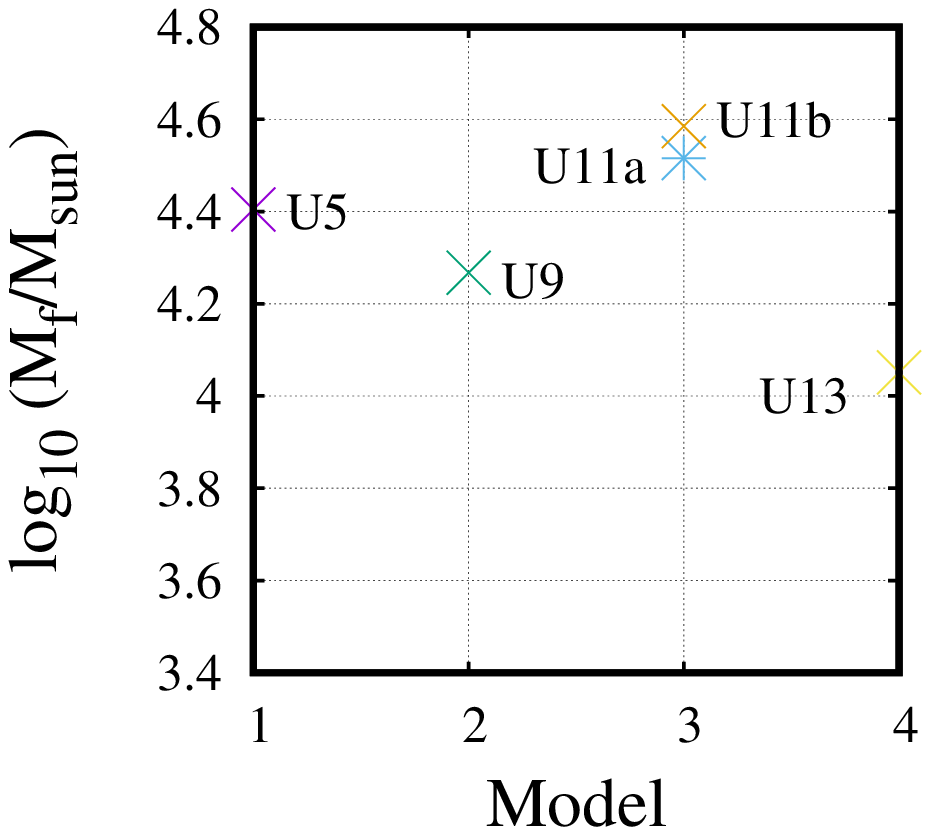} &
\includegraphics[width=2.5 in]{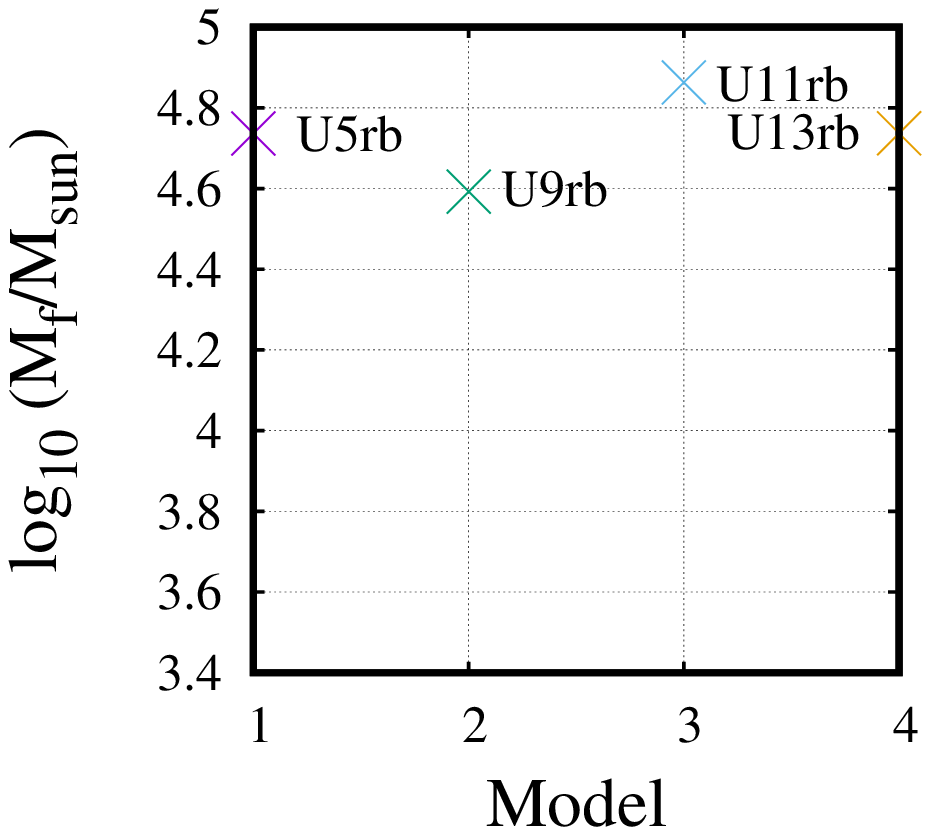}
\end{tabular}
\caption{\label{fig:NLogMassFrags} The log of the mass of the cloudlets defined
in Section \ref{subsec:prop}, at the same time that the snapshots shown
in Fig.\ref{Mosps}, Fig. \ref{Mospsb2}, Fig.\ref{MospsRot} and Fig.\ref{MospsRotb}, respectively.
(top left-hand) $U$ with a low level of turbulence;
(top right-hand) $Ur$ with a low azimuthal velocity;
(bottom left-hand) $Ub$ with a high of turbulence and
(bottom right-hand) $Urb$ with a high level of turbulence.}
\end{center}
\end{figure}

We show the mass associated with the cloudlets of the models on the vertical axis of Fig.\ref{fig:NLogMassFrags}. 
We have considered the mass of only those particles that entered
into the calculation of the physical properties shown in Section \ref{subsec:charac}. In the
horizontal axis, we show model numbers from 1 to 4, ordering the models in the following
way: $U5$, $U9$, $U11$ and $U13$, respectively.

The mass of the cloudlets for models $U5$, $U9$ and $U13$
are shown in the top left-hand panel, so that the mass ranges from
$\log \left( M_{\rm f}/ M_{\odot} \right)=$3.6 to 3.8. Two cloudlets of model $U11$ have 
the largest masses, around $\log \left( M_{\rm f}/ M_{\odot} \right)=4.6$.

In contrast to what we have observed when compared to other physical properties of models $U$ 
with $Ub$, the behavior is different, in the case of the mass of the cloudlets, without any trend.

The mass of the cloudlets for model $Ur$ is shown in the top right-hand panel of
Fig.\ref{fig:NLogMassFrags}. We see that the models $Ur5$, $Ur9$ and $Ur11$ all have 
similar cloudlet masses, which are around $\log \left( M_{\rm f}/ M_{\odot} \right)=4.5$. Meanwhile,
the cloudlet mass for model $Ur13$ is a little smaller mass than the mass of the other models,
$\approx \log \left( M_{\rm f}/ M_{\odot} \right)=4.4$.

Given that models $Urb$ have the highest in-fall radial velocity of the all the models, as can be seen in
\ref{fig:VelrMosCen2}, then their mass accretion rate must be the highest too; for this, the
mass enclosed is systematically higher than in the other models, although there is not
much difference in the mass values observed in the panels of Fig.\ref{fig:NLogMassFrags}
when compared to the big difference in the in-fall radial velocity.

It must be emphasized that the mass scale shown in Fig.\ref{fig:NLogMassFrags} is in
agreement with the mass observed for an open cluster of stars, which is
around $10^{4} \ M_{\odot}$, while the mass scale of a globular cluster of stars
is around $10^{5} \ M_{\odot}$, see \citet{kumai}.

The results of the collapse of a gas core (e.g., in the so called "the standard
isothermal simulation") are identified as protostars (\citet{boss95}, \citet{boss2000}, \citet{burkert}
and \citet{arreaga2007}). In the same sense, the gas structures obtained from
the simulations of the present paper and whose properties are shown
in Section \ref{subsec:charac} can be called a proto-cluster of protostars. As is well-known
for simulations of the collapse of a gas core, the mass of the proto-stars
depends on the mass of the parent cloud, see for instance \citet{arreaga2016}. The same
is expected to be true for proto-clusters.

%%%%%%%%%%%%%%%%%%%%%%%%%%%%%%%%%%%%%%%%%%%%%%%%%%%%%%%%%%%%%%%%%%%%%%%%%%%%%%%%%%%%%%%%%%%%

Finally, it must be recalled that the results displayed in Fig.\ref{fig:AlphavsBetaFragsCen2MosCen2} and
Fig.\ref{fig:NLogMassFrags} are taken when the collapse is still ongoing, so that the peak density is
around $\log \left( \rho_{\rm max}/\rho_0 \right) \approx 5$. As we have seen in
Section \ref{subsec:peak}, the final state of the collapse reached a peak
density around $\log \left( \rho_{\rm max}/\rho_0 \right) \approx 8$.
%%%%%%%%%%%%%%%%%%%%%%%%%%%%%%%%%%%%%%%%%%%%%%%%%%%%%%%%%%%%%%%%%%%%%%%%%%%%%%%%%%%%%%%%%%%%
%%%%%%%%%%%%%%%%%%%%%%%%%%%%%%%%%%%%%%%%%%%%%%%%%%%%%%%%%%%%%%%%%%%%%%%%%%%%%%%%%%%%%%%%%%%%%%%
%%%%%%%%%%%%%%%%%%%%%%%%%%%%%%%%%%%%%%%%%%%%%%%%%%%%%%%%%%%%%%%%%%%%%%%%%%%%%%%%%%%%%%%%%%%%%%%
%%%%%%%%%%%%%%%%%%%%%%%%%%%%%%%%%%%%%%%%%%%%%%%%%%%%%%%%%%%%%%%%%%%%%%%%%%%%%%%%%%%%%%%%%%%%%%%
\section{Discussion}
\label{sec:dis}

Although the dynamics of an isolated turbulent cloud is well-known- see for instance,
\citet{goodwina}, \citet{goodwinb} and \citet{goodwin2006}, in Section \ref{subsec:nubeaislada}- we 
begin by describing the evolution of the isolated turbulent cloud, to discuss its
influence on the collision models considered in Sections \ref{subsec:turb}, \ref{subsec:levelturb}
and \ref{subsec:wellapprox}. After this, we then commence the discussion about the most important features
of the collision models presented in Section \ref{sec:results}.

%%%%%%%%%%%%%%%%%%%%%%%%%%%%%%%%%%%%%%%%%%%%%%%%%%%%%%%%%%%%%%%%%%%%%%%%%%%%%%%%%%%%%%%%%%%%%%%%%%
\subsection{The collapse of the isolated turbulent cloud}
\label{subsec:nubeaislada}

We mentioned in Sections \ref{sec:int} and
\ref{subs:energies} that the initial conditions of the isolated cloud are chosen to favor
its gravitational collapse. The curve of the peak density for the isolated cloud 
develops a small peak at a time smaller than $t/t_{ff}=0.1$. This increase of
density happen because of the multitude of gas lumps formed by the collisions between gas particles that occur
simultaneously throughout the cloud. This density 
peak does not appear for low levels of initial kinetic
energies, which is measured by the $\beta$ ratio defined in Eq.\ref{defbeta}; for high values
of the $\beta$ ratio, this early peak is quite noticeable. 

In contrast, the time required by the isolated cloud to reach the highest
density values, for instance $\log \left( \rho_{\rm max}/\rho_0 \right) \approx 6$, does not
depend significantly on the level of the initial energy, at least for a wide range of the initial
$\beta$ ratio. This is due to the fact that almost all of the kinetic energy available is equally
dissipated by means of the random collision of particles, as described in the
previous paragraph. The time required by the cloud to reach its highest peak density
is around $t/t_{ff}=2.5$, which is of the same order of time that can be seen
in Fig.\ref{fig:DenMax} for the collapse of the collision models.

In the time interval between $0.1 < \, t/t_{ff} \, < 2.0$, a relaxation of the small gas lumps occurs
throughout the cloud, so that the peak density curve decreases quickly. From there, it increases
very slowly, up to times $t/t_{ff} \, > \,2.0$, in which the final collapse takes place very quickly.

\begin{figure}
\begin{center}
\begin{tabular}{cc}
\includegraphics[width=2.0 in]{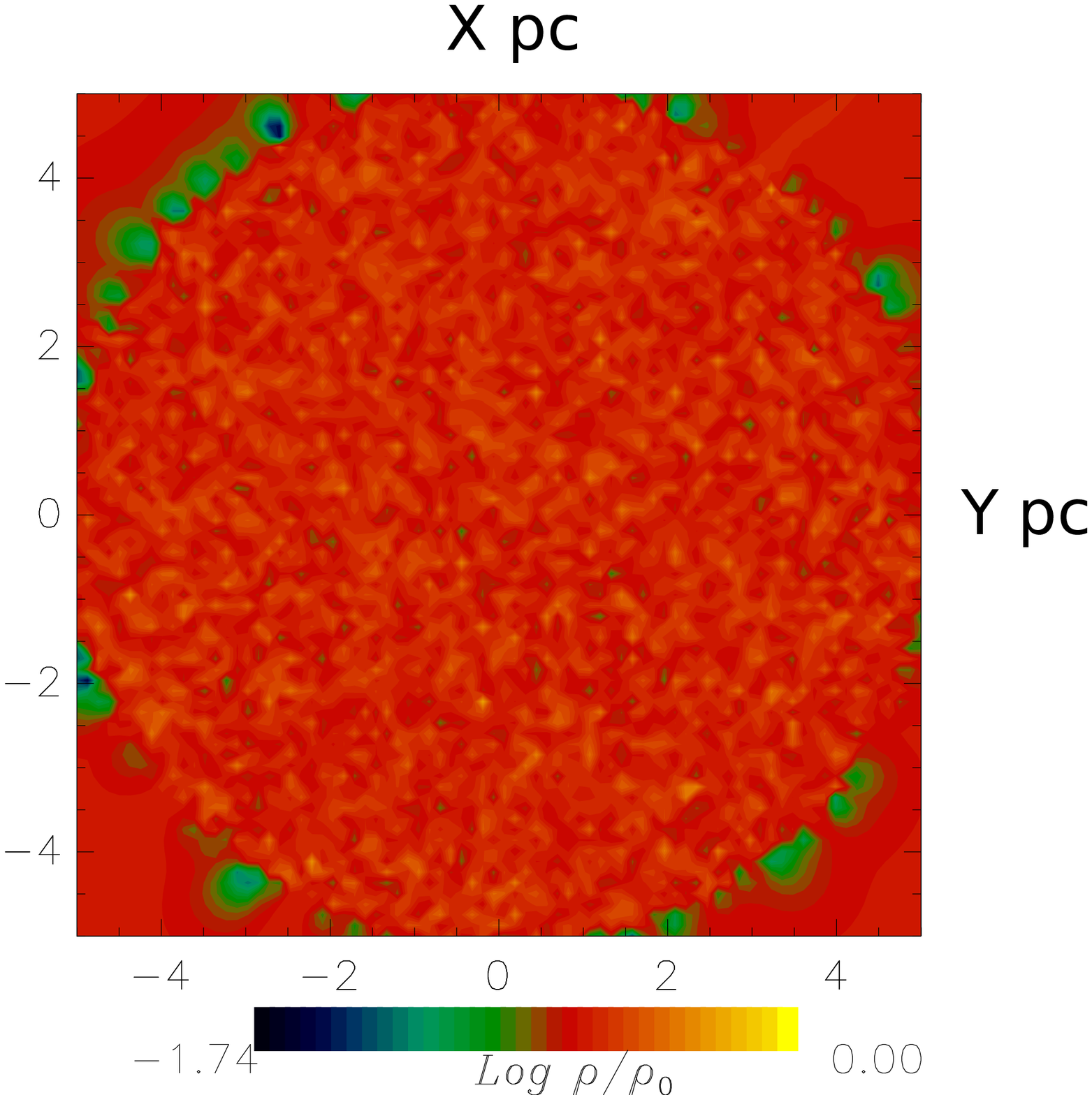} &
\includegraphics[width=2.14 in]{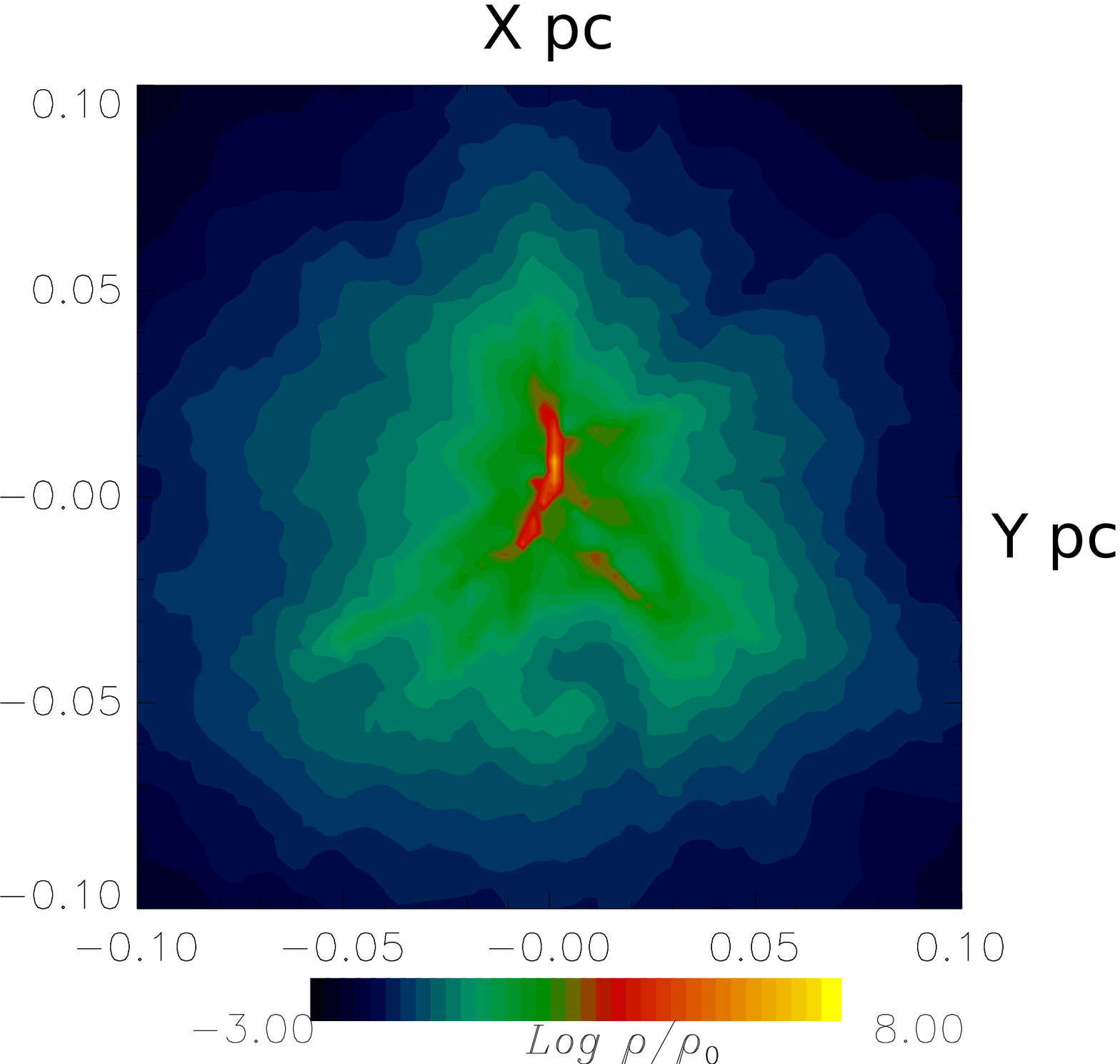}\\
\end{tabular}
\caption{\label{MosNubeinicial} Column density plots of the isolated turbulent cloud, 
for a thin slice of gas parallel to the x-y plane. The unit 
of length is one parsec. The plots are shown in panels as follows:
(left-hand) at time $t/t_{ff}= 0.02$ and peak density $\log \left( \rho_{\rm max}/\rho_0 \right)=0.53$;
(right-hand) at time $t/t_{ff}=2.5$ and peak density $\log \left( \rho_{\rm max}/\rho_0 \right)=8.0$.}
\end{center}
\end{figure}

From the point of view of the column density plots, the occurrence of collisions between gas
particles, as a consequence of turbulent velocity field implemented initially, is seen as
a random formation of many over-dense lumps of gas, which are homogeneously distributed
across the entire cloud volume, see the left-hand panel of Figure \ref{MosNubeinicial}. Later, when 
the initial kinetic energy of the cloud is dissipated, the cloud gets a physical state similar 
to a free-fall collapse, which is seen as a clear tendency
to a global collapse towards its central region. However, at the final evolution stage that could
be followed in this paper, the mass accretion with spherical symmetry is lost, so that a central
dense filamentary structure forms that is highly anisotropic and with a high possibility of
fragmenting, see right-hand panel of Figure \ref{MosNubeinicial}. The behavior described in this section 
is paradigmatic of turbulence.

%%%%%%%%%%%%%%%%i

%%%%%%%%%%%%%%%%%%%%%%%%%%%%%%%%%%%%%%%%%%%%%%%%%%%%%%%%%%%%%%%%%%%%%%%%%%%%%%%%%%%%%%%%%%%%%%%%%%%%%%%
\subsection{Does the turbulence make a difference in the collision simulations ?}
\label{subsec:turb}

The occurrence of the collision between the sub-clouds induced by the translation
velocity $v_{L}:v_{R}$ prevents the gas
particles from forming small lumps of gas throughout the cloud by means of early random collisions;
as explained in the previous Section \ref{subsec:nubeaislada}.

There is over-dense gas in the contact region between the colliding sub-clouds. 
This over-density accelerates the collapse of the remaining gas of the cloud, so
that the turbulence does not have time enough to get relaxed by dissipation of the kinetic energy. For
this reason, the turbulence does not play a fundamental role in the outcome of the
simulations, such as $U$ and $Ub$. In fact, if one turns off the turbulence and keeps
only the collision process of the sub-clouds in these models, then the results are
basically the same.

%%%%%%%%%%%%%%%%%%%%%%%%%%%%%%%%%%%%%%%%%%%%%%%%%%%%%%%%%%%%%%%%%%%%%%%%%%%%%%%%%%%%%%%%%%%%%%%%%%%%%%%
\subsection{Does the level of turbulence make a difference in the collision simulations ?}
\label{subsec:levelturb}

We recall that the level of turbulence in the simulations can be modified by introducing an
arbitrary multiplicative constant in Eq.\ref{velPhi}. As we mentioned in Section
\ref{justi}, there is interest in considering models of turbulent clouds with extreme initial
kinetic energy in addition to those clouds with low-level of turbulence, which
are more favored statistically. Consequently, we have studied the effect of the level
of turbulence on the simulations (i.e., models $Ub$), as can be seen in Table \ref{tab:models}.

The average Mach velocity ${\cal M}_p$ of the gas particles for the low-level of turbulence models 
$U$ is around ${\cal M}_p \approx 2.9$. For the radial component of the velocity (calculated with respect
to the origin of coordinates of the simulation box), the average Mach number is ${\cal M}_r \approx -0.13$.
For the tangential component of the velocity, the average Mach number is ${\cal M}_t \approx 0.01$. In the
meantime, the translational velocities (around 15 km/s) given to the particles that are to collide
are of order ${\cal M}_c \approx 6.6$. Then, for models $U$, ${\cal M}_c \gg {\cal M}_{p,r,t}$.

For the high-level of turbulence models $Ub$, we have an average ${\cal M}_p \approx 25$. Despite 
this significant increase of the
magnitude of the velocity, the average radial and tangential components do not change appreciably
with respect to those of models $U$; that is, ${\cal M}_r \approx -0.12$ and
${\cal M}_t \approx 0.01$. In contrast, for models $Ub$ we have ${\cal M}_c \ll {\cal M}_{p}$.

In spite of this opposite features in the relation of Mach numbers for models $U$ and $Ub$ with respect to
the translational velocity, the outcome of the simulations $U$ and $Ub$ do not show any
significant difference with respect to the final configuration of models $U5$, $U9$ and $U13$. The
only differences that can be observed are: (i) that the double bridge of gas formed in model $U11$ becomes only one bridge
in model $U11b$; and (ii) that the density peak curves, shown in Fig.\ref{fig:DenMax}, for the $Ub$ models
are displaced to right-hand side at large free-fall times, so that the collapse takes a little longer than for
models $U$.

%%%%%%%%%%%%%%%%%%%%%%%%%%%%%%%%%%%%%%%%%%%%%%%%%%%%%%%%%%%%%%%%%%%%%%%%%%%%%%%%%%%%%%%%%%%%%%%%%%%%%%%%
\subsection{How useful is the approximation of an azimuthal velocity to 
mimic tidal forces in the collision simulations ?}
\label{subsec:wellapprox}

There is an obvious problem with the approximation of a velocity instead of a tidal 
force, as described in Section \ref{subs:Vcir}, which is that the azimuthal velocity entered only once
in the simulations, as an initial condition of the gas particles. Obviously, this
is a severe limitation of the model, similar to that of the turbulence, which is not 
replenished continually during a simulation. Therefore, the effect of the tidal interaction must 
be activated during all the simulation time.

However, we observe in Fig. \ref{MospsRot} that the immediate effect of the azimuthal velocity 
on the simulated cloud is a strong tendency for the gas to be accumulated quickly at the cloud's 
center. In this case, if the azimuthal velocity terms given in Eq. \ref{velcomponentssol} were 
implemented at every time step of the simulation, then to model
more appropriately the tidal force at all the simulation time one would 
expect this tendency to accelerate the central collapse of the cloud.

It should be emphasized that the previous statement is based on the results of a very naive model, in
which the only information about the massive center exerting a gravitational force on the cloud,
is by means of the circular velocity, which is given by $\sqrt{2\, G \, M(R) / R}$, as described
in Section \ref{subs:Vcir}.

According to Section \ref{subs:Vcir}, the approximation of the azimuthal velocity is valid as long
as the ratio between the cloud radius to the distance to the gravitational center is quite small.
A way to check the applicability of this approximation is obviously to make the calculation
without the approximation. However, this is not an immediate calculation. The main difficulty 
is the difference in length and mass scales when considering a small
cloud (with a very few parsecs of radius) near a massive
object (probably with a scale of kpc in radius, separated from the cloud by several
hundreds of pc o even a kpc and whose mass can quite greater than that
of the cloud), both of which must have evolved together in the same simulation code.

For instance, \citet{gnedin} resorts to a
re-simulation technique, so that a low-resolution simulation of the massive
object (e.g., a central dwarf galaxy) is first simulated to obtain an
approximate gravitational potential. This is then used in a second high-resolution simulation of the
cloud, in which this potential is taken into account as an external time-varying field on the gas 
particles. However, applying this technique to the problem presented in this work would require a future paper.

%%%%%%%%%%%%%%%%%%%%%%%%%%%%%%%%%%%%%%%%%%%%%%%%%%%%%%%%%%%%%%%%%%%%%%%%%%%%%%%%%%%%%%%%%%%%%%%%%%%%%%%%%%%%%%%%
\subsection{A brief review of the literature on this subject}
\label{subsec:briefrev}

A lot of papers have simulated isolated
clouds and followed their collisions. However, simulations of clouds under the influence of an
external gravitational potential are limited in number.

Let us now mention briefly some results of more accurate calculation methods of the tidal effects
on clouds, which is a subject that has a long history. For instance, \citet{sigalotti92} used a time-varying
gravitational potential to calculate the equal-sized cloud-cloud tidal interaction of clouds that are in an elliptic
orbit, and reported configuration transformations on the clouds in their course to collapse.

More recently, \citet{longmore13b} proposed that the collapse of the Brick is a progenitor of a
star cluster, whose collapse was triggered as a consequence of the tidal compression exerted by Sgr B2 during
the most recent peri-center passage of the Brick.

\citet{Kruijssen15} determined a realistic orbit of a dense gas streams in the CMZ. In a subsequent paper,
\citet{Kruijssen19} calculated the tidal interaction of the galactic
center on the orbit followed by the dense gas streams of the CMZ. They found that the tidal interaction
acting upon the clouds makes a compression on the vertical direction, which causesthe clouds to become
pancake-like structures.

Later, \citet{dale19} simulated the evolution of turbulent clouds in orbit
at the CMZ. The authors were prepared to have similar magnitude of the kinetic energy to the gravitational energy
and found that the clouds collapse rapidly. This paper is a mature way of simulating tidal 
force in cloud dynamic evolutions by introducing the tidal force potential.

%%%%%%%%%%%%%%%%%%%%%%%%%%%%%%%%%%%%%%%%%%%%%%%%%%%%%%%%%%%%%%%%%%%%%%%%%%%%%%%%%%%%%%%%%%%%%%%
\subsection{Applicability of these simulations to represent the evolved clouds of the CMZ}
\label{subs:applica}

In view of Sections \ref{subsec:turb} and \ref{subsec:levelturb}, the collision process (and its
parameters) is clearly the dominant physical mechanism in shaping the appearance of the cloud in the
simulation outcome. It is possible that this collision of sub-clouds and with the cloud's
self-gravity, are the dominant processes of the cloud evolution, even over the tidal interaction
with the massive center, and above all, for the small scale of the circular velocity that is induced, 
as compared to the magnitude of the other velocities involved, which are the turbulent and 
the translational velocities, see Section \ref{subs:Vcir}.

For models $U5$, $U9$, and $U13$, the geometry of the resulting configuration can be
well characterized by defining a center and a radius of a cloudlet. For model $U11$, this spherical
structure does not make sense, as can be seen in Figs. \ref{Mosps} and \ref{VisMoseps}. The outcome of model
$U11$ is a complex, structured molecular gas cloud that exhibits an interconnected network of components. This is
the only model that can be compared or approximated to the cloud configuration called the Brick.

In fact, for the Brick, a shell-like structure with radius of $1.3$ pc has been revealed
from observations in the integrated intensity map of SO. For instance, see Figs.1 and 3
of \citet{longmore}; Figs.1 and 3 of \citet{kauffmann}; Figs. 1 and 2 of \citet{higuchi}. \citet{Kruijssen19}
presented three panels in their Figure 6, to compare the results of ALMA observations
of the Brick to a synthetic observations obtained from a numerical simulation. A complex gas structure can
be seen in these panels, in which the gas condenses in a persistent diagonal direction with many twisted and
bending filaments connected in a messy way. Model $U11$ of this paper clearly shows a similar diagonal
direction of the dense gas.

It may seem that there is a huge problem with models $Ur$, given that all of the different structures
obtained as a result of the collision process in model $U$ are destroyed because of the azimuthal
velocity of models $Ur$. However, the configuration obtained in these models $Ur$ can be well recognized as
the final outcome of the formation process of  a YMC, which are observed
to be a strongly centrally condensation of gas, with a mass enclosed of stars about
$10^{4}\, M_{\odot}$ in a size of one pc, see for instance \citet{rathborne}. The Arches cloud is an 
example of this kind of observed configuration, see \citet{por}.

%%%%%%%%%%%%%%%%%%%%%%%%%%%%%%%%%%%%%%%%%%%%%%%%%%%%%%%%%%%%%%%%%%%%%%%%%%%%%%%%%%%%%%%%%%%%%%%
%%%%%%%%%%%%%%%%%%%%%%%%%%%%%%%%%%%%%%%%%%%%%%%%%%%%%%%%%%%%%%%%%%%%%%%%%%%%%%%%%%%%%%%%%%%%%%%
\section{Concluding remarks}
\label{sec:conclu}

We examined models with three kinds of velocities, which are turbulent, translational
and azimuthal. These velocity kinds were introduced as initial conditions of
the simulation particles. Then, the particles are left to evolve as a self-gravitating gas
by using the public hydrodynamic code Gadget2.

The role played by these velocities determines the subsequent evolution of the cloud.
It must be emphasized that all of the models that are considered in this paper include the same
turbulent velocity spectrum (calibrated to fix the initial energies and physical
properties which favors the global collapse of the cloud) and the same translational
velocity (which produces the collision between two dissimilar sub-clouds).

In models $U$ with a low level of turbulence, we observed the coalescence of the
sub-clouds, enriched by the asymmetry in radii and translational velocities of the
sub-clouds. When the impact parameter was introduced, the model
produced a binary system with interconnected arms and with a complex structure. In models $Ub$, with a high
level of turbulence, we obtained a similar structure to that observed in models $U$. However, in model $Ub$, the arms
and tails are larger than those of models $U$. The most significant change between these simulations was
observed in models $U11$ and $U11b$, such that the double bridge of gas found in model $U11$ becomes a single
bridge in model $U11b$.

The free parameters of these models $U$ and $Ub$ (i.e., the impact parameter, the radii 
and the translational velocities of the sub-clouds) have been kept fixed. If these parameters were allowed
to vary, then it will certainly produce more interesting configuration outcomes.

In addition to the turbulent and the translational velocities, the last
models, $Ur$ and $Urb$, also include a low and high azimuthal velocity, respectively. The purpose of
this azimuthal velocity was to mimic, at least initially, the effect of the tidal force on the cloud.
The magnitude of the azimuthal velocity induced in the cloud depends explicitly on the
distance $R$ from the cloud and the mass $M$ of the gravitational center by means of the circular
velocity $\sqrt{2\, G \, M(R) / R}$.

We observed that the presence of this azimuthal velocity in the simulation always induces a
centrally located lump of gas in the cloud. If $V_{\rm cir}$ is small, then the collision process of the
sub-clouds dominates the dynamics of the cloud, even over the turbulence and therefore the cloud evolution
changes little of the that observed without the azimuthal velocity. However, if $V_{\rm cir}$ is large
compared with the other velocities involved, then the cloud evolution changes significantly: reducing
too much the collapsing time, suppressing any sign of the collision of the sub-clouds, and producing a central
condensation,  so that the different structures obtained as a result of the collision process are
destroyed. In fact, we observed that the cloud collapses faster when the azimuthal velocity is larger. 

We recall that this approximation is only valid when the ratio between the cloud radius to the distance to the
gravitational center is quite small; in other words, only for spatially compact clouds. For this kind of
cloud, the observation described above is in good agreement with the known fact that clouds in the CMZ
are observed to be denser than clouds in the ISM. Furthermore, because gravity is a ubiquitous force,
this azimuthal velocity approximation allows us to explain why centrally condensed clouds are
more abundant that uniform clouds in the ISM, see \citet{ward} and \citet{andre}.

In addition to information on shapes, we have also provided information about the physical properties
of the final collapse products and their surrounding region, which include the density, mass
and velocity profile. As mentioned in Section \ref{subs:applica}, we have found
proto-cluster structures that are still in their formation process. Furthermore, by the mass scale and the
radius of the resulting centrally condensed configurations, the outcomes of models $Ur$ and $Urb$ can be
identified with the final process of the formation of a young massive proto-cluster. Basically, the models 
show a strong flow of particles towards the
cloud center, at different radial velocities ( a few Mach ) and
with some bending trajectories.
%%%%%%%%%%%%%%%%%%%%%%%%%%%%%%%%%%%%%%%%%%%%%%%%%%%%%%%%%%%%%%%%%%%%%%%%%%%%%%%%%%%%%%%%%%%%%%%%%%%
\begin{acknowledgements}
The authors gratefully acknowledges the computer resources, technical expertise, and
support provided by the Laboratorio Nacional de Superc\'omputo del Sureste de M\'exico
through grant number 202201010N.
\end{acknowledgements}
%%%%%%%%%%%%%%%%%%%%%%%%%%%%%%%%%%%%%%%%%%%%%%%%%%%%%%%%%%%%%%%%%%%%%%%%%%%%%%%%%%%%%%%%%%%%%%%%%%%%%%%%%% 

\end{document}